\documentclass[a4paper,11pt]{article}
\pdfoutput=1 

\usepackage{jheppub} 
\usepackage{tikz}
\usepackage{dsfont}
\usepackage{extarrows}
\usepackage{hyperref}
\usepackage{cleveref}
\usepackage{supertabular}
\usepackage{todonotes}
\usepackage{mathrsfs}
\usepackage{array}
\usepackage{lipsum}
\usepackage{breqn}
\usepackage{physics}
\usepackage{tikz-cd}
\usepackage{slashed}
\usetikzlibrary{decorations.markings}

\tikzset{line/.style={line width=0.25mm},
curve/.style={line,smooth,tension=1},
->-/.style={decoration={
  markings,
  mark=at position #1 with {\arrow[>=stealth]{>}}},postaction={decorate}},
-<-/.style={decoration={
  markings,
  mark=at position #1 with {\arrow[>=stealth]{<}}},postaction={decorate}},
}

\usepackage{helvet} 
\usepackage{amsmath}
\usepackage{amssymb} 

\newcommand{\newreptheorem}[2]{%
\newenvironment{rep#1}[1]{%
 \def\rep@title{#2 \ref{##1}}%
 \begin{rep@theorem}}%
 {\end{rep@theorem}}}
\makeatother

\newreptheorem{lemma}{Lemma}

\newreptheorem{conj}{Conjecture}

\usepackage{lscape} 
\usepackage{braket}

\usepackage[T1]{fontenc} 

\newcommand{\Hom}{{\text{Hom}}}

\newcommand{\be}{\begin{equation}}
\newcommand{\ee}{\end{equation}}
\newcommand{\ba}{\begin{aligned}}
\newcommand{\ea}{\end{aligned}}
\newcommand{\bea}{\begin{eqnarray}}
\newcommand{\eea}{\end{eqnarray}}

\def\bp{\begin{pmatrix}}
\def\ep{\end{pmatrix}}

\def\bbR{\mathbb{R}}

\def\bbZ{\mathbb{Z}}

\def\Hom{\textrm{Hom}}

\title{\boldmath Flat Gauging of Continuous (Non-invertible) Symmetries and Non-compact BF SymTFT for Compact Boson}

\author[a]{Qiang Jia}
\author[b]{and Yi Zhang}
\affiliation[a]{Department of Physics, Korea Advanced Institute of Science and Technology, \\
Daejeon 34141, Korea}
\affiliation[b]{Kavli IPMU (WPI), UTIAS, The University of Tokyo, \\
Kashiwa, Chiba 277-8583, Japan}

\abstract{
We study flat gauging of continuous symmetries by summing over flat gauge-field configurations. We focus on the two-dimensional compact boson and construct the torus partition function with general flat $U(1)_M\times U(1)_W$ backgrounds. We show that flat gauging either $U(1)_M$ or $U(1)_W$ decompactifies the theory to the non-compact free boson, and that the dual $\mathbb{Z}$ background combines with the remaining $U(1)$ background into a non-compact $\mathbb{R}$ symmetry background due to the mixed anomaly. We also revisit the self-dual radius, where flat gauging the diagonal $SO(3)\subset (SU(2)_L\times SU(2)_R)/\mathbb{Z}_2$, first pointed out by Gaberdiel and Suchanek, gives the continuous orbifold which lies outside the usual $c=1$ moduli space. On the orbifold branch, we study finite and continuous non-invertible flat gaugings and explain why the continuous case requires a prescription for zero-measure fixed loci on the moduli space. Finally, we formulate the SymTFT of torus sigma models as a non-compact BF theory, whose topological boundary states encode the Narain moduli space and the $O(D,D;\mathbb{Z})$ T-duality action.
}

\begin{document}
\maketitle
\flushbottom

\section{Introduction}

Global symmetries are among the most important structures in quantum field theory (QFT). They constrain the dynamics and organize the spectrum of local operators and Hilbert spaces. From a modern point of view, global symmetries can also be described by topological defects supported on submanifolds of spacetime. Depending on the choice of submanifolds, the same defect can be viewed either as an operator acting on the Hilbert space, when it extends along the spatial direction, or as a defect background defining a twisted Hilbert space, when it extends along the temporal direction. In this paper, we usually refer to both descriptions as topological defects. This description naturally generalizes the notion of symmetry to higher-form symmetries, higher-group/categorical symmetries, non-invertible symmetries, subsystem symmetries, and so on~\cite{Gaiotto:2014kfa,Kong:2015flk,Cordova:2018cvg,Benini:2018reh,Ji:2019jhk,Kong:2020cie,Choi:2021kmx,Kaidi:2021xfk,Bhardwaj:2022yxj,Choi:2022zal,Antinucci:2022eat,Kaidi:2022uux,Cao:2023doz,Schafer-Nameki:2023jdn,Bhardwaj:Lecture,Luo:2023ive,Shao:2023gho}. When certain topological defects are preserved under the renormalization group (RG) flow, they will survive in the infrared and therefore provide useful tools to study infrared physics~\cite{Gaiotto:2012np,Chang:2018iay,Komargodski:2020mxz,Tavares:2024vtu,Cordova:2024goh}. For example, in two-dimensional rational conformal field theory (CFT), the Verlinde lines generate a non-invertible symmetry, and the primary operators are charged under them~\cite{Verlinde:1988te}. If the CFT is deformed by relevant operators and flows to a gapped phase, the surviving topological lines, namely those commuting with the relevant deformation, put strong constraints on the gapped vacuum~\cite{Chang:2018iay} and spectrum of particle and soliton excitations~\cite{Cordova:2024vsq,Cordova:2024iti,Chen:2025qub} and $S$-matrix~\cite{Copetti:2024rqj,Copetti:2024dcz} in the infrared.

A symmetry can be studied by coupling the theory to background gauge fields. For a finite group symmetry, the background fields are necessarily flat and characterized by their holonomies around loops. Equivalently, after triangulating spacetime, the background can be represented by a compatible network of codimension-one topological defects on the dual triangulation, which we call a \emph{topological defect network}. For a finite non-invertible symmetry, the defect language is even more useful, since there is in general no ordinary gauge-field description. Gauging a finite symmetry is then realized by summing over topologically inequivalent defect networks with the proper weights~\cite{Bhardwaj:2017xup,Tachikawa:2017gyf,Gaiotto:2020iye,Diatlyk:2023fwf,PerezLona:2023gauging,PerezLona:2024gauging,Yu:2025subfactors}. This is the sense in which the usual orbifold construction can be formulated purely in terms of topological defects.

For a continuous group-like symmetry, such as a Lie group $G$, the defect network prescription only describes flat gauge backgrounds. Indeed, the corresponding transition functions defining the $G$-bundle are taken to be constant functions with values in $G$.
Therefore gauging by summing over defect networks is in general inequivalent to the usual dynamical gauging of a continuous symmetry. Instead, one may restrict the gauge field to be flat and sum, or integrate, only over the moduli space of flat connections. We will call this operation continuous orbifolding, or equivalently \emph{flat gauging}~\cite{Gaberdiel:2011aa,Jia:2025vrj,Jia:2026jmt,HoTatLam_toappear_1,HoTatLam_toappear_2}. For a continuous non-invertible symmetry~\cite{Bhardwaj:2022yxj,GarciaEtxebarria:2022jky,Antinucci:2022eat,Damia:2023gtc,Hsin:2024aqb,Hsin:2025ria,Delmastro:2025ksn}, the analogous operation should be formulated as an integral over topologically inequivalent defect networks. 

A subtle point is that the partition function need not be continuous on the moduli space of flat connections, or more generally on the moduli space of topologically inequivalent defect networks. Thus the naive integral can discard zero-measure loci even when they carry physical data, leading to an incorrect gauged partition function. This issue appears in two examples below: the flat gauging of the diagonal $SO(3)$ symmetry of the $c=1$ WZW model, and the flat gauging of continuous non-invertible symmetry on the orbifold branch of the compact boson.

Flat gauging is simple enough to be performed exactly in many examples, but it can still change the theory in a nontrivial way. For example, Gaberdiel and Suchanek studied the flat gauging of $N-1$ free bosons by the diagonal continuous group $SU(N)/\mathbb{Z}_N$~\cite{Gaberdiel:2011aa}. In the special case $N=2$, the resulting theory lies outside the usual $c=1$ moduli space of circle and orbifold theories~\cite{Ginsparg:1987eb, Ginsparg:1988ui}. This example shows that flat gauging can produce genuinely new CFTs, and we will revisit it below.

The main purpose of this paper is to study flat gauging of continuous symmetries in two-dimensional conformal field theory, mainly at the level of torus partition functions. Our basic example is the 2d compact boson, which is useful because almost everything can be written explicitly. At a generic radius, the compact boson has two continuous symmetries, the momentum symmetry $U(1)_M$ and the winding symmetry $U(1)_W$, with a mixed 't Hooft anomaly between them. As a result, the torus partition function in the presence of flat $U(1)_M\times U(1)_W$ backgrounds is not an ordinary function of the holonomies. Rather, it is a section of a line bundle over the space of flat backgrounds. We will construct this background partition function explicitly and check its anomalous transformation laws and modular transformations.

After this preparation, we flat gauge one of the two continuous $U(1)$ symmetries on the circle branch, say the momentum symmetry $U(1)_M$, as an illustration, and turn on the background of the dual $\mathbb{Z}$ symmetry. We show that the dual $\mathbb{Z}$ symmetry background combines with the background gauge field for the remaining $U(1)_W$ symmetry into a non-compact $\mathbb{R}_W$ gauge field, due to the mixed anomaly between $U(1)_M$ and $U(1)_W$. Therefore the $U(1)_M$ flat gauging decompactifies the compact boson, with $\mathbb{R}_W$ interpreted as the translation symmetry of the dual boson. The flat gauging of $U(1)_W$ is obtained by exchanging momentum and winding.

Flat gauging does not always simply give a non-compact boson. At the self-dual radius, the compact boson is equivalent to the $\widehat{\mathfrak{su}(2)}_1$ WZW model, and the symmetry is enhanced to $(SU(2)_L\times SU(2)_R)/\mathbb{Z}_2$. We revisit the flat gauging of the diagonal $SO(3)$ subgroup by separating the untwisted and twisted sectors. After integrating over the continuous holonomy, the twisted-sector contribution looks very similar to that of the orbifolded dual non-compact boson, but the full theories differ because of the surviving untwisted sector. As discussed in~\cite{Gaberdiel:2011aa}, the twisted and untwisted sectors are related to the Runkel-Watts limit~\cite{Runkel:2001ng} and the Roggenkamp-Wendland limit~\cite{Roggenkamp:2003qp} of Virasoro minimal models, respectively.

We also study the orbifold branch of the compact boson, obtained by gauging the $\bbZ_2$-reflection symmetry $C:X\rightarrow -X$ of the compact boson field $X$~\cite{Ginsparg:1988ui}. The continuous $U(1)_M\times U(1)_W$ symmetry defects on the circle branch do not remain invertible after the orbifolding. Instead, they become continuous families of non-invertible defects $L_M\times L_W$. We first discuss finite subcategories of the continuous non-invertible symmetry $L_M$, and write the partition functions with a generic background associated to a topological defect network. We then study the gauging of these finite non-invertible symmetries at the level of torus partition functions, which changes the radius of the orbifold theory.

The flat gauging of continuous non-invertible symmetry on the orbifold branch is more subtle. If we first flat gauge $U(1)_M$ on the circle branch and then gauge the reflection $C$, the answer should be the orbifolded (dual) non-compact boson $\mathbb{R}/\mathbb{Z}_2$, which has a single fixed point at the origin. However, if we start directly from the orbifold branch and naively replace the finite sum over simple non-invertible lines by an integral, the theta-function terms contributed by the fixed points are supported on loci of zero measure on the moduli space and are lost. Moreover, taking the continuous limit of the finite non-invertible symmetry gauging gives another answer, since at finite radius there are always two $C$-fixed points instead of one. We will compare these procedures carefully. We will also introduce a resolution parameter for the zero-measure loci on the moduli space. One value reproduces the continuous limit of the finite non-invertible gauging, while another value reproduces the non-compact orbifold partition function. We leave the intrinsic definition of the continuous non-invertible measure, including the treatment of these zero-measure fixed loci, as an open question.

Finally, we discuss the Symmetry Topological Field Theory (SymTFT) interpretation. The SymTFT is a topological field theory in one higher dimension whose boundary conditions and topological operators encode the symmetry, anomaly, gauging, and duality data of the physical theory~\cite{Witten:1998wy,Apruzzi:2021nmk,Bhardwaj:2022yxj,Kaidi:2022cpf,Kaidi:2023maf,Bhardwaj:2023ayw,Apruzzi:2023uma,Cao:2023rrb,Antinucci:2024zjp,Brennan:2024fgj,Bonetti:2024cjk,Bhardwaj:2024igy,Choi:2024tri,Chen:2024ulc,Gagliano:2024off,Argurio:2024ewp,Jia:2025jmn,Jia:2025vrj,Delmastro:2025ksn,Bonetti:2025dvm,Jia:2025uun,Jia:2026vcr,Jia:2026jmt}. For the compact boson, the relevant three-dimensional topological theory is a non-compact BF theory. Its line operators are labeled by real charges, and different compactification radii are encoded by different Lagrangian algebras, or equivalently by different topological boundary conditions~\cite{Benini:2022hzx,Argurio:2024ewp}. From this point of view, changing the radius is changing the topological boundary. The decompactified theory is described by a different topological boundary, where one condenses a continuous family of lines. We then generalize this construction to $D$ compact bosons. The Lagrangian algebra is determined by the target-space metric $G_{IJ}$ and the antisymmetric $B$-field $B_{IJ}$, and the overlap between the topological boundary state and the physical boundary state reproduces the Narain partition function. The redundancy of this parametrization gives the Narain moduli space and the action of the $T$-duality group.

Compared with previous studies of continuous orbifolds, the Runkel-Watts theory, the symmetry structure of the $c=1$ moduli space, finite non-invertible gauging, and the SymTFT description of compact bosons~\cite{Gaberdiel:2011aa,Runkel:2001ng,Chang:2020imq,Thorngren:2021yso,Diatlyk:2023fwf,PerezLona:2023gauging,PerezLona:2024gauging,Yu:2025subfactors,Argurio:2024ewp,Damia:2024xju,Bharadwaj:2024gpj}, our emphasis is on flat gauging of continuous symmetry as an operation on torus partition functions with general defect backgrounds. This viewpoint lets us track how gauging reorganizes the remaining symmetry and how the same operation is represented in the non-compact BF SymTFT.

The organization of the paper is as follows. In section 2, we review the theory of compact boson, emphasizing the mixed anomaly between $U(1)_M$ and $U(1)_W$, and construct the partition function with general flat $U(1)_M\times U(1)_W$ backgrounds. In section 3, we flat gauge one of the two $U(1)$ symmetries, namely $U(1)_M$, and obtain the (dual) non-compact boson, together with its non-compact $\mathbb{R}_W$ symmetry background. In section 4, we revisit the flat gauging of the diagonal $SO(3)$ symmetry at the self-dual radius. In section 5, we turn to the orbifold branch and discuss finite and continuous non-invertible flat gaugings. In section 6, we formulate the corresponding SymTFT in terms of non-compact BF theory and explain how the Narain moduli space is encoded by topological boundary states. Some details of modular transformations are collected in the appendix.

\paragraph{Note added:} While this work was being completed, the independent work~\cite{Yu:2026gdf} appeared, with overlap with our SymTFT discussion of compact bosons.

\section{Compact boson with general flat $U(1)_M\times U(1)_W$ backgrounds}

This section aims to provide the basic setup for the remaining discussion in the paper. We will first review the theory of a 2D non-compact and compact free real boson $X$. The theory has a $U(1)_M\times U(1)_W$ global symmetry with a mixed 't Hooft anomaly. The moduli space of the flat connection $\mathcal{M}$ is isomorphic to $T^4$, parametrized by the $U(1)$-valued holonomies of $U(1)_M$ and $U(1)_W$ along the two 1-cycles of the torus. However, due to the mixed anomaly, the torus partition function $Z$ is not a single-valued function on the moduli space $\mathcal{M}$, but rather a section of the line bundle over $\mathcal{M}$ twisted by the anomaly. We will write down the torus partition function $Z[(\alpha,\beta),(\mu,\nu)]$ with a general flat $U(1)_M$ background $(\alpha,\beta)$ and $U(1)_W$ background $(\mu,\nu)$, satisfying the proper twisted boundary condition on the moduli space $\mathcal{M}$. We then check that this partition function is covariant under modular transformations.

Let us begin with a real free boson with action~\cite{DiFrancesco:1997nk}
    \begin{equation}
        S= \frac{g}{2}\int dt dx\, \partial_{\mu} X \partial^{\mu} X \,,
    \end{equation}
where we keep the coupling constant $g$ explicitly. If $X\in \mathbb{R}$ is non-compact, the torus partition function is
    \begin{equation}
        Z(\tau) = \frac{\text{Vol}(X)}{\sqrt{2\pi g \tau_2} |\eta(\tau)|^2} \,,
    \end{equation}
where $\tau$ is the complex structure of the worldsheet torus. Here $\text{Vol}(X)$ is the infinite volume of the target space, contributed by the zero mode $\int dX_0$ of the path integral. One can either regularize the partition function by removing the zero mode contribution, or introduce a cut-off length $\Lambda$ so that $\text{Vol}(X)=\Lambda$. For a compact boson with radius $r$, we impose
    \begin{equation}
        X \sim X+2\pi r\,,
    \end{equation}
On a spatial worldsheet circle of length $L$, the Hilbert space decomposes into sectors labeled by momentum and winding numbers. In our convention, the mode expansion is
    \begin{equation}
        X(x,t) = X_0 + \frac{m}{g rL}t + \frac{2\pi rn}{L}x +\frac{i}{\sqrt{4\pi g}}\sum_{k\neq 0}\frac{1}{k}\left(\hat{a}_ke^{2\pi i k(x-t)/L} -\hat{\bar{a}}_{-k} e^{2\pi i k(x+t)/L} \right)\,,
    \end{equation}
so that $X(x+L,t)=X(x,t)+2\pi n r$. Here $m,n\in \mathbb{Z}$ are the momentum and winding numbers, and the torus partition function is
    \begin{equation}
        Z_R(\tau,\bar{\tau}) = \textrm{Tr}\, q^{L_0-\frac{1}{24}} \bar{q}^{\overline{L}_0-\frac{1}{24}} = \frac{1}{|\eta(\tau)|^2} \sum_{m,n\in \mathbb{Z}} q^{2\pi g \left(\frac{m}{4\pi g r} +\frac{1}{2}nr \right)^2} \bar{q}^{2\pi g \left(\frac{m}{4\pi g r} -\frac{1}{2}n r \right)^2}\,,
    \end{equation}
where $q=e^{2\pi i \tau}$ and $\eta(\tau)$ is the Dedekind $\eta$-function.

The compact boson has two continuous $U(1)$ symmetries. The first one is the momentum symmetry $U(1)_M$, which shifts the scalar $X$ by a constant. Its charge is
    \begin{equation}
        Q_M = g\int_{0}^L dx\, \partial_t X = \frac{m}{r}\,.
    \end{equation}
The second one is the winding symmetry $U(1)_W$, which shifts the dual scalar $X^{\vee}$ related by $d X^{\vee} \sim *_2dX$. The corresponding charge is the winding number
    \begin{equation}
        Q_W = \frac{1}{2\pi} \int_0^{L} dx\, \partial_x X = nr\,.
    \end{equation}
Thus the physical charge lattices are $\frac{1}{r} \mathbb{Z}$ and $r\mathbb{Z}$. 

In the following, we will also use $\widetilde{m}=m/r$ and $\widetilde{n}=nr$ to label the physical momentum and winding charges, which depend on the radius $r$, in order to distinguish them from the integer-valued momentum and winding numbers $m,n$. Moreover, we will choose the convention $\pi g=1$ so that the partition function reads
    \begin{equation}\label{eq:compact-boson-partition-function}
        Z_r(\tau,\bar{\tau}) = \frac{1}{|\eta(\tau)|^2} \sum_{\widetilde{m}\in \mathbb{Z}/r,\widetilde{n}\in \mathbb{Z} r} q^{\frac{1}{4} p_L^2} \bar{q}^{\frac{1}{4} p_R^2}=\frac{1}{|\eta(\tau)|^2} \sum_{\widetilde{m}\in \mathbb{Z}/r,\widetilde{n}\in \mathbb{Z} r} q^{\frac{1}{2} \left(\frac{\widetilde{m}}{2} +\widetilde{n} \right)^2} \bar{q}^{\frac{1}{2} \left(\frac{\widetilde{m}}{2} -\widetilde{n} \right)^2}\,,
    \end{equation}
with
    \begin{equation}\label{eq:left_right_momenta_1}
        p_L = \sqrt{2}\widetilde{n}+\frac{\widetilde{m}}{\sqrt{2}}\,,\quad p_R = \sqrt{2}\widetilde{n}-\frac{\widetilde{m}}{\sqrt{2}}\,.
    \end{equation}
We will use the integer-quantized charges $n,m$ and physical charges $\widetilde{n},\widetilde{m}$ interchangeably.

\subsection{Mixed anomaly between $U(1)_M$ and $U(1)_W$}

The two continuous symmetries are not completely independent in the presence of background fields, and there is a mixed anomaly between them. The anomaly polynomial is given by
    \begin{equation}
         I^{(0)}_4=\int F_{M} \wedge F_W\,,
    \end{equation}
where $F_M,F_W$ are the field strengths of the $U(1)_M$ and $U(1)_W$ symmetries, and we normalize the field strength as $\int_{\Sigma_2} F \in \mathbb{Z}$ for any closed 2-cycle $\Sigma_2$.

In the following, we will evaluate the twisted boundary conditions on $\mathcal{M}$ of the torus partition function using the 3D mapping cylinder $T^2 \times [0,1]$. To begin with, let us denote the temporal and spatial directions of the worldsheet torus as $x_1,x_2\in [0,1)$, and the interval of the 3D mapping cylinder is parametrized by $\tau\in [0,1]$. Suppose we turn on the general flat $U(1)_M\times U(1)_W$ background on the worldsheet as
\begin{equation}
    A^e_M = \alpha dx^1 + \beta dx^2\,, \quad A^e_W = \mu dx^1 + \nu dx^2\,,
\end{equation}
and the partition function is denoted as $Z[A^e_M,A^e_W]$. Under a gauge transformation so that $(A^e_M,A^e_W) \rightarrow (A^g_M,A^h_W)$, with $g,h$ arbitrary gauge functions for $U(1)_M$ and $U(1)_W$, the ratio of $Z[A^g_M,A^h_W]$ to $Z[A^e_M,A^e_W]$ can be obtained by evaluating the Chern-Simons integral on the mapping cylinder, namely
\begin{equation}\label{eq:mapping-cylinder}
    \frac{Z[A^g_M,A^h_W]}{Z[A^e_M,A^e_W]} = \exp \left( 2\pi i \int_{T^2 \times [0,1]} I^{(1)}_3 (A_M,A_W) \right)\,,
\end{equation}
where the Chern-Simons density $I^{(1)}_3$ satisfies $d I^{(1)}_3 = I^{(0)}_4$, and is chosen to be
    \begin{equation}
        I^{(1)}_3 (A_M,A_W) =  \frac{1}{2} A_M \wedge dA_W + \frac{1}{2} A_W \wedge d A_M\,,
    \end{equation}
which is written in a symmetric way. The gauge fields $A_M(\tau),A_W(\tau)$ in the bulk satisfy the boundary condition
    \begin{equation}
        \begin{gathered}
            A_M(0) = A_M^e\,, \quad A_W(0)=A_W^e\,,\\
            A_M(1)=A_M^g\,,\quad A_W(1)=A_W^h\,.
        \end{gathered}
    \end{equation}

In the present case, we are interested in the large gauge transformation shifting the holonomy of $U(1)_M$ and $U(1)_W$ by integers. For example, let us consider
    \begin{equation}
        g= e^{2\pi i x^1}\,, \quad h=1\,,
    \end{equation}
which shifts the holonomy of $A_M$ along the $x^1$ direction as
    \begin{equation}
        A_M^g = (\alpha+1) dx^1 + \beta dx^2\,, \quad A^h_W = \mu dx^1 + \nu dx^2\,.
    \end{equation}
We then build the connecting gauge fields $A_M^g(\tau),A_W^h(\tau)$ as
     \begin{equation}
        A_M(\tau) = (\alpha+\tau) dx^1 + \beta dx^2\,, \quad A_W(\tau) = \mu dx^1 + \nu dx^2\,,
    \end{equation}
so that the Chern-Simons integral on the mapping cylinder is evaluated to be
    \begin{equation}
        \int_{T^2 \times [0,1]} I^{(1)}_3 (A_M,A_W) = \frac{1}{2} \int_{T^2 \times [0,1]} \nu dx^2 \wedge d\tau \wedge dx^1 = \frac{1}{2}\nu\,,
    \end{equation}
where we take the convention of orientation such that $\int d\tau \wedge dx^1\wedge dx^2=1$. Then the partition function transform according to \eqref{eq:mapping-cylinder} as
    \begin{equation}\label{eq:moduli-boundary-condition-1}
        Z[(\alpha+1,\beta),(\mu,\nu)] = e^{\pi i \nu} Z[(\alpha,\beta),(\mu,\nu)]\,.
    \end{equation}
Similarly, we can apply the mapping cylinder method to evaluate the anomalous phase for other three holonomy variables $\beta,\mu,\nu$, and we have
    \begin{equation}\label{eq:moduli-boundary-condition-2}
        \begin{split}
            Z[(\alpha,\beta+1),(\mu,\nu)]\, =& \, e^{-\pi i \mu} Z[(\alpha,\beta),(\mu,\nu)]\,,\\
            Z[(\alpha,\beta),(\mu+1,\nu)]\, =& \, e^{\pi i \beta} Z[(\alpha,\beta),(\mu,\nu)]\,,\\
            Z[(\alpha,\beta),(\mu,\nu+1)] \, =& \, e^{-\pi i \alpha} Z[(\alpha,\beta),(\mu,\nu)]\,.
        \end{split}
    \end{equation}
We shall use the anomalous phase of the partition function derived above as a guiding property to write down the general torus partition function of the compact boson.

\subsection{Partition function with general flat background}

In this subsection, we will write down the torus partition function $Z[(\alpha,\beta),(\mu,\nu)]$ for a general background satisfying the boundary condition \eqref{eq:moduli-boundary-condition-1} and \eqref{eq:moduli-boundary-condition-2}. We will use the tilde version of holonomies similar to $\widetilde{m},\widetilde{n}$
    \begin{equation}
        \widetilde{\alpha} = r\alpha\,, \quad \widetilde{\beta} = r\beta\,, \quad \widetilde{\mu}=\frac{1}{r}\mu\,,\quad \widetilde{\nu}=\frac{1}{r}\nu\,,
    \end{equation}
with $\widetilde{\alpha},\widetilde{\beta} \in [0,r)$ and $\widetilde{\mu},\widetilde{\nu}\in[0,1/r)$. In the rest of the paper, we will often use the unit-period holonomies $\alpha,\beta,\mu,\nu$ and the radius-dependent holonomies $\widetilde{\alpha},\widetilde{\beta},\widetilde{\mu},\widetilde{\nu}$ interchangeably, with the above rescaling understood.

Begin with the partition function \eqref{eq:compact-boson-partition-function} with zero background field 
    \begin{equation}
        Z_r[(0,0),(0,0)] = \frac{1}{|\eta(\tau)|^2} \sum_{\widetilde{m}\in \mathbb{Z}/r,\widetilde{n}\in \mathbb{Z} r} q^{\frac{1}{2} \left(\frac{\widetilde{m}}{2} +\widetilde{n} \right)^2} \bar{q}^{\frac{1}{2} \left(\frac{\widetilde{m}}{2} -\widetilde{n} \right)^2}\,.
    \end{equation}
If we turn on the $U(1)_M\times U(1)_W$ holonomies along the temporal direction, or equivalently, inserting $U(1)_M\times U(1)_W$ symmetry generators along the spatial directions, the partition function is then
    \begin{equation}
        Z_r[(\widetilde{\alpha},0),(\widetilde{\mu},0)] = \textrm{Tr}_{\mathcal{H}} e^{2\pi i \widetilde{\alpha} Q_M + 2\pi i \widetilde{\mu} Q_W} q^{L_0-\frac{1}{24}} \bar{q}^{\overline{L}_0-\frac{1}{24}}\,,
    \end{equation}
therefore the whole tower of the sector $|\widetilde{n},\widetilde{m}\rangle$ with momentum and winding number $\widetilde{n}$ and $\widetilde{m}$ are multiplied by the phase factor $e^{2\pi i \widetilde{\alpha} \widetilde{m}+2\pi i \widetilde{\mu} \widetilde{n}}$, and we have
    \begin{equation}
        Z_r[(\widetilde{\alpha},0),(\widetilde{\mu},0)] = \frac{1}{|\eta(\tau)|^2} \sum_{\widetilde{m}\in \mathbb{Z}/r,\widetilde{n}\in \mathbb{Z} r} e^{2\pi i \widetilde{\alpha} \widetilde{m}+2\pi i \widetilde{\mu} \widetilde{n}} q^{\frac{1}{2} \left(\frac{\widetilde{m}}{2} +\widetilde{n} \right)^2} \bar{q}^{\frac{1}{2} \left(\frac{\widetilde{m}}{2} -\widetilde{n} \right)^2}\,.
    \end{equation}
Doing an modular $S$-transformation, we can also obtain the defect partition function
    \begin{equation}
        Z_r[(0,\widetilde{\beta}),(0,\widetilde{\nu})] = \frac{1}{|\eta(\tau)|^2} \sum_{\widetilde{m}\in \mathbb{Z}/r,\widetilde{n}\in \mathbb{Z} r} q^{\frac{1}{2} \left(\frac{\widetilde{m}+\widetilde{\nu}}{2} +\widetilde{n}+\widetilde{\beta} \right)^2} \bar{q}^{\frac{1}{2} \left(\frac{\widetilde{m}+\widetilde{\nu}}{2} -\widetilde{n}-\widetilde{\beta} \right)^2}\,.
    \end{equation}
The spatial holonomies of $U(1)_W$ and $U(1)_M$ twist the boundary conditions for $X^{\vee}$ and $X$, therefore the momentum and winding charges $\widetilde{m},\widetilde{n}$ are shifted $\widetilde{\nu},\widetilde{\beta}$, respectively. 

It is then straightforward to generalize the partition function to the general backgrounds as
    \begin{equation}\label{eq:partition-function-full-background-old}
    \begin{split}
        Z[(\widetilde{\alpha},\widetilde{\beta}),(\widetilde{\mu},\widetilde{\nu})] \equiv \frac{1}{\eta \bar{\eta}} &\sum_{n,m={-\infty}}^{\infty} e^{-\pi i (\widetilde{\alpha}\widetilde{\nu}+\widetilde{\mu} \widetilde{\beta})}e^{2\pi i \widetilde{\alpha} (\widetilde{m}+\widetilde{\nu})+2\pi i \widetilde{\mu}(\widetilde{n}+\widetilde{\beta})}\\ &\times  q^{\frac{1}{2}(\frac{(\widetilde{m}+\widetilde{\nu})}{2}+\widetilde{n}+\widetilde{\beta})^2}\bar{q}^{\frac{1}{2}(\frac{(\widetilde{m}+\widetilde{\nu})}{2}-\widetilde{n}-\widetilde{\beta})^2}\,,
    \end{split}
    \end{equation}
where $e^{2\pi i \widetilde{\alpha} (\widetilde{m}+\widetilde{\nu})+2\pi i \widetilde{\mu}(\widetilde{n}+\widetilde{\beta})}$ is the eigenvalues of the symmetry generators $e^{2\pi i \widetilde{\alpha} Q_M + 2\pi i \widetilde{\mu} Q_W}$ in the twisted sector, and the first phase factor $e^{-\pi i (\widetilde{\alpha}\widetilde{\nu}+\widetilde{\mu} \widetilde{\beta})}$ is determined by the requirement that the partition function transform properly according to boundary condition \eqref{eq:moduli-boundary-condition-1} and \eqref{eq:moduli-boundary-condition-2}. Indeed, one can check the partition function satisfies
    \begin{equation}
    \begin{split}
    Z[(\widetilde{\alpha}+r,\widetilde{\beta}),(\widetilde{\mu},\widetilde{\nu})]\, =\, &e^{\pi i \widetilde{\nu}r}Z[(\widetilde{\alpha},\widetilde{\beta}),(\widetilde{\mu},\widetilde{\nu})]\,, \\
    Z[(\widetilde{\alpha},\widetilde{\beta}),(\widetilde{\mu},\widetilde{\nu}+\frac{1}{r})]\, =\, &e^{-\pi i \widetilde{\alpha}/r}Z[(\widetilde{\alpha},\widetilde{\beta}),(\widetilde{\mu},\widetilde{\nu})]\,,\\
    Z[(\widetilde{\alpha},\widetilde{\beta}+r),(\widetilde{\mu},\widetilde{\nu})]\, =\, &e^{-\pi i \widetilde{\mu}r}Z[(\widetilde{\alpha},\widetilde{\beta}),(\widetilde{\mu},\widetilde{\nu})]\,, \\ Z[(\widetilde{\alpha},\widetilde{\beta}),(\widetilde{\mu}+\frac{1}{r},\widetilde{\nu})]\, =\, &e^{\pi i \widetilde{\beta}/r}Z[(\widetilde{\alpha},\widetilde{\beta}),(\widetilde{\mu},\widetilde{\nu})]\,,
    \end{split}
    \end{equation}
which reflects the projective phase due to the mixed anomaly between the two $U(1)$ symmetries. Moreover, with that addition phase factor, the partition function is modular covariant and satisfies
    \begin{equation}\label{eq:modular-covariance}
    \begin{split}
        Z[(\widetilde{\alpha},\widetilde{\beta}),(\widetilde{\mu},\widetilde{\nu})](\tau+1)=&Z[(\widetilde{\alpha}+\widetilde{\beta},\widetilde{\beta}),(\widetilde{\mu}+\widetilde{\nu},\widetilde{\nu})](\tau)\,,\\
        Z[(\widetilde{\alpha},\widetilde{\beta}),(\widetilde{\mu},\widetilde{\nu})](-1/\tau)=&Z[(\widetilde{\beta},-\widetilde{\alpha}),(\widetilde{\nu},-\widetilde{\mu})](\tau)\,,
    \end{split}
    \end{equation}
which are exactly the transformation of flat holonomies under the mapping class group of the torus. We leave the proof to the appendix.

\section{Flat gauging of continuous $U(1)$ symmetry and the non-compact free boson}

We now perform the flat gauging of $U(1)$ explicitly. By flat gauging we mean that we sum over the flat holonomies of the symmetry background. There are two points we should stress here. First, before summing over the holonomies, the background partition function is only projectively well-defined because of the mixed anomaly discussed in the previous section. Therefore, if we want to gauge one of the $U(1)$ symmetries, we should choose a representative of the line bundle on $\mathcal{M}$ such that this particular $U(1)$ acts without anomaly. Second, after gauging the $U(1)$, there exists a dual discrete symmetry $\mathbb{Z}$, which is the Fourier partner of $U(1)$. In the present case, the discrete background of $\mathbb{Z}$ will combine with the remaining $U(1)$ symmetry into a non-compact continuous background valued in $\mathbb{R}$, which is the translation symmetry of the non-compact scalar field $X$ or $X^{\vee}$. This is the precise sense in which the flat gauging decompactifies the compact boson.\footnote{For a related SymTFT discussion of the corresponding non-compact BF description, see~\cite{Yu:2026gdf}.}

\subsection{Flat gauging of continuous $U(1)$ symmetry}

Let us consider flat gauging $U(1)_M$ symmetry in the presence of a $U(1)_W$ background. We consider the gauged partition function with a dual background field as the Fourier transformation
    \begin{equation}\label{eq:gauge_U(1)_M}
        Z[(\widetilde{p},\widetilde{q}),(\widetilde{\mu},\widetilde{\nu})]=\frac{1}{|U(1)_M|}\int_{0}^{r}d\widetilde{\alpha} \int_0^{r} d\widetilde{\beta} e^{2\pi i (\widetilde{p} \widetilde{\beta}-\widetilde{q} \widetilde{\alpha})}Z[(\widetilde{\alpha},\widetilde{\beta}),(\widetilde{\mu},\widetilde{\nu})]\,.
    \end{equation}
Here $(\widetilde{p},\widetilde{q})\in \mathbb{Z}$ label the background for the dual $\mathbb{Z}$ symmetry, the exponential factor is the standard Fourier kernel, and $|U(1)_M|=r$ is the period of $U(1)_M$. Notice that unlike the discrete case, the continuous flat gauging is sensitive to the normalization of the volume of the symmetry group, and the result will differ by a constant if we choose a different normalization. In this section, we will normalize the momentum and winding $U(1)$ groups so that their corresponding charges are the physical charges $\widetilde{m},\widetilde{n}$ depending on $r$.

However, the integral in \eqref{eq:gauge_U(1)_M} is ill-defined since the partition function $Z[(\widetilde{\alpha},\widetilde{\beta}),(\widetilde{\mu},\widetilde{\nu})]$ is not a single-valued function of $\widetilde{\alpha}$ and $\widetilde{\beta}$. In order to gauge $U(1)_M$, we need a representative in which the large gauge transformations of $U(1)_M$ are non-anomalous. This can be achieved by adding the topological phase
    \begin{equation}
        e^{-\pi i \widetilde{\alpha} \widetilde{\nu} +\pi i \widetilde{\mu} \widetilde{\beta}}\,,
    \end{equation}
which corresponds to adding a local counterterm 
    \begin{equation}
        \Delta S = -\pi i \int A_M \wedge A_W\,,
    \end{equation}
to the worldsheet action. With this choice of local counterterm, the background partition function \eqref{eq:partition-function-full-background-old} becomes
\begin{equation}\label{eq:partition-function-full-background}
    Z[(\widetilde{\alpha},\widetilde{\beta}),(\widetilde{\mu},\widetilde{\nu})]=\frac{1}{\eta \bar{\eta}}\sum_{m,n} e^{2\pi i \alpha m+2\pi i \mu(n+\beta)} q^{\frac{1}{2}(\frac{(m+\nu)}{2r}+(n+\beta)r)^2}\bar{q}^{\frac{1}{2}(\frac{(m+\nu)}{2r}-(n+\beta)r)^2}\,,
\end{equation}
and one can check the new partition function is indeed a single-valued function of $\widetilde{\alpha},\widetilde{\beta}$ 
    \begin{equation}
    \begin{split}
    Z[(\widetilde{\alpha}+r,\widetilde{\beta}),(\widetilde{\mu},\widetilde{\nu})]\,=\,&Z[(\widetilde{\alpha},\widetilde{\beta}),(\widetilde{\mu},\widetilde{\nu})]\,, \\ Z[(\widetilde{\alpha},\widetilde{\beta}+r),(\widetilde{\mu},\widetilde{\nu})]\,=\,&Z[(\widetilde{\alpha},\widetilde{\beta}),(\widetilde{\mu},\widetilde{\nu})]\,,
    \end{split}
    \end{equation}
and the anomalous phases for $\widetilde{\mu},\widetilde{\nu}$ are doubled compared to \eqref{eq:moduli-boundary-condition-2}
    \begin{equation}
    \begin{split}
    Z[(\widetilde{\alpha},\widetilde{\beta}),(\widetilde{\mu}+1/r,\widetilde{\nu})]\,=\,&e^{2\pi i \widetilde{\beta}/r}Z[(\widetilde{\alpha},\widetilde{\beta}),(\widetilde{\mu},\widetilde{\nu})]\,,\\ Z[(\widetilde{\alpha},\widetilde{\beta}),(\widetilde{\mu},\widetilde{\nu}+1)]\,=\,&e^{-2\pi i \widetilde{\alpha}/r}Z[(\widetilde{\alpha},\widetilde{\beta}),(\widetilde{\mu},\widetilde{\nu})]\,.
    \end{split}
    \end{equation}
Thus the $U(1)_M$ holonomies can be integrated over, and we obtain
    \begin{equation}
    \begin{split}
        Z[(\widetilde{p},\widetilde{q}),(\widetilde{\mu},\widetilde{\nu})] =&r \int_{0}^1 d \alpha \int_0^1 d \beta e^{2\pi i (p \beta - q \alpha)} \frac{1}{\eta \bar{\eta}}\sum_{m,n} e^{2\pi i \alpha m+2\pi i \mu(n+\beta)} q^{\frac{1}{2}(\frac{(m+\nu)}{2r}+(n+\beta)r)^2}\bar{q}^{\frac{1}{2}(\frac{(m+\nu)}{2r}-(n+\beta)r)^2}\\
        =&r \int_0^1 d\beta e^{2\pi i p \beta} \frac{1}{\eta \bar{\eta}}\sum_{m,n} \delta_{m-q,0}e^{2\pi i \mu(n+\beta)} q^{\frac{1}{2}(\frac{(m+\nu)}{2r}+(n+\beta)r)^2}\bar{q}^{\frac{1}{2}(\frac{(m+\nu)}{2r}-(n+\beta)r)^2}\\
        =&r \int_0^1 d\beta e^{2\pi i p \beta} \frac{1}{\eta \bar{\eta}}\sum_n e^{2\pi i \mu(n+\beta)} q^{\frac{1}{2}(\frac{(q+\nu)}{2r}+(n+\beta)r)^2}\bar{q}^{\frac{1}{2}(\frac{(q+\nu)}{2r}-(n+\beta)r)^2}\,.
    \end{split}
    \end{equation}
where the exponential terms are collected as
    \begin{equation}
    \begin{split}
        &\exp\left[2\pi i \left(p \beta + (n+\beta)\mu+\frac{1}{2}\left(\tau - \bar{\tau} \right)\left(\frac{(q+\nu)^2}{4r^2}+(n+\beta)^2r^2 \right) +\frac{1}{2}(\tau+\bar{\tau})\left(\frac{2(q+\nu)}{2r} \right)(n+\beta)r \right) \right]\\
        =&\exp \left[2\pi i \left(i \tau_2 r^2(n+\beta)^2+(p+\mu+(q+\nu)\tau_1)(n+\beta)+\frac{i(q+\nu)^2}{4r^2}\tau_2 \right) \right]\,.
    \end{split}
    \end{equation}
Here the integral over $\alpha$ imposes the constraint $m=q$. This is the usual projection onto the $U(1)_M$-invariant states, but in the presence of the dual background, the projected momentum is shifted by the integer $q$. The remaining integral over $\beta$ is more interesting: it turns the shifted winding variable $n+\beta$ into a real variable. This is where the compact charge lattice is replaced by a continuum. 

Let us further write $\sum_{n}\int_0^1 d \beta$ as $\int_{-\infty}^{\infty} dx$ with $x = n+\beta$, and the integral becomes
    \begin{equation}\label{eq:partition_function_gauge_U1}
    \begin{split}
        Z[(\widetilde{p},\widetilde{q}),(\widetilde{\mu},\widetilde{\nu})]=& r \int_{-\infty}^{\infty} dx \frac{1}{\eta \bar{\eta}} e^{2\pi i \left(i \tau_2 r^2x^2+(p+\mu+(q+\nu)\tau_1)x+\frac{i(q+\nu)^2}{4r^2}\tau_2 \right)}\\
        =& r \int_{-\infty}^{\infty} dx \frac{1}{\eta \bar{\eta}}e^{-2\pi r^2 \tau_2 \left(x-\frac{2\pi i (p+\mu+(q+\nu)\tau_1)}{4\pi r^2 \tau_2} \right)^2-\frac{\pi}{2r^2 \tau_2}|p+\mu+(q+\nu)\tau|^2}\\
        =&\frac{1}{\eta \bar{\eta}\sqrt{2 \tau_2}}e^{-\frac{\pi}{2r^2 \tau_2}|(p+\mu)+(q+\nu)\tau|^2} = \frac{1}{\eta \bar{\eta}\sqrt{2 \tau_2}}e^{-\frac{\pi}{2 \tau_2}|x+\tau y|^2}\equiv Z_{\mathbb{R}_W}[x,y] \,,
    \end{split}
    \end{equation}
with $x=\frac{1}{r}(p+\mu)$ and $y=\frac{1}{r}(q+\nu)$, and both of them take values in $\mathbb{R}$. 

We see that the holonomies of the dual $\mathbb{Z}$ symmetry combine with the original $U(1)_W$ holonomies into the holonomies of a continuous non-compact $\mathbb{R}_W$ symmetry. Since $U(1)_W$ is the momentum symmetry for the dual compact boson, we obtain the partition function of the dual non-compact boson in a flat $\mathbb{R}_W$ background.

As mentioned before, for non-compact boson there exists non-compact zero mode $X_0$, which should be regularized by introducing a cutoff $\Lambda$. In the present case, by fixing the period of $U(1)_M$ and $U(1)_W$ to be $\frac{1}{r}$ and $r$, we have normalized the cutoff $L$ to be one. If we fix the period of $U(1)_M$ and $U(1)_W$ to be one instead, then we will get
    \begin{equation}
        \frac{1}{\eta \bar{\eta}r\sqrt{2 \tau_2}}e^{-\frac{\pi}{2 \tau_2}|x+\tau y|^2} = \frac{\sqrt{2}r'}{\eta \bar{\eta}\sqrt{ \tau_2}}e^{-\frac{\pi}{2 \tau_2}|x+\tau y|^2}\,,
    \end{equation}
where the cutoff is set to be proportional to the dual radius $r'=\frac{1}{2r}$.

On the other hand, if we flat gauge $U(1)_W$ symmetry, then the remaining $U(1)_M$ should be extended to the $\mathbb{R}_M$ instead, and the resulting theory is the (not dual) non-compact boson. Similarly, we can evaluate the partition function as
    \begin{equation}
        Z_{\mathbb{R}_M}[x',y']=\frac{1}{\eta \bar{\eta}}\sqrt{\frac{2}{\tau_2}}e^{-\frac{2\pi}{ \tau_2}|x'+\tau y'|^2} \quad \text{or} \quad \frac{\sqrt{2}r}{\eta \bar{\eta}\sqrt{\tau_2}}e^{-\frac{2\pi}{ \tau_2}|x'+\tau y'|^2}\,,
    \end{equation}
for the two normalizations. Compared to \eqref{eq:partition_function_gauge_U1}, the mismatch of the exponent is because we choose the dual radius $r'$ to satisfy $rr'=\frac{1}{2}$, so that the self-dual radius is $r=r'=\frac{1}{\sqrt{2}}$. Recall the momentum and winding charges are $\frac{m}{r}$ and $nr$, and in the dual theory their roles get exchanged so that the dual momentum charge is $\frac{n}{2r'}$. This still holds when we extend $U(1)_W$ to $\mathbb{R}_W$. After we rescale both $x',y'$ by a factor $2$ so that the dual momentum charge is quantized by $\frac{1}{r'}$, the exponent is exactly the same as that in \eqref{eq:partition_function_gauge_U1}.

In the rest of this paper, unless stated otherwise, we will mainly consider flat gauging operations associated with the momentum symmetry $U(1)_M$. Therefore the decompactified theory should be understood as the dual non-compact boson, whose translation symmetry is the non-compact symmetry $\mathbb{R}_W$.

\subsection{Dual symmetry from the group-extension point of view}
The computation for the compact boson above shows that, after flat gauging $U(1)_M$, the dual $\bbZ$ symmetry does not remain independent of the original $U(1)_W$ symmetry. Instead, the two combine into
\begin{equation}
    0\rightarrow \bbZ\rightarrow \bbR_W\rightarrow U(1)_W\rightarrow 0.
\end{equation}
The issue is not the existence of this extension itself, which is elementary,\footnote{
This is an extension of topological groups: $\bbZ$ has the discrete topology, while $\bbR$ and $U(1)$ are equipped with the standard topologies. The extension classes are isomorphic to $H^2(BU(1),\bbZ)$ and for compact boson the class is exactly $[\frac{F_W}{2\pi}]\in H^2(BU(1)_W,\bbZ) \cong \bbZ$.} but the mechanism by which the anomaly forces the dual background and the remaining background to combine after the gauging. Note also that in the previous compact boson computation, we turned on a flat background gauge field for $U(1)_W$, while the following analysis allows generic non-flat backgrounds.

Recall the $D=2$ finite-group discussion of~\cite{Tachikawa:2017gyf}. Suppose that a two-dimensional theory has finite symmetry group
\begin{equation}
    \Gamma=A\times G,
\end{equation}
where $A$ is abelian. Let $X$ be the spacetime and let $Y$ be a three-dimensional bulk with boundary $X$. We assume that the background fields on $X$ extend to $Y$. Consider the mixed anomaly
\begin{equation}
    \int_Y a_A\cup g^*(\hat e)\in \bbR/\bbZ,
    \label{eq:finitegroupanomaly}
\end{equation}
where $a_A$ is an $A$ gauge field, $g$ is a $G$ gauge field, and
\begin{equation}
    \hat e\in H^2(BG,\hat A),\qquad \hat A=\Hom(A,U(1)).
\end{equation}
In the finite case both backgrounds are flat and can be represented by defect networks. Gauging $A$ means summing over $A$ domain-wall networks. The mixed anomaly implies that the $G$ background sources the dual $\hat A$ background. Thus, after gauging $A$, a background is not an independent pair of a $\hat A$ background and a $G$ background, but a pair $(\hat a,g)$ satisfying
\begin{equation}
    \delta \hat a=g^*(\hat e).
    \label{eq:relationaftergauging}
\end{equation}
Equivalently, the symmetry after gauging is the extension
\begin{equation}
    0 \rightarrow \hat A \rightarrow \tilde \Gamma \rightarrow G \rightarrow 0,
    \label{eq:finitegroupsequence}
\end{equation}
with extension class $\hat e$.

For the compact boson we use the dictionary
\begin{equation}
    A=U(1)_M,\qquad G=U(1)_W,\qquad \hat A=\Hom(U(1)_M,U(1))=\bbZ.
\end{equation}
The finite-group domain-wall argument cannot be copied literally. In the present flat gauging, the $U(1)_M$ variable being summed over is flat and can be described by topological defect lines. The remaining $U(1)_W$ background is a general connection and may have curvature, so it is not described by a topological defect network. We therefore replace the domain-wall derivation by a direct analysis of the background gauge fields.

A $U(1)_W$ gauge field on $X$, and similarly after extension to $Y$, can be represented by the data\footnote{This is the Hopkins--Singer model~\cite{Hopkins:2002rd} for differential characters introduced by Cheeger and Simons~\cite{CheegerSimons}. See also the physics discussions in~\cite{Hsieh:2020jpj,GarciaEtxebarria:2024fuk}.}
\begin{equation}
    (n_W,a_W,\omega_W),
\end{equation}
where
\begin{equation}
    n_W\in C^2(X,\bbZ),\qquad
    a_W\in C^1(X,\bbR),\qquad
    \omega_W\in \Omega^2(X),
\end{equation}
satisfying\footnote{Here and below we suppress the standard map from differential forms to real cochains; thus $\omega_W$ denotes its image in $C^2(X,\bbR)$ whenever it appears in a cochain equation.}
\begin{equation}
    \delta n_W=0,\qquad
    d\omega_W=0,\qquad
    \delta a_W= \omega_W -n_W.
\end{equation}
The class $[n_W]\in H^2(X,\bbZ)$ is the first Chern class of the $U(1)_W$ bundle. The holonomy of this gauge field on a loop $\mathcal{C}$ is given by
\begin{equation}
    \exp 2\pi i \int_{\mathcal{C}} a_W \,.
\end{equation}

A flat $U(1)_M$ gauge field can be described by the reduced data
\begin{equation}
        (n_M,a_M, 0) \in C^2(X,\bbZ) \times C^1(X,\bbR) \times \Omega^2(X)\,,
\end{equation}
such that 
\begin{equation}
    \delta a_M = -n_M.
\end{equation}
This determines a closed cochain $\tilde a_M \in C^1(X,\bbR/\bbZ)$ satisfying $\delta\tilde a_M= 0$.

The continuous analogue of \eqref{eq:finitegroupanomaly} is the inflow term
\begin{equation}
   \exp\left(2 \pi i \int_Y \tilde a_M \cup n_W\right)\in U(1),
\end{equation}
using the assumed extensions to $Y$. In continuum notation this is
\begin{equation}
   \exp\left(2\pi i\int_Y A_M\wedge \frac{dA_W}{2\pi}\right).
    \label{eq:continuumanomaly}
\end{equation}

After gauging $U(1)_M$, the dual symmetry is $\widehat{U(1)_M}=\bbZ$. Let its background be
\begin{equation}
    \hat a\in C^1(X,\bbZ).
\end{equation}
In the absence of the mixed anomaly, $\hat a$ would be closed. To keep track of this dual background during the gauging, we include the Fourier pairing in the partition function,
\begin{equation}
   \exp 2 \pi i  \int_X \tilde a_M \cup \hat a.
\end{equation}
This boundary term gives (using $\delta \tilde a_M =0$)
\begin{equation}
    \int_X \tilde a_M\cup\hat a
    =
    -\int_Y \tilde a_M \cup\delta\hat a .
\end{equation}
Thus the total dependence on the flat $U(1)_M$ variable is
\begin{equation}
    \int_Y \tilde a_M\cup (n_W-\delta\hat a).
\end{equation}
Since the flat $U(1)_M$ gauge field is summed over, the allowed background data are constrained by
\begin{equation}
    \delta\hat a=n_W,
\end{equation}
for the Fourier pairing convention chosen above. This is the continuous replacement of the relation~\eqref{eq:relationaftergauging}. 

Now the combination of $U(1)_W$ and $\bbZ$ backgrounds is explicit. The post-gauging background data are
\begin{equation}
    (n_W,a_W,\omega_W;\hat a),
\end{equation}
with
\begin{equation}
    \delta a_W=\omega_W -n_W,\qquad
    \delta\hat a=n_W.
\end{equation}
Adding the two equations gives
\begin{equation}
    \delta(a_W+\hat a)= \omega_W.
\end{equation}
We define
\begin{equation}
    \tilde a :=a_W+\hat a\in C^1(X,\bbR)
\end{equation} with curvature represented by $\omega = \omega_W$:
\begin{equation}
    (\tilde a,\omega),
    \qquad
    \delta \tilde a = \omega.
\end{equation}
The relation $\delta\tilde a=\omega$ is stronger than simply restricting $U(1)_W$ to the topologically trivial sector $[n_W]=0\in H^2(X,\bbZ)$. Indeed, the cochain $\hat a$ is part of the background data and its coboundary trivializes the characteristic class of $U(1)_W$. In this sense, it lifts the compact $U(1)_W$ differential cocycle through the universal cover $\bbR_W\rightarrow U(1)_W$.

The gauge equivalences give the same conclusion. Changing the representative of the $U(1)_W$ field by
\begin{equation}
    n_W\mapsto n_W+\delta m,\qquad
    a_W\mapsto a_W-m+\delta f,
\end{equation}
with
\begin{equation}
    m\in C^1(X,\bbZ),\qquad f\in C^0(X,\bbR),
\end{equation}
must be accompanied by
\begin{equation}
    \hat a\mapsto \hat a+m+\delta\ell,
    \qquad
    \ell\in C^0(X,\bbZ),
\end{equation}
so that $\delta\hat a=n_W$ is preserved. Then
\begin{equation}
    \tilde a=a_W+\hat a
\end{equation}
transforms as
\begin{equation}
    \tilde a\mapsto \tilde a+\delta(f+\ell),
\end{equation}
which is the ordinary $\bbR$ gauge transformation. Hence the data
\begin{equation}
    \left((n_W,a_W,\omega_W),\hat a\right),
    \qquad
    \delta a_W= \omega_W -n_W,\qquad
    \delta\hat a=n_W,
\end{equation}
modulo these equivalences, is exactly the data of an $\bbR_W$ gauge field $\tilde a$.

The mixed anomaly has therefore been absorbed into the constraint $\delta\hat a=n_W$. Once the background is written in terms of the single $\bbR_W$ gauge field $\tilde a$, there is no residual anomaly coming from this mixed term~\eqref{eq:continuumanomaly}. A separate local inflow term for the original $U(1)_W$ symmetry, e. g. $2\pi i\int_Y A_W\wedge \frac{dA_W}{2\pi},$ if present in other cases, would have to be treated separately.

 For the compact boson on the torus, this reduces to the holonomy statement already visible in the partition function: a $U(1)_W$ holonomy $\mu\in \bbR/\bbZ$ and a dual $\bbZ$ holonomy $p\in\bbZ$ enter only through the real lift $p+\mu\in\bbR$, with the large-gauge identification $\mu\mapsto\mu+1$, $p\mapsto p-1$.

\subsection{Matching the vertex operators}

The match of partition functions is still not sufficient to conclude the gauged theory is the non-compact boson. In this section, we will also examine the vertex operators of the gauged theory, and see an explicit match with those of the non-compact boson. 

Consider the vertex operators in the compact boson
    \begin{equation}
        V_{m,n}(z,\bar{z}) =\, : \exp \left(i\sqrt{2}p_LX_L(z)+i\sqrt{2}p_RX_R(z) \right):\,,
    \end{equation}
where $:\cdots:$ is the normal ordering, and the left and right moving momenta are
    \begin{equation}
        p_L= \sqrt{2}rn+\frac{m}{\sqrt{2}r}\,, \quad p_R = \sqrt{2}rn-\frac{m}{\sqrt{2}r}\,,
    \end{equation}
with the conformal weights
    \begin{equation}
        h_{n,m} = \frac{1}{4} p_L^2\,,\quad \bar{h}_{n,m}=\frac{1}{4}p_R^2\,.
    \end{equation}
The OPE between vertex operators are
    \begin{equation}
        V_{m,n}(z,\bar{z}) V_{m',n'}(0,0)\sim z^{\frac{1}{2}p_L p'_L} \bar{z}^{\frac{1}{2}p_R p'_R} V_{m+m',n+n'}(0)+ \cdots\,.
    \end{equation}
The branch cut in this OPE records the mutual locality of two vertex operators. For integral $m,n$ and $m',n'$, the two vertex operators $V_{m,n}$ and $V_{m',n'}$ are mutually local. However, once a defect line is attached to one of the operator, the branch cut will remember the holonomy caused by the defect, as we will see soon.

When gauging the $U(1)_M$ symmetry, we should also include sectors with a $U(1)_M$ defect attached to the local operator. This shifts the winding number and gives the defect operator
    \begin{equation}
        V_{m,n+\frac{\theta}{2\pi}} = \, : \exp \left( i\left(\frac{m}{r}+2 r \left(n+\frac{\theta}{2\pi}\right) \right)X_L(z) + i\left(-\frac{m}{r}+2 r \left(n+\frac{\theta}{2\pi}\right)  \right)X_R(z)  \right): \,,
    \end{equation}
where the winding charge $n$ is shifted by an amount of $\frac{\theta}{2\pi}$. Label the twisted momenta as
    \begin{equation}
        p_L(\theta) = \sqrt{2}r \left(n+\frac{\theta}{2\pi}\right)+\frac{m}{\sqrt{2}r}\,, \quad p_R(\theta) = \sqrt{2}r \left(n+\frac{\theta}{2\pi}\right)-\frac{m}{\sqrt{2}r}\,,
    \end{equation}
and one has the general OPE
    \begin{equation}
        V_{m,n+\frac{\theta}{2\pi}} V_{m',n'+\frac{\theta'}{2\pi}} \sim z^{\frac{1}{2}p_L(\theta) p'_L(\theta')} \bar{z}^{\frac{1}{2}p_R(\theta) p'_R(\theta')} V_{m+m',n+n'+\frac{\theta+\theta'}{2\pi}}+\cdots \,.
    \end{equation}
Notice that one can write
\begin{equation}
    z^{\frac{1}{2}p_L(\theta) p'_L(\theta')} \bar{z}^{\frac{1}{2}p_R(\theta) p'_R(\theta')} = r^{\frac{1}{2}(p_L(\theta) p'_L(\theta')+p_R(\theta) p'_R(\theta'))} e^{\frac{i}{2}(p_L(\theta) p'_L(\theta')-p_R(\theta) p'_R(\theta')) \varphi}\,,
\end{equation}
with $z=|z|e^{i \varphi}$. Using 
    \begin{equation}
        p_L(\theta) p'_L(\theta')-p_R(\theta) p'_R(\theta')=2m \left(n'+\frac{\theta'}{2\pi} \right) + 2m' \left(n+\frac{\theta}{2\pi} \right)\,,
    \end{equation}
one finds the holonomies
    \begin{equation}
         V_{m,n+\frac{\theta}{2\pi}}(ze^{2\pi i},\bar{z}e^{-2\pi i}) V_{m',n'+\frac{\theta'}{2\pi}}(0,0) = e^{im \theta' + im' \theta} V_{m,n+\frac{\theta}{2\pi}}(z,\bar{z}) V_{m',n'+\frac{\theta'}{2\pi}}(0,0)\,,
    \end{equation}
which implies the operator $V_{m,n+\frac{\theta}{2\pi}}(z,\bar{z})$ is attached to the defect line $e^{i\theta} \in U(1)_M$.

After the $U(1)_M$ projection, the surviving local operators have zero momentum number. Thus we keep $V_{0,n+\frac{\theta}{2\pi}}$. Relabeling $x=r(n+\frac{\theta}{2\pi}) \in \mathbb{R}$, the surviving vertex operators $V^W_x\equiv V_{0,r(n+\frac{\theta}{2\pi})}$ satisfy the OPE
    \begin{equation}
        V^W_x(z,\bar{z}) V^W_y(0,0) = |z|^{2xy} V^W_{x+y}(0,0) + \cdots\,.
    \end{equation}
The label $x$ is now continuous because $\theta$ is continuous. This is exactly the local operator algebra of a (dual) non-compact free boson, written in the momentum basis~\cite{DiFrancesco:1997nk}.  On the other hand, if we gauge $U(1)_W$ we would get
    \begin{equation}
        V^M_x(z,\bar{z}) V^M_y(0,0) = |z|^{\frac{xy}{2}} V^M_{x+y}(0,0) + \cdots\,,
    \end{equation}
where the factor $2$ mismatch in the exponent of $|z|$ is again due to the same reason as discussed below \eqref{eq:partition_function_gauge_U1}.

\subsection{Gauging $\mathbb{Z}$ symmetry of the non-compact boson}

The previous construction should have an inverse operation. Starting with the dual non-compact boson with the $\mathbb{R}_W$-symmetry, we can gauge a discrete $\mathbb{Z}$ subgroup of this symmetry and recover a compact boson, which is the familiar statement that quotienting a line by a lattice gives a circle. Here we will see the statement directly from the level of partition function.

Let us begin with the non-compact boson partition function $Z[a_1,a_2]$, where $a_1,a_2\in \mathbb{R}$ are the holonomies of $\mathbb{R}$-symmetry, and gauge a $\mathbb{Z}$ subgroup with $a_1,a_2 \in  \frac{1}{r} \mathbb{Z}$. Firstly, we need to fix the normalization when summing over $\mathbb{Z}$. When we gauge a $U(1)$ symmetry with period $r$, the flat gauging is normalized as
    \begin{equation}
        \frac{1}{r}\int_0^r d \widetilde{\alpha} \int_0^r d \widetilde{\beta} e^{2\pi i (\widetilde{\alpha}a_2 + \widetilde{\beta}a_1)} \times Z[\widetilde{\alpha},\widetilde{\beta}] \,,
    \end{equation}
where $r$ is the volume of $U(1)$. The inverse operation, namely summing over the dual $\mathbb{Z}$ background, should therefore be normalized as $\frac{1}{r}\sum_{a_1,a_2\in \frac{1}{r}\mathbb{Z}}$, so that they combine to
    \begin{equation}
        \frac{1}{r^2}\sum_{a_1,a_2\in \frac{1}{r}\mathbb{Z}}\int_0^r d \widetilde{\alpha} \int_0^r d \widetilde{\beta} e^{2\pi i (\widetilde{\alpha}a_2 + \widetilde{\beta}a_1)} \times Z[\widetilde{\alpha},\widetilde{\beta}]=Z[0,0]\,.
    \end{equation}

Therefore the $\mathbb{Z}$ gauged partition function is written as
    \begin{equation}
        Z=\frac{1}{r}\sum_{p,q} \frac{1}{\eta \bar{\eta}\sqrt{2 \tau_2}}e^{-\frac{\pi}{2 \tau_2}|\frac{p}{r}-\tau \frac{q}{r}|^2} = \frac{1}{r}\sum_{p,q} \frac{1}{\eta \bar{\eta}\sqrt{2 \tau_2}}  e^{-\frac{\pi}{2 r^2 \tau_2} \left(p^2 + |\tau|^2 q^2-2\tau_1 p q \right)}\,.
    \end{equation}
Using the trick of Poisson resummation, we write
    \begin{equation}
    \begin{split}
        Z=&\frac{1}{r}\sum_{p,q} \int_{-\infty}^{\infty} dx  \delta(x-p)\frac{1}{\eta \bar{\eta}\sqrt{2 \tau_2}}  e^{-\frac{\pi}{2 r^2 \tau_2} \left(x^2 + |\tau|^2 q^2-2\tau_1  qx \right)}\\
        =&\frac{1}{r}\sum_{s,q} \int_{-\infty}^{\infty} dx  e^{2\pi i s x}\frac{1}{\eta \bar{\eta}\sqrt{2 \tau_2}}  e^{-\frac{\pi}{2 r^2 \tau_2} \left(x^2 + |\tau|^2 q^2-2\tau_1  qx \right)}\,,
    \end{split}
    \end{equation}
where we use
    \begin{equation}
        \sum_{p\in \mathbb{Z}} \delta(x-p) = \sum_{s\in \mathbb{Z}} e^{2\pi i s x}\,.
    \end{equation}
Then we have
    \begin{equation}
    \begin{split}
        Z=&\frac{1}{r}\sum_{s,q} \int_{-\infty}^{\infty} dx  e^{2\pi i s x}\frac{1}{\eta \bar{\eta}\sqrt{2 \tau_2}}  e^{-\frac{\pi}{2 r^2 \tau_2} \left(x^2 + |\tau|^2 q^2-2\tau_1  qx \right)}\\
        =&\frac{1}{r}\sum_{s,q} \int_{-\infty}^{\infty} dx\frac{1}{\eta \bar{\eta}\sqrt{2 \tau_2}} e^{-\frac{\pi}{2r^2 \tau_2}(x-\tau_1 q -2i \tau_2 r^2 s)^2} e^{2\pi i \left(\frac{1}{2}(\frac{q}{2r}+rs)^2\tau - \frac{1}{2}(\frac{q}{2r}-rs)^2\bar{\tau} \right)}\\
        =& \frac{1}{\eta \bar{\eta}} \sum_{s,q} q^{\frac{1}{2}(\frac{q}{2r}+rs)^2} \bar{q}^{\frac{1}{2}(\frac{q}{2r}-rs)^2}\,,
    \end{split}
    \end{equation}
and we recover the partition function of the compact boson.

\section{Flat gauging of diagonal $SO(3) \subset (SU(2)_L \times SU(2)_R)/\mathbb{Z}_2$ at the self-dual radius of the compact boson}

In the previous section, we have seen that the flat gauging of $U(1)_M$ (or $U(1)_W$) in the compact boson effective decompactifies the theory. We have checked that from both partition function level and vertex operator level. In this section, we will consider another example of flat gauging $SO(3)$ symmetry of compact boson at the self-dual radius, which is first discussed in \cite{Gaberdiel:2011aa}. Interestingly, the resulting theory lies outside the usual $c=1$ moduli space of circle and orbifold theories. As discussed in~\cite{Gaberdiel:2011aa}, the twisted and untwisted sectors are related to the Runkel-Watts limit~\cite{Runkel:2001ng} and the Roggenkamp-Wendland limit~\cite{Roggenkamp:2003qp} of Virasoro minimal models, respectively. This example shows that flat gauging can produce genuinely new CFTs.

At the self-dual radius $r=\frac{1}{\sqrt{2}}$, the compact boson partition function is
    \begin{equation}
        Z(\tau,\bar{\tau})=\frac{1}{\eta \bar{\eta}} \sum_{n,m={-\infty}}^{\infty} q^{\frac{1}{4}(m+n)^2}\bar{q}^{\frac{1}{4}(m-n)^2}\,.
    \end{equation}
At this point, the chiral algebra is enhanced from $\widehat{\mathbf{u}(1)}$ to $\widehat{\mathfrak{su}(2)}$, which means that the same CFT admits two useful descriptions: In the compact boson description the states are organized by the momentum and winding lattice, and in the WZW description the states are organized by $\widehat{\mathfrak{su}(2)}$ representations. The equality of the two descriptions is already visible at the level of the partition function. The partition function can be rewritten using the two $\widehat{\mathfrak{su}(2)}_1$ characters as
    \begin{equation}
        Z(\tau,\bar{\tau}) = \left|\chi_0^{\widehat{\mathfrak{su}(2)}_1}(q) \right|^2 + \left|\chi_{\frac{1}{2}}^{\widehat{\mathfrak{su}(2)}_1}(q) \right|^2\,,
    \end{equation}
with the affine characters
    \begin{equation}
    \begin{split}
        \chi_0^{\widehat{\mathfrak{su}(2)}_1}(q)& = \sum_{n\in \mathbb{Z}} \frac{q^{n^2}}{\eta(q)}=\sum_{\ell=0,1,2,\cdots} (2\ell +1) \chi_{\ell}^{\text{Vir}}(q)\,,\\
        \quad \chi_{\frac{1}{2}}^{\widehat{\mathfrak{su}(2)}_1}(q)& = \sum_{n\in \mathbb{Z}+\frac{1}{2}} \frac{q^{n^2}}{\eta(q)}=\sum_{\ell=\frac{1}{2},\frac{3}{2},\frac{5}{2},\cdots} (2\ell +1) \chi_{\ell}^{\text{Vir}}(q)\,,
    \end{split}
    \end{equation}
with the irreducible Virasoro characters defined by
    \begin{equation}
        \chi_{\ell}^{\text{Vir}}(q) = \frac{q^{\ell^2}-q^{(\ell+1)^2}}{\eta(q)}\,.
    \end{equation}
Equivalently, the affine Hilbert spaces decompose into Virasoro modules as
    \begin{equation}
        \mathcal{H}_0^{\widehat{\mathfrak{su}(2)}_1}=\bigoplus_{j=0,1,2,\cdots} D_j \otimes \mathcal{H}^{\text{Vir}}_j\,, \quad \mathcal{H}_{\frac{1}{2}}^{\widehat{\mathfrak{su}(2)}_1}=\bigoplus_{j=\frac{1}{2},\frac{3}{2},\frac{5}{2},\cdots} D_j \otimes \mathcal{H}^{\text{Vir}}_j\,,
    \end{equation}
where $D_j$ is the $SU(2)$ representation with dimension $2j+1$. The total Hilbert space is therefore
    \begin{equation}
        \mathcal{H}^{\widehat{\mathfrak{su}(2)}_1} = \left(\bigoplus_{j,j'\in \mathbb{Z}_{\geq 0}} D_j \otimes D_{j'} \otimes \mathcal{H}_j^{\text{Vir}} \otimes \overline{\mathcal{H}}_{j'}^{\text{Vir}}\right) \oplus \left(\bigoplus_{j,j'\in \mathbb{Z}_{\geq 0}+\frac{1}{2}} D_j \otimes D_{j'} \otimes \mathcal{H}_j^{\text{Vir}} \otimes \overline{\mathcal{H}}_{j'}^{\text{Vir}}\right)\,,
    \end{equation}
where we use the fact that all $SU(2)$ irreducible representations are self-conjugate.

In the following, we consider the flat gauging of $SO(3)$ separately in the untwisted sector and in the twisted sectors. We label the defect Hilbert space by a spatial holonomy $h\in SO(3)$. On the torus, a flat connection is specified by a commuting pair of holonomies $(g,h)$, with $gh=hg$, where $g$ is the temporal holonomy. Therefore, once the spatial holonomy $h$ is fixed, the temporal holonomy must lie in the centralizer $C(h)$. This distinction is important: for the untwisted sector $h=e$, the centralizer is the full group, $C(e)=SO(3)$, while for a generic twisted sector $h\neq e$, the centralizer is only a maximal torus, $C(h)=U(1)$. Because of this, the projection in the two sectors is different. We will first discuss the twisted sectors and then return to the untwisted sector.

\subsection{Flat gauging of $SO(3)$ : twisted sector}
The $SO(3)$ subgroup is the diagonal subgroup of $SU(2)_L \times SU(2)_R$, and it acts by conjugation on the $SU(2)$-valued WZW field $g$
    \begin{equation}
        g \rightarrow h g h^{-1}\,, \quad h \in SU(2)\,,
    \end{equation}
Since $g$ is left invariant for $h=-1$, the faithful group acting in this way is $SO(3)$ rather than $SU(2)$. 

Let us first consider the contribution from the $SO(3)$ twisted sectors. Suppose we turn on a spatial holonomy of $h\in SO(3)$, and consider the partition function $Z[g,h]$. Here $g \in SO(3)$ is the temporal holonomy satisfying $gh=hg$. Since the partition function is invariant under the simultaneous conjugation 
    \begin{equation}
        Z[kgk^{-1},khk^{-1}]=Z[g,h]\,,
    \end{equation}
we can always fix $h\in T/W$ where $T$ is the Cartan torus of $SO(3)$ and $W=\mathbb{Z}_2$ is the Weyl group. In the present case, the Cartan torus is simply the $U(1)_M$ parametrized by $U_M(\theta)$ with $\theta\in [0,r)$, where $r=1/\sqrt{2}$. The Weyl reflection map $\theta$ to $-\theta$, so that we can choose $\theta\in[0,r/2)$ as representative of the conjugacy class.

Let us set $h=U_M(\widetilde{\beta})\in U(1)_M$, and the centralizer group $C_G(h)=T=U(1)_M$ is also the Cartan torus. Therefore for a fixed twisted sector labeled by $U_M(\widetilde{\beta})$, the gauging gives the partition function 
    \begin{equation}
        Z_{\widetilde{\beta}}=\sqrt{2} \int_{0}^{\frac{1}{\sqrt{2}}} d\widetilde{\alpha} \,Z[(\widetilde{\alpha},\widetilde{\beta}),(0,0)] =\sqrt{2}\int_{0}^{\frac{1}{\sqrt{2}}} d\widetilde{\alpha} \, \frac{1}{\eta \bar{\eta}} \sum_{m,n} e^{2\pi i \alpha m}q^{\frac{1}{4}\left(m+(n+\beta) \right)^2} \bar{q}^{\frac{1}{4}\left( m-(n+\beta)\right)^2}\,,
    \end{equation}
where $Z[(\widetilde{\alpha},\widetilde{\beta}),(0,0)]$ is read from \eqref{eq:partition-function-full-background}. Next, we need to sum over twisted sectors labeled by different conjugacy classes, which gives
    \begin{equation}\label{eq:SO(3)_gauge_twisted}
        [Z/SO(3)]_{\text{twisted}} =\sqrt{2}{\int_0'}^{\frac{1}{2\sqrt{2}}}d\widetilde{\beta} \int_{0}^{\frac{1}{\sqrt{2}}} d\widetilde{\alpha} \, \frac{1}{\eta \bar{\eta}} \sum_{m,n} e^{2\pi i \alpha m}q^{\frac{1}{4}\left(m+(n+\beta) \right)^2} \bar{q}^{\frac{1}{4}\left( m-(n+\beta)\right)^2}=\frac{1}{2}Z_{\mathbb{R}_W}\,,
    \end{equation}
with $Z_{\mathbb{R}_W}=\frac{1}{\eta \bar{\eta}\sqrt{2 \tau_2}}$\,.
Here we use $\int'$ to emphasize that we remove the point $\widetilde{\beta}=0$ from the integral. However, since a point has zero measure, removing a point does not affect the result of the integral. Therefore we can treat the integral as if the integrand at $\widetilde{\beta}=0$ is the same as at other points. The untwisted sector corresponding to $\widetilde{\beta}=0$ will be discussed later.

The gauged partition function is half of the partition function of the (dual) non-compact boson, because the range of $\widetilde{\beta}$ is $\widetilde{\beta}\in [0,\pi/2)$ rather than $[0,\pi)$. Physically, this is because the twisted sectors labeled by $U_M(-\widetilde{\beta})$ and $U_M(\widetilde{\beta})$ are related by the Weyl group of $SO(3)$, and they should be identified after gauging. On the other hand, for $U(1)$ gauging all twisted sectors contribute independently to the gauged partition function.

As a remark, the gauged partition function resembles the partition function of the orbifolded (dual) non-compact boson, since the latter also has the contribution $\frac{1}{2}Z_{\mathbb{R}_W}$ as its target space is $\mathbb{R}_W/\mathbb{Z}_2$. However, we emphasize that they are actually different theories. First, the orbifolded non-compact boson also receives oscillator contributions from the single $\mathbb{Z}_2$ fixed point $X=0$, and the full partition function is
    \begin{equation}
        Z/(U(1)_M\rtimes C)
        =
        \frac{1}{2}Z_{\mathbb{R}_W}
        +
        \frac{|\vartheta_3 \vartheta_4|}{4|\eta|^2}
        +\frac{|\vartheta_2 \vartheta_3|}{4|\eta|^2}
        +\frac{|\vartheta_2 \vartheta_4|}{4|\eta|^2}\,,
    \end{equation}
where the last three terms are absent in the $SO(3)$ gauged partition function. Second, as we will see in the next section, their gauged partition functions from the untwisted sector are different.

\subsection{Flat gauging of $SO(3)$ : untwisted sector}

Let us move to the untwisted sector. To gauge this $SO(3)$ subgroup, we need to project to the $SO(3)$ singlet states. Using the decomposition
    \begin{equation}
        D_j \otimes D_{j'} = D_{|j-j'|}\oplus D_{|j-j'|+1}\oplus \cdots \oplus D_{|j+j'|}\,,
    \end{equation}
one finds that a singlet appears only when $j=j'$ with multiplicity one. Therefore the untwisted sector after the $SO(3)$ projection is
    \begin{equation}\label{eq:SO(3)_gauge_untwisted}
        \left[Z/SO(3)\right]_{\text{untwisted}} = \sum_{\ell=0,\frac{1}{2},1,\frac{3}{2},\cdots} \left|\chi^{\text{Vir}}_{\ell}(q)\right|^2\,.
    \end{equation}

In the following, we will compare it to the untwisted sector contribution of the orbifolded (dual) non-compact boson, namely the $U(1)_M\rtimes C$ gauged partition function of the untwisted sector of the compact boson. If we project only to the $U(1)_M$ invariant states, then in $D_j\otimes D_{j'}$ the number of weight-zero components is $2\min(j,j')+1$. Therefore one obtains
    \begin{equation}
        \left[Z/(U(1)_M\rtimes C)\right]_{\text{untwisted}} = \sum_{\substack{j,j'=0,\frac{1}{2},1,\cdots\\j+j'\in \mathbb{Z}}} (2 \min(j,j') + 1) \chi^{\text{Vir}}_{j}(q) \overline{\chi}^{\text{Vir}}_{j'}(\bar{q})\,.
    \end{equation}
Moreover, under the symmetry $C$, the affine $SU(2)$ generators transforms according to
    \begin{equation}
        J^{\pm } \rightarrow J^{\mp}\,, \quad J^3 \rightarrow -J^3\,,
    \end{equation}
which implies we can identify 
    \begin{equation}
        C=e^{i\pi J_x}\,.
    \end{equation}
For a state $|j,m\rangle$ in the representation $D_j$, the action of $C$-operators can be read from the Wigner $D$-matrix as
    \begin{equation}
        C|j,m\rangle = e^{i\pi J_x} |j,m\rangle = (-1)^{j+m}|j,-m\rangle\,.
    \end{equation}
Setting $m=0$ for even $j$, one has $C|j,0\rangle = (-1)^j$ acting on the $U(1)$ invariant states. Therefore the $U(1)\rtimes C$ gauged partition function of the untwisted sector is
    \begin{equation}
        \left[Z/(U(1)_M\rtimes C)\right]_{\text{untwisted}} = \frac{1}{2}\sum_{\substack{j,j'=0,\frac{1}{2},1,\cdots\\j+j'\in \mathbb{Z}}}  \left(2 \min(j,j') + 1+\delta_{j\in \mathbb{Z}} \delta_{j'\in \mathbb{Z}}(-1)^{j+j'}\right) \chi^{\text{Vir}}_{j}(q) \overline{\chi}^{\text{Vir}}_{j'}(\bar{q})\,,
    \end{equation}
where the delta function is defined by $\delta_{j\in \mathbb{Z}}=\sum_{n\in \mathbb{Z}}\delta_{j,n}$. Here we see the untwisted sector partition function is indeed different from that of $SO(3)$ gauging.

Therefore, the total partition function should be the combination of both twisted part and untwisted part, and we write
\begin{equation}\label{eq:SO(3)_gauge_full}
    Z/SO(3)=\sum_{\ell=0,\frac{1}{2},1,\frac{3}{2},\cdots} \left|\chi^{\text{Vir}}_{\ell}(q)\right|^2 + \frac{1}{2} Z'_{\mathbb{R}_W}\,,
\end{equation}
where we use $Z'_{\mathbb{R}_W}$ to emphasize that we should remove the states with integer winding number in \eqref{eq:SO(3)_gauge_twisted}. This set has measure zero and does not affect the result of the integral.

\section{Flat gauging of continuous non-invertible symmetry on the orbifold branch}\label{sec:non-invertible}

In this section, we will move to the orbifold branch of the compact boson by gauging the charge conjugation $\mathbb{Z}_2$ symmetry $C$. Denote the $U(1)_M$ and $U(1)_W$ generators on the circle branch as $U_M(\theta)$ and $U_W(\varphi)$, with $\theta,\varphi \in [0,2\pi)$, the conjugation symmetry $C$ acts on the $U(1)$ groups according to
    \begin{equation}
        CU_M(\theta)C^{-1} = U_M(-\theta)\,, \quad CU_W(\varphi)C^{-1} = U_W(-\varphi)\,.
    \end{equation}
When we gauge $C$, each fixed points $U_M(0),U_M(\pi),U_W(0),U_W(\pi)$ will split into two simple symmetry operators
    \begin{equation}
        U_M(0)=U_W(0)= 1+\eta\,, \quad U_M(\pi) = \eta_m + \eta_m\,\eta\,, \quad U_W(\pi) = \eta_w + \eta_w\, \eta\,,
    \end{equation}
where $\eta$ is the quantum $\mathbb{Z}_2$ symmetry that distinguishes the operators from twisted/untwisted sectors in the circle branch, and $\eta_m$ and $\eta_w$ acts on the vertex operators in the same way as $U_M(\pi)$ and $U_W(\pi)$    
\begin{equation}
    \hat{\eta}_m(V_{m,n}) = (-1)^m V_{m,n}\,, \quad \hat{\eta}_w (V_{m,n}) = (-1)^n V_{m,n}\,.
\end{equation}
For generic $U(1)$ generators, the surviving symmetry operators after $C$-gauging are the 2-dimensional $\mathbb{Z}_2$ orbits
    \begin{equation}
        L_M(\theta) = U_M(\theta) + U_M(-\theta)\,, \quad L_W(\varphi)=U_W(\varphi) + U_W(-\varphi)\,, \quad \theta,\varphi\in (0,\pi)\,,
    \end{equation}
and they satisfy the non-invertible fusion rule inherited from the $U(1)_M\times U(1)_W$~\cite{Chang:2020imq,Thorngren:2021yso}
    \begin{equation}
    \begin{split}
        L_M(\theta_1) \times L_M(\theta_2) =& L_M(\theta_1+\theta_2) + L_M(\theta_1-\theta_2)\,,\\
        L_W(\varphi_1) \times L_W(\varphi_2) =& L_W(\varphi_1+\varphi_2) + L_W(\varphi_1-\varphi_2)\,,
    \end{split}
    \end{equation}
where we identify $\theta\sim  \theta+2\pi\sim -\theta$ in $L_M$, and similar to $L_W$. We also have
    \begin{equation}
        L_M(0)=L_W(0)=1+\eta \,, \quad L_M(\pi) = \eta_m + \eta_m\, \eta \,, \quad L_W(\pi) = \eta_w + \eta_w\, \eta\,.
    \end{equation}
Therefore, the symmetry $(U(1)_M \times U(1)_W)\rtimes C$ on the circle branch becomes a continuous non-invertible symmetry on the orbifold branch after gauging.

In this section, we will consider gauging the non-invertible symmetry on the orbifold branch at the level of partition function. We will first construct the partition functions associated with a general non-invertible symmetry background, which is depicted as a defect network. Then we will first consider the gauging of finite non-invertible symmetry at the level of partition function, following the method in \cite{Diatlyk:2023fwf}. After that, we will move to the flat gauging of continuous non-invertible symmetry.

\subsection{Partition function for general background of non-invertible symmetry}

The partition function in the orbifold branch with zero background takes the form
\begin{equation}
    Z_{\text{orb}} = \frac{1}{2} \left(Z(\tau) +\frac{|\vartheta_3 \vartheta_4|}{|\eta|^2} \right) + \frac{1}{2}\left(\frac{|\vartheta_2 \vartheta_3|}{|\eta|^2} + \frac{|\vartheta_2 \vartheta_4|}{|\eta|^2}\right)\,,
\end{equation}
where we omit the label of radius. The first term is from the untwisted sector of the circle branch partition function
    \begin{equation}
    Z_{\text{circ}} 
        \,\begin{gathered}
\begin{tikzpicture}[scale=.75]
\node at (0, 1.45) {};
\draw[line] (-.15,-.2)--(-.25, -.2) -- (-.25, 1.4)-- (-.15,1.4);
\draw[very thick] (0,0) rectangle (1.2,1.2);
\draw[line] (1.35, -.2)--(1.45, -.2) -- (1.45, 1.4)--(1.35, 1.4);
\end{tikzpicture}
\end{gathered} = Z_r(\tau)\,, \quad Z_{\text{circ}}
        \,\begin{gathered}
\begin{tikzpicture}[scale=.75]
\node at (0, 1.45) {};
\draw[line] (-.15,-.2)--(-.25, -.2) -- (-.25, 1.4)-- (-.15,1.4);
\draw[very thick] (0,0) rectangle (1.2,1.2);
\draw[line, red, thick] (0,.6)--(1.2,.6);
\draw[line] (1.35, -.2)--(1.45, -.2) -- (1.45, 1.4)--(1.35, 1.4);
\end{tikzpicture}
\end{gathered} = \frac{|\vartheta_3 \vartheta_4|}{|\eta|^2}\,,
    \end{equation}
and the second term is from the twisted sector
    \begin{equation}
        Z_{\text{circ}}
        \,\begin{gathered}
\begin{tikzpicture}[scale=.75]
\node at (0, 1.45) {};
\draw[line] (-.15,-.2)--(-.25, -.2) -- (-.25, 1.4)-- (-.15,1.4);
\draw[very thick] (0,0) rectangle (1.2,1.2);
\draw[line, red, thick] (.6,0)--(.6,1.2);
\draw[line] (1.35, -.2)--(1.45, -.2) -- (1.45, 1.4)--(1.35, 1.4);
\end{tikzpicture}
\end{gathered} = \frac{|\vartheta_2 \vartheta_3|}{|\eta|^2}\,, \quad Z_{\text{circ}} 
        \,\begin{gathered}
\begin{tikzpicture}[scale=.75]
\node at (0, 1.45) {};
\draw[line] (-.15,-.2)--(-.25, -.2) -- (-.25, 1.4)-- (-.15,1.4);
\draw[very thick] (0,0) rectangle (1.2,1.2);
\draw[red,line] (0.6,0)--(0.6,1.2);
\draw[red,line] (0,0.6)--(1.2,0.6);
\draw[line] (1.35, -.2)--(1.45, -.2) -- (1.45, 1.4)--(1.35, 1.4);
\end{tikzpicture}
\end{gathered} = \frac{|\vartheta_2 \vartheta_4|}{|\eta|^2}\,,
    \end{equation}
so that we have
    \begin{equation}\label{eq:partition_function_circ_orb}
Z_{\text{orb}} 
        \,\begin{gathered}
\begin{tikzpicture}[scale=.75]
\node at (0, 1.45) {};
\draw[line] (-.15,-.2)--(-.25, -.2) -- (-.25, 1.4)-- (-.15,1.4);
\draw[very thick] (0,0) rectangle (1.2,1.2);
\draw[line] (1.35, -.2)--(1.45, -.2) -- (1.45, 1.4)--(1.35, 1.4);
\end{tikzpicture}
\end{gathered} = \frac{1}{2} \left(
Z_{\text{circ}} 
        \,\begin{gathered}
\begin{tikzpicture}[scale=.75]
\node at (0, 1.45) {};
\draw[line] (-.15,-.2)--(-.25, -.2) -- (-.25, 1.4)-- (-.15,1.4);
\draw[very thick] (0,0) rectangle (1.2,1.2);
\draw[line] (1.35, -.2)--(1.45, -.2) -- (1.45, 1.4)--(1.35, 1.4);
\end{tikzpicture}
\end{gathered} + 
        Z_{\text{circ}}
        \,\begin{gathered}
\begin{tikzpicture}[scale=.75]
\node at (0, 1.45) {};
\draw[line] (-.15,-.2)--(-.25, -.2) -- (-.25, 1.4)-- (-.15,1.4);
\draw[very thick] (0,0) rectangle (1.2,1.2);
\draw[line, red, thick] (0,.6)--(1.2,.6);
\draw[line] (1.35, -.2)--(1.45, -.2) -- (1.45, 1.4)--(1.35, 1.4);
\end{tikzpicture}
\end{gathered}
+
Z_{\text{circ}}
        \,\begin{gathered}
\begin{tikzpicture}[scale=.75]
\node at (0, 1.45) {};
\draw[line] (-.15,-.2)--(-.25, -.2) -- (-.25, 1.4)-- (-.15,1.4);
\draw[very thick] (0,0) rectangle (1.2,1.2);
\draw[line, red, thick] (.6,0)--(.6,1.2);
\draw[line] (1.35, -.2)--(1.45, -.2) -- (1.45, 1.4)--(1.35, 1.4);
\end{tikzpicture}
\end{gathered}
+
Z_{\text{circ}} 
        \,\begin{gathered}
\begin{tikzpicture}[scale=.75]
\node at (0, 1.45) {};
\draw[line] (-.15,-.2)--(-.25, -.2) -- (-.25, 1.4)-- (-.15,1.4);
\draw[very thick] (0,0) rectangle (1.2,1.2);
\draw[red,line] (0.6,0)--(0.6,1.2);
\draw[red,line] (0,0.6)--(1.2,0.6);
\draw[line] (1.35, -.2)--(1.45, -.2) -- (1.45, 1.4)--(1.35, 1.4);
\end{tikzpicture}
\end{gathered}
\right)\,.
    \end{equation}
Compared to the previous sections where we label the partition function $Z[(\theta_x,\theta_y),\cdots]$ using the holonomy pair $(\theta_x,\theta_y)$, here for the non-invertible symmetry, it is better to use a defect network to depict the background field. We will use the box $\begin{gathered}
    \begin{tikzpicture}[scale=0.5]
        \node at (0, 1.25) {};
        \draw[very thick] (0,0) rectangle (1.2,1.2);
    \end{tikzpicture}
\end{gathered}$ for the worldsheet torus, and lines for symmetry operators/defects. For example, $\begin{gathered}
\begin{tikzpicture}[scale=.5]
\node at (0, 1.25) {};
\draw[very thick] (0,0) rectangle (1.2,1.2);
\draw[line, red, thick] (0,.6)--(1.2,.6);
\end{tikzpicture}
\end{gathered}$ represents the background with the insertion of the $C$-symmetry operator along the spatial direction;  $\begin{gathered}
\begin{tikzpicture}[scale=.5]
\node at (0, 1.25) {};
\draw[very thick] (0,0) rectangle (1.2,1.2);
\draw[line, red, thick] (.6,0)--(.6,1.2);
\end{tikzpicture}
\end{gathered}$ represents the background with the insertion of the $C$-defect operator along the temporal direction, and $\begin{gathered}
\begin{tikzpicture}[scale=.5]
\node at (0, 1.35) {};
\draw[very thick] (0,0) rectangle (1.2,1.2);
\draw[red,line] (0.6,0)--(0.6,1.2);
\draw[red,line] (0,0.6)--(1.2,0.6);
\end{tikzpicture}
\end{gathered}$ represents the background with the insertion of the $C$-lines along both cycle. The last one can be also drawn as $\begin{gathered}
\begin{tikzpicture}[scale=.5]
\node at (0, 1.35) {};
\draw[very thick] (0,0) rectangle (1.2,1.2);
\draw[red,thick]  (.6,0) arc (0:90:.6); 
\draw[red,thick] (.6,1.2) arc (-180:-90:.6);
\end{tikzpicture}
\end{gathered}$ after resolving the four-way junction. Throughout this section, we will use the red line for $C$ symmetry operator/defect.

For a non-invertible symmetry $\mathscr{C}$ with the general fusion rule
    \begin{equation}
        L_1 \times L_2 = \sum_{L_3}N_{L_1 L_2}^{L_2} L_3\,,\quad (N_{L_1 L_2}^{L_3} \in \mathbb{Z}_{\geq 0})
    \end{equation}
where $L_i\in \mathscr{C}$ are simple topological defect lines (TDL) in the theory. Then the general defect network on the torus is depicted as
    
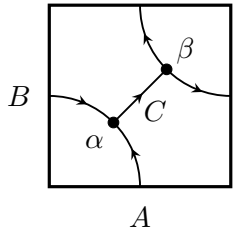
\begin{equation}\label{eq:defect_network}
        \begin{gathered}
\begin{tikzpicture}[scale=2]
\draw[very thick] (0,0) rectangle (1.2,1.2);
\draw[line,->-=.55]  (.6,0) arc (0:45:.6); 
\draw[line,->-=.55]  (0,0.6) arc (90:45:.6); 
\draw[line,-<-=.55] (.6,1.2) arc (-180:-135:.6);
\draw[line,-<-=.55] (1.2,0.6) arc (-90:-135:.6);
\draw[line,->-=.55] (0.6/1.414,0.6/1.414)--(1.2-0.6/1.414,1.2-0.6/1.414);
\node at (-0.2,0.6) {$B$};
\node at (0.6,-0.2) {$A$};
\node at (0.6+0.1,0.6-0.1) {$C$};
\filldraw[black] (0.6/1.414,0.6/1.414) circle (1pt);
\filldraw[black] (1.2-0.6/1.414,1.2-0.6/1.414) circle (1pt);
\node at (0.3,0.3) {$\alpha$};
\node at (0.9,0.9) {$\beta$};
\end{tikzpicture}
\end{gathered}\nonumber
    \end{equation}
where $A,B,C\in \mathscr{C}$ are simple TDLs such that $C$ is one of the fusion channel of $A\times B$, namely $N_{AB}^C \neq 0$. The junctions $\alpha,\beta$ are attached to the complex vector spaces $\mathbb{C}^{N_{AB}^C}$, so that we can label $\alpha,\beta=1,\cdots,N_{AB}^C$ after we choose a basis for the vector space. In general, for the fixed input $A,B$, there exists $\sum_{C} (N_{AB}^C)^2$ independent defect networks on the torus.

In order to gauge the non-invertible symmetry at the level of partition function, one needs to write down partition function for each independent defect network on the torus, and sum over them with proper weight factors. In general, that is a difficult task. In the present case, since the non-invertible symmetry on the orbifold branch arise by gauging the $\mathbb{Z}_2$ symmetry of an invertible symmetry on the circle branch, we can always go back to the circle branch and do the computation therein. According to \eqref{eq:partition_function_circ_orb}, we need to consider the contributions from all the four sectors on the circle branch, twisted by the $C$-lines. In the following, we will evaluate the partition function with a general non-invertible symmetry background in this way.

We begin with simplest background by inserting a single $L_M(\theta)$ operator along the spatial circle of the torus. Recall that $L_M(\theta) = U_M(\theta) + U_M(-\theta)$, we need to evaluate the following four diagrams on the circle branch
\begin{equation}
    \begin{gathered}
        \begin{tikzpicture}
            \draw[very thick] (0,0) rectangle (1.8,1.8);
            \draw[line] (0,0.9)--(1.8,0.9);
            \node at (0.45,1.1) {$\pm \theta$};
        \end{tikzpicture}
    \end{gathered}\qquad 
    \begin{gathered}
        \begin{tikzpicture}
            \draw[very thick] (0,0) rectangle (1.8,1.8);
            \draw[line] (0,0.9)--(1.8,0.9);
            \draw[line,red] (0,0.7)--(1.8,0.7);
            \node at (0.45,1.1) {$\pm \theta$};
        \end{tikzpicture}
    \end{gathered}\qquad
    \begin{gathered}
        \begin{tikzpicture}
            \draw[very thick] (0,0) rectangle (1.8,1.8);
            \draw[line] (0,0.9)--(1.8,0.9);
            \draw[line,red] (1.1,0)--(1.1,1.8);
            \node at (0.45,1.1) {$\pm \theta$};
        \end{tikzpicture}
    \end{gathered}\qquad
        \begin{gathered}
        \begin{tikzpicture}
            \draw[very thick] (0,0) rectangle (1.8,1.8);
            \draw[line] (0,0.9)--(1.8,0.9);
            \draw[line,red] (1.1,0)--(1.1,1.8);
            \draw[line,red] (0,0.7)--(1.8,0.7);
            \node at (0.45,1.1) {$\pm \theta$};
        \end{tikzpicture}
    \end{gathered}
\end{equation}
for $U_M(\theta)$ and $U_M(-\theta)$ separately. In the following, we will simply use $\pm\theta$ on the diagram for short. In the untwisted sector, the first diagram contributes 
    \begin{equation}
        \frac{1}{2} \left( \lambda_{\theta,0} + \lambda_{-\theta,0}\right)\,,
    \end{equation}
where the half factor is from \eqref{eq:partition_function_circ_orb}, and we introduce $\lambda_{\theta_1,\theta_2}$ as the circle branch partition function twisted by the $U(1)_M$ holonomies
    \begin{equation}
        \lambda_{\theta_1,\theta_2} = \frac{1}{|\eta(\tau)|^2} \sum_{m,n} e^{i \theta_1 m} q^{\frac{1}{2}\left(\frac{m}{2r} + (n+\frac{\theta_2}{2\pi})r \right)^2} \bar{q}^{\frac{1}{2}\left(\frac{m}{2r}-(n+\frac{\theta_2}{2\pi})r \right)^2}\,.
    \end{equation}
Since $U_M(\theta)$ only acts on the momentum modes, and leaves the oscillators invariant. Therefore the second diagram survive and gives another contribution
    \begin{equation}
        \frac{|\vartheta_3 \vartheta_4|}{2|\eta|^2}\,.
    \end{equation}

For the twisted sector, we need to examine whether given configuration is legal on the torus, and there are several ways to think of that. Denote the holonomies along the temporal and spatial directions as $g,h$, then one must have $gh=hg$ in order for the background to be flat. Equivalently, the symmetry operator $U_g$ should preserve the symmetry defect $U_h$, otherwise the defect cannot form a loop on the torus. From the level of partition function, the symmetry operator $U_g$ should map the twisted Hilbert space $\mathcal{H}_{h}$ to itself, otherwise it will give zero after taking the trace.

For the third diagram, the symmetry operator is $U(\pm \theta)$ and the symmetry defect is $C$, since one has
    \begin{equation}
        C U(\theta) C^{-1} = U(-\theta)\,, \quad CU(-\theta) C^{-1} = U(\theta)\,,
    \end{equation}
it is inconsistent to insert the $U_M(\theta)$ and $U_M(-\theta)$ loop operators with the presence of the $C$-defect. Similarly, for the fourth diagram, the symmetry operator is $U(\pm \theta)C$, and we have
    \begin{equation}
        C \left( U(\theta)C \right)C^{-1} = U(-\theta)C\,, \quad C \left( U(-\theta)C \right)C^{-1} = U(\theta)C\,,
    \end{equation}
which do not commute with $C$. Therefore the contribution of the twisted sector is zero, and we have
    \begin{equation}
        Z_{\text{orb}}\,
        \begin{gathered}
\begin{tikzpicture}[scale=.75]
\node at (0, 1.45) {};
\draw[line] (-.15-.4,-.2-.4)--(-.25-.4, -.2-.4) -- (-.25-.4, 1.4)-- (-.15-.4,1.4);
\draw[very thick] (0,0) rectangle (1.2,1.2);
\draw[line, thick] (0,.6)--(1.2,.6);
\draw[line] (1.35, -.2-.4)--(1.45, -.2-.4) -- (1.45, 1.4)--(1.35, 1.4);
\node at (-.15-.2,-.2+0.8) {$\theta$};
\end{tikzpicture}
\end{gathered}
=\frac{1}{2} \left( \lambda_{\theta,0} + \lambda_{-\theta,0} \right) +\frac{|\vartheta_3 \vartheta_4|}{|\eta|^2}\,.
    \end{equation}

We then consider inserting a single $L_M(\theta)$ defect along the spatial circle of the torus, where we have the following four diagrams on the circle branch
\begin{equation}
    \begin{gathered}
        \begin{tikzpicture}
            \draw[very thick] (0,0) rectangle (1.8,1.8);
            \draw[line] (0.9,0)--(0.9,1.8);
            \node at (0.55,1.1) {$\pm \theta$};
        \end{tikzpicture}
    \end{gathered}\qquad 
    \begin{gathered}
        \begin{tikzpicture}
            \draw[very thick] (0,0) rectangle (1.8,1.8);
            \draw[line] (0.9,0)--(0.9,1.8);
            \node at (0.55,1.1) {$\pm \theta$};
            \draw[line,red] (0,0.7)--(1.8,0.7);
        \end{tikzpicture}
    \end{gathered}\qquad
    \begin{gathered}
        \begin{tikzpicture}
            \draw[very thick] (0,0) rectangle (1.8,1.8);
            \draw[line] (0.9,0)--(0.9,1.8);
            \node at (0.55,1.1) {$\pm \theta$};
            \draw[line,red] (1.1,0)--(1.1,1.8);
        \end{tikzpicture}
    \end{gathered}\qquad
        \begin{gathered}
        \begin{tikzpicture}
            \draw[very thick] (0,0) rectangle (1.8,1.8);
            \draw[line] (0.9,0)--(0.9,1.8);
            \node at (0.55,1.1) {$\pm \theta$};
            \draw[line,red] (1.1,0)--(1.1,1.8);
            \draw[line,red] (0,0.7)--(1.8,0.7);
        \end{tikzpicture}
    \end{gathered}
\end{equation}
In the untwisted sector, the first diagram contributes
    \begin{equation}
        \frac{1}{2}(\lambda_{0,\theta}+\lambda_{0,-\theta})\,,
    \end{equation}
and since $C$ does not commute with $U(\pm \theta)$, the second diagram gives zero. In the twisted sector, for the third diagram the defect $U_M(\pm\theta)$ merge $C$ and will impose the twisted boundary conditions
    \begin{equation}
    X(\sigma+2\pi )= -X(\sigma)\pm \theta r\,,
    \end{equation}
along the spatial circle of the worldsheet. Thus it merely shifts the location of the two fixed points to $\pm \frac{1}{2}\theta r$ and $\pm \frac{1}{2}\theta r + \pi r$, and does not change the contributions from the oscillators localized at each fixed point, and we still have
    \begin{equation}
        \frac{|\vartheta_2 \vartheta_3|}{|\eta|^2}\,.
    \end{equation}
The last diagram does not contribute since $U(\pm \theta)C$ does not commute with $C$. Therefore we have 
    \begin{equation}
        Z_{\text{orb}}\,
        \begin{gathered}
\begin{tikzpicture}[scale=.75]
\node at (0, 1.45) {};
\draw[line] (-.15-.4,-.2-.4)--(-.25-.4, -.2-.4) -- (-.25-.4, 1.4)-- (-.15-.4,1.4);
\draw[very thick] (0,0) rectangle (1.2,1.2);
\draw[line, thick] (.6,0)--(.6,1.2);
\draw[line] (1.35, -.2-.4)--(1.45, -.2-.4) -- (1.45, 1.4)--(1.35, 1.4);
\node at (.6,-.35) {$\theta$};
\end{tikzpicture}
\end{gathered}
=\frac{1}{2}(\lambda_{0,\theta}+\lambda_{0,-\theta})+\frac{|\vartheta_2 \vartheta_3|}{|\eta|^2}\,.
    \end{equation}

We now turn to the general cases where we insert both symmetry operator $L_M(\theta_1)$ and symmetry defect $L_M(\theta_2)$. The non-invertible symmetry background is depicted as the following two defect networks
    
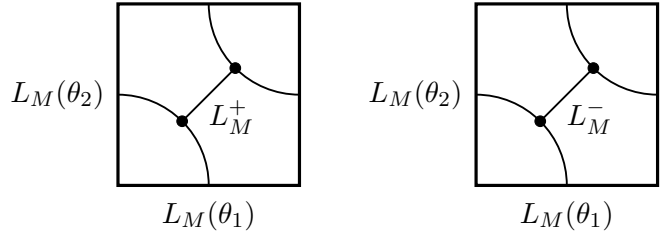
\begin{equation}
        \begin{gathered}
\begin{tikzpicture}[scale=2]
\draw[very thick] (0,0) rectangle (1.2,1.2);
\draw[line]  (.6,0) arc (0:45:.6); 
\draw[line]  (0,0.6) arc (90:45:.6); 
\draw[line] (.6,1.2) arc (-180:-135:.6);
\draw[line] (1.2,0.6) arc (-90:-135:.6);
\draw[line] (0.6/1.414,0.6/1.414)--(1.2-0.6/1.414,1.2-0.6/1.414);
\node at (-0.4,0.6) {$L_M(\theta_2)$};
\node at (0.6,-0.2) {$L_M(\theta_1)$};
\node at (0.6+0.15,0.6-0.15) {$L_M^+$};
\filldraw[black] (0.6/1.414,0.6/1.414) circle (1pt);
\filldraw[black] (1.2-0.6/1.414,1.2-0.6/1.414) circle (1pt);
\end{tikzpicture}
\end{gathered}\qquad
        \begin{gathered}
\begin{tikzpicture}[scale=2]
\draw[very thick] (0,0) rectangle (1.2,1.2);
\draw[line]  (.6,0) arc (0:45:.6); 
\draw[line]  (0,0.6) arc (90:45:.6); 
\draw[line] (.6,1.2) arc (-180:-135:.6);
\draw[line] (1.2,0.6) arc (-90:-135:.6);
\draw[line] (0.6/1.414,0.6/1.414)--(1.2-0.6/1.414,1.2-0.6/1.414);
\node at (-0.4,0.6) {$L_M(\theta_2)$};
\node at (0.6,-0.2) {$L_M(\theta_1)$};
\node at (0.6+0.15,0.6-0.15) {$L_M^-$};
\filldraw[black] (0.6/1.414,0.6/1.414) circle (1pt);
\filldraw[black] (1.2-0.6/1.414,1.2-0.6/1.414) circle (1pt);
\end{tikzpicture}
\end{gathered}\nonumber
    \end{equation}
with $L_M^+ = L_M(\theta_1+\theta_2)$ and $L_M^-=L_M(\theta_1-\theta_2)$, and all junction vector spaces are $\mathbb{C}$, and we choose the trivial unit basis and omit the labels. 

For the $L_M^+$ diagram, the untwisted sector contributions from the circle branch can be drawn as the following diagrams
    
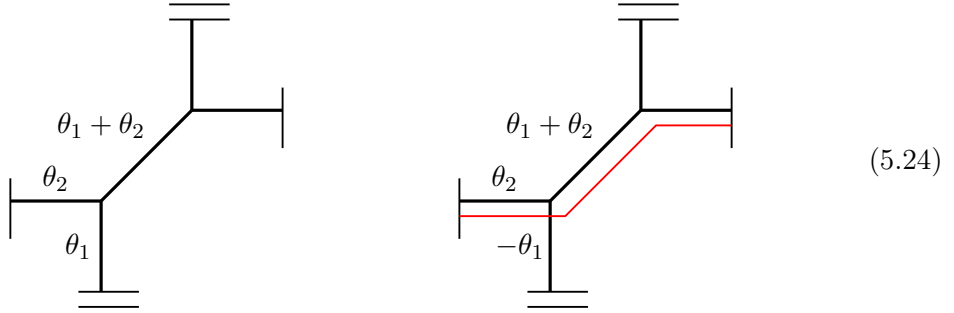
\begin{equation}
    \begin{gathered}
        \begin{tikzpicture}
                \draw[line,very thick] (0,1.2)--(1.2,1.2);
                \draw[line,very thick] (1.2,0)--(1.2,1.2);
                \draw[line,very thick] (1.2,1.2)--(2.4,2.4);
                \draw[line,very thick] (2.4,2.4)--(3.6,2.4);
                \draw[line,very thick] (2.4,2.4)--(2.4,3.6);
                \node at (0.6,1.5) {$\theta_2$};
                \node at (0.9,0.6) {$\theta_1$};
                \node at (1.8-0.6,1.8+0.4) {$\theta_1+\theta_2$};
                \draw[line] (0,0.7)--(0,1.5);
                \draw[line] (3.6,0.7+1.2)--(3.6,1.5+1.2);
                \draw[line] (0.9,0)--(1.7,0);
                \draw[line] (0.9,0-0.2)--(1.7,0-0.2);
                \draw[line] (0.9+1.2,3.6)--(1.7+1.2,3.6);
                \draw[line] (0.9+1.2,0+0.2+3.6)--(1.7+1.2,0+0.2+3.6);
            \end{tikzpicture}
        \end{gathered}\qquad \qquad \qquad
        \begin{gathered}
        \begin{tikzpicture}
                \draw[line,very thick] (0,1.2)--(1.2,1.2);
                \draw[line,very thick] (1.2,0)--(1.2,1.2);
                \draw[line,very thick] (1.2,1.2)--(2.4,2.4);
                \draw[line,very thick] (2.4,2.4)--(3.6,2.4);
                \draw[line,very thick] (2.4,2.4)--(2.4,3.6);
                \node at (0.6,1.5) {$\theta_2$};
                \node at (0.8,0.6) {$-\theta_1$};
                \node at (1.8-0.6,1.8+0.4) {$\theta_1+\theta_2$};
                \draw[line] (0,0.7)--(0,1.5);
                \draw[line] (3.6,0.7+1.2)--(3.6,1.5+1.2);
                \draw[line] (0.9,0)--(1.7,0);
                \draw[line] (0.9,0-0.2)--(1.7,0-0.2);
                \draw[line] (0.9+1.2,3.6)--(1.7+1.2,3.6);
                \draw[line] (0.9+1.2,0+0.2+3.6)--(1.7+1.2,0+0.2+3.6);
                \draw[line,red] (0,1)--(1.2+0.2,1)--(1.2+0.2+1.2,1+1.2)--(1.2+0.2+1.2+1,1+1.2);
            \end{tikzpicture}
        \end{gathered}
    \end{equation}
and another two by flipping the sign of both $\theta_1$ and $\theta_2$. Here the endpoints of the vertical and horizontal lines are identified, separately, and we omit the arrow in \eqref{eq:defect_network} to simplify the notation. Notice that in the second diagram, one starts with $U_M(-\theta_1)$ which is flipped to $U_M(\theta_1)$ after crossing the $C$-lines. Similarly, for the twisted sector we have the following diagrams
    
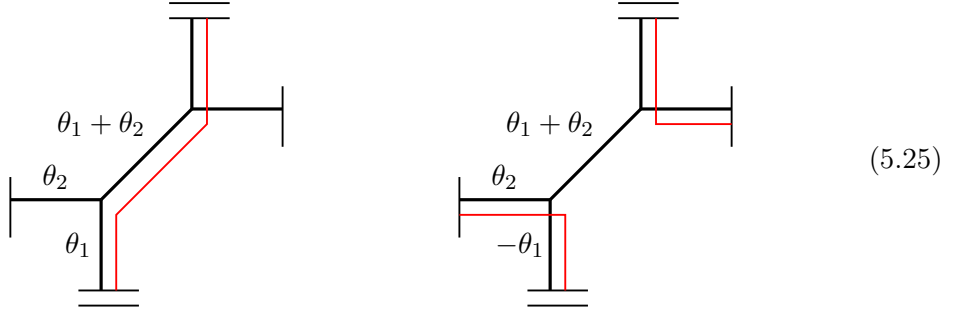
\begin{equation}
    \begin{gathered}
        \begin{tikzpicture}
                \draw[line,very thick] (0,1.2)--(1.2,1.2);
                \draw[line,very thick] (1.2,0)--(1.2,1.2);
                \draw[line,very thick] (1.2,1.2)--(2.4,2.4);
                \draw[line,very thick] (2.4,2.4)--(3.6,2.4);
                \draw[line,very thick] (2.4,2.4)--(2.4,3.6);
                \node at (0.6,1.5) {$\theta_2$};
                \node at (0.9,0.6) {$\theta_1$};
                \node at (1.8-0.6,1.8+0.4) {$\theta_1+\theta_2$};
                \draw[line] (0,0.7)--(0,1.5);
                \draw[line] (3.6,0.7+1.2)--(3.6,1.5+1.2);
                \draw[line] (0.9,0)--(1.7,0);
                \draw[line] (0.9,0-0.2)--(1.7,0-0.2);
                \draw[line] (0.9+1.2,3.6)--(1.7+1.2,3.6);
                \draw[line] (0.9+1.2,0+0.2+3.6)--(1.7+1.2,0+0.2+3.6);
                \draw[line,red] (1.2+0.2,0)--(1.2+0.2,1)--(1.2+0.2+1.2,1+1.2)--(1.2+0.2+1.2,1+1.2+1.4);
            \end{tikzpicture}
        \end{gathered}\qquad \qquad \qquad
        \begin{gathered}
        \begin{tikzpicture}
                \draw[line,very thick] (0,1.2)--(1.2,1.2);
                \draw[line,very thick] (1.2,0)--(1.2,1.2);
                \draw[line,very thick] (1.2,1.2)--(2.4,2.4);
                \draw[line,very thick] (2.4,2.4)--(3.6,2.4);
                \draw[line,very thick] (2.4,2.4)--(2.4,3.6);
                \node at (0.6,1.5) {$\theta_2$};
                \node at (0.8,0.6) {$-\theta_1$};
                \node at (1.8-0.6,1.8+0.4) {$\theta_1+\theta_2$};
                \draw[line] (0,0.7)--(0,1.5);
                \draw[line] (3.6,0.7+1.2)--(3.6,1.5+1.2);
                \draw[line] (0.9,0)--(1.7,0);
                \draw[line] (0.9,0-0.2)--(1.7,0-0.2);
                \draw[line] (0.9+1.2,3.6)--(1.7+1.2,3.6);
                \draw[line] (0.9+1.2,0+0.2+3.6)--(1.7+1.2,0+0.2+3.6);
                \draw[line,red] (0,1)--(1.2+0.2,1)--(1.2+0.2,0);
                \draw[line,red] (1.2+0.2+1.2,1+1.2+1.4)--(1.2+0.2+1.2,1+1.2)--(1.2+0.2+1.2+1,1+1.2);
            \end{tikzpicture}
        \end{gathered}
    \end{equation}
and another two by flipping the sign of both $\theta_1$ and $\theta_2$.

Let us write down the partition function for each diagram. For the first diagram, we simply have
    \begin{equation}
        \frac{1}{2} \left(\lambda_{\theta_1,\theta_2}+\lambda_{-\theta_1,-\theta_2} \right)\,,
    \end{equation}
and for the second diagram, since $U_M(\theta_2)C$ do not commute with $U_M(-\theta_1)$, the contribution vanishes. The third diagram vanishes due to the same reason. For the fourth diagram, notice that
    \begin{equation}
        (U_M(\theta_2)C) (U_M(-\theta_1)C)(U_M(\theta_2)C)^{-1}=U_M(2\theta_2+\theta_1) C\,,
    \end{equation}
therefore $U_M(\theta_2)C$ commute with $U_M(-\theta_1)C$ if and only if
    \begin{equation}
        2\theta_2 +2\theta_1 = 0\quad \text{mod} \quad 2\pi\,,
    \end{equation}
which implies $\theta_2 = \pi - \theta_1$ for $\theta_1,\theta_2 \in (0,\pi)$. With the insertion of $U_M(-\theta_1)C$ defect, the fixed points are
    \begin{equation}
        \sigma_0 = -\frac{\theta_1}{2} r \,, \quad \text{and} \quad \sigma_1=-\frac{\theta_1}{2}r+\pi r\,.
    \end{equation}
However, the symmetry operator $U_M(\theta_2)C = U_M(-\theta_1+\pi)C$ will exchange the two fixed points $\sigma_0 \leftrightarrow \sigma_1$, and thus exchange the oscillators localized on the two fixed points. After taking the trace, the result will vanish. As a summary, the partition function is
    \begin{equation}
        Z_{\text{orb}}\,
        \begin{gathered}
\begin{tikzpicture}[scale=.75]
\node at (0, 1.45) {};
\draw[line] (-.15-.4,-.2-.4)--(-.25-.4, -.2-.4) -- (-.25-.4, 1.4)-- (-.15-.4,1.4);
\draw[very thick] (0,0) rectangle (1.2,1.2);
\draw[line]  (.6,0) arc (0:90:.6); 
\draw[line] (.6,1.2) arc (-180:-90:.6);
\draw[line] (0.6/2^0.5,0.6/2^0.5)--(1.2-0.6/2^0.5,1.2-0.6/2^0.5);
\draw[line] (1.35, -.2-.4)--(1.45, -.2-.4) -- (1.45, 1.4)--(1.35, 1.4);
\node at (-.15-.2,-.2+0.8) {$\theta_2$};
\node at (.6,-.35) {$\theta_1$};
\node at (0.6+0.15,0.6-0.15) {$+$};
\end{tikzpicture}
\end{gathered} = \frac{1}{2} \left(\lambda_{\theta_1,\theta_2}+\lambda_{-\theta_1,-\theta_2} \right)\,,
\end{equation}
where $L_M^{+}$ should be understood as $\eta_m$ when $\theta_1+\theta_2=\pi$.

Finally, for the $L_M^-$ diagram, the untwisted sector contributions from the circle branch can be drawn as the following diagrams
    \begin{equation}
    \begin{gathered}
        \begin{tikzpicture}
                \draw[line,very thick] (0,1.2)--(1.2,1.2);
                \draw[line,very thick] (1.2,0)--(1.2,1.2);
                \draw[line,very thick] (1.2,1.2)--(2.4,2.4);
                \draw[line,very thick] (2.4,2.4)--(3.6,2.4);
                \draw[line,very thick] (2.4,2.4)--(2.4,3.6);
                \node at (0.6,1.5) {$\theta_2$};
                \node at (0.8,0.6) {$-\theta_1$};
                \node at (1.8-0.6,1.8+0.4) {$-\theta_1+\theta_2$};
                \draw[line] (0,0.7)--(0,1.5);
                \draw[line] (3.6,0.7+1.2)--(3.6,1.5+1.2);
                \draw[line] (0.9,0)--(1.7,0);
                \draw[line] (0.9,0-0.2)--(1.7,0-0.2);
                \draw[line] (0.9+1.2,3.6)--(1.7+1.2,3.6);
                \draw[line] (0.9+1.2,0+0.2+3.6)--(1.7+1.2,0+0.2+3.6);
            \end{tikzpicture}
        \end{gathered}\qquad \qquad \qquad
        \begin{gathered}
        \begin{tikzpicture}
                \draw[line,very thick] (0,1.2)--(1.2,1.2);
                \draw[line,very thick] (1.2,0)--(1.2,1.2);
                \draw[line,very thick] (1.2,1.2)--(2.4,2.4);
                \draw[line,very thick] (2.4,2.4)--(3.6,2.4);
                \draw[line,very thick] (2.4,2.4)--(2.4,3.6);
                \node at (0.6,1.5) {$\theta_2$};
                \node at (0.9,0.6) {$\theta_1$};
                \node at (1.8-0.6,1.8+0.4) {$-\theta_1+\theta_2$};
                \draw[line] (0,0.7)--(0,1.5);
                \draw[line] (3.6,0.7+1.2)--(3.6,1.5+1.2);
                \draw[line] (0.9,0)--(1.7,0);
                \draw[line] (0.9,0-0.2)--(1.7,0-0.2);
                \draw[line] (0.9+1.2,3.6)--(1.7+1.2,3.6);
                \draw[line] (0.9+1.2,0+0.2+3.6)--(1.7+1.2,0+0.2+3.6);
                \draw[line,red] (0,1)--(1.2+0.2,1)--(1.2+0.2+1.2,1+1.2)--(1.2+0.2+1.2+1,1+1.2);
            \end{tikzpicture}
        \end{gathered}
    \end{equation}
and another two by flipping the sign of both $\theta_1$ and $\theta_2$. Similarly, for the twisted sector we have the following diagrams
    \begin{equation}
    \begin{gathered}
        \begin{tikzpicture}
                \draw[line,very thick] (0,1.2)--(1.2,1.2);
                \draw[line,very thick] (1.2,0)--(1.2,1.2);
                \draw[line,very thick] (1.2,1.2)--(2.4,2.4);
                \draw[line,very thick] (2.4,2.4)--(3.6,2.4);
                \draw[line,very thick] (2.4,2.4)--(2.4,3.6);
                \node at (0.6,1.5) {$\theta_2$};
                \node at (0.8,0.6) {$-\theta_1$};
                \node at (1.8-0.6,1.8+0.4) {$-\theta_1+\theta_2$};
                \draw[line] (0,0.7)--(0,1.5);
                \draw[line] (3.6,0.7+1.2)--(3.6,1.5+1.2);
                \draw[line] (0.9,0)--(1.7,0);
                \draw[line] (0.9,0-0.2)--(1.7,0-0.2);
                \draw[line] (0.9+1.2,3.6)--(1.7+1.2,3.6);
                \draw[line] (0.9+1.2,0+0.2+3.6)--(1.7+1.2,0+0.2+3.6);
                \draw[line,red] (1.2+0.2,0)--(1.2+0.2,1)--(1.2+0.2+1.2,1+1.2)--(1.2+0.2+1.2,1+1.2+1.4);
            \end{tikzpicture}
        \end{gathered}\qquad \qquad \qquad
        \begin{gathered}
        \begin{tikzpicture}
                \draw[line,very thick] (0,1.2)--(1.2,1.2);
                \draw[line,very thick] (1.2,0)--(1.2,1.2);
                \draw[line,very thick] (1.2,1.2)--(2.4,2.4);
                \draw[line,very thick] (2.4,2.4)--(3.6,2.4);
                \draw[line,very thick] (2.4,2.4)--(2.4,3.6);
                \node at (0.6,1.5) {$\theta_2$};
                \node at (0.9,0.6) {$\theta_1$};
                \node at (1.8-0.6,1.8+0.4) {$-\theta_1+\theta_2$};
                \draw[line] (0,0.7)--(0,1.5);
                \draw[line] (3.6,0.7+1.2)--(3.6,1.5+1.2);
                \draw[line] (0.9,0)--(1.7,0);
                \draw[line] (0.9,0-0.2)--(1.7,0-0.2);
                \draw[line] (0.9+1.2,3.6)--(1.7+1.2,3.6);
                \draw[line] (0.9+1.2,0+0.2+3.6)--(1.7+1.2,0+0.2+3.6);
                \draw[line,red] (0,1)--(1.2+0.2,1)--(1.2+0.2,0);
                \draw[line,red] (1.2+0.2+1.2,1+1.2+1.4)--(1.2+0.2+1.2,1+1.2)--(1.2+0.2+1.2+1,1+1.2);
            \end{tikzpicture}
        \end{gathered}
    \end{equation}
and another two by flipping the sign of both $\theta_1$ and $\theta_2$. The contributions to the partition function can be read in exactly the same way as before. The first diagram gives
    \begin{equation}
        \frac{1}{2} \left(\lambda_{-\theta_1,\theta_2}+\lambda_{-\theta_1,\theta_2} \right)\,,
    \end{equation}
and the second and third diagram vanish. For the last diagram, using the fact
    \begin{equation}
        (U_M(\theta_2)C) (U_M(\theta_1)C)(U_M(\theta_2)C)^{-1}=U_M(2\theta_2-\theta_1) C\,,
    \end{equation}
therefore $U_M(\theta_2)C$ commute with $U_M(\theta_1)C$ if and only if
    \begin{equation}
        2\theta_2 -2\theta_1 = 0\quad \text{mod} \quad 2\pi\,,
    \end{equation}
which implies $\theta_2 = \theta_1$ for $\theta_1,\theta_2 \in (0,\pi)$. With the insertion of $U_M(\theta_1)C$ defect, the fixed points are
    \begin{equation}
        \sigma_0 = \frac{\theta_1}{2} r \,, \quad \text{and} \quad \sigma_1=\frac{\theta_1}{2}r+\pi\,.
    \end{equation}
Compared to the $L_M^+$ case, now the symmetry operator $U_M(\theta_2)C = U_M(\theta_1)C$ leaves both fixed points invariant, so that the fourth diagram gives
\begin{equation}
    \frac{|\vartheta_2 \vartheta_4|}{|\eta|^2}\,,
\end{equation}
when $\theta_1=\theta_2$. Therefore the partition functions read
    \begin{equation}
        Z_{\text{orb}}\,
        \begin{gathered}
\begin{tikzpicture}[scale=.75]
\node at (0, 1.45) {};
\draw[line] (-.15-.4,-.2-.4)--(-.25-.4, -.2-.4) -- (-.25-.4, 1.4)-- (-.15-.4,1.4);
\draw[very thick] (0,0) rectangle (1.2,1.2);
\draw[line]  (.6,0) arc (0:90:.6); 
\draw[line] (.6,1.2) arc (-180:-90:.6);
\draw[line] (0.6/2^0.5,0.6/2^0.5)--(1.2-0.6/2^0.5,1.2-0.6/2^0.5);
\draw[line] (1.35, -.2-.4)--(1.45, -.2-.4) -- (1.45, 1.4)--(1.35, 1.4);
\node at (-.15-.2,-.2+0.8) {$\theta_2$};
\node at (.6,-.35) {$\theta_1$};
\node at (0.6+0.15,0.6-0.15) {$-$};
\end{tikzpicture}
\end{gathered} = \frac{1}{2} \left(\lambda_{-\theta_1,\theta_2}+\lambda_{-\theta_1,\theta_2} \right) + \delta_{\theta_1,\theta_2} \frac{|\vartheta_2 \vartheta_4|}{|\eta|^2}\,,
\end{equation}
where $L_M^{-}$ should be understood as the identity when $\theta_1=\theta_2$, and the corresponding partition function is related to the $L_M$-defect partition function via a modular $T$-transformation.

We also need to consider the partition function with insertion of $\eta_m$ symmetry operator or/and defect. Since $\eta_m$ is invertible, we have three sectors labeled by the diagrams
\begin{equation}
    \begin{gathered}
        \begin{tikzpicture}
            \draw[very thick] (0,0) rectangle (1.8,1.8);
            \draw[line,dashed] (0,0.9)--(1.8,0.9);
        \end{tikzpicture}
    \end{gathered}\qquad 
    \begin{gathered}
        \begin{tikzpicture}
            \draw[very thick] (0,0) rectangle (1.8,1.8);
            \draw[line,dashed] (0.9,0)--(0.9,1.8);
        \end{tikzpicture}
    \end{gathered}\qquad
    \begin{gathered}
        \begin{tikzpicture}
            \draw[very thick] (0,0) rectangle (1.8,1.8);
            \draw[line,dashed]  (.9,0) arc (0:90:.9); 
            \draw[line,dashed] (.9,1.8) arc (-180:-90:.9);
        \end{tikzpicture}
    \end{gathered}
\end{equation}
where we use a dashed line to represent $\eta_m$. The computation is exactly the same since we can treat $\eta_m$ as $U(\pi)$ in the circle branch, and we will omit the details and give the results
\begin{equation}
\begin{split}
    Z_{\text{orb}}
        \,\begin{gathered}
\begin{tikzpicture}[scale=.75]
\node at (0, 1.45) {};
\draw[line] (-.15,-.2)--(-.25, -.2) -- (-.25, 1.4)-- (-.15,1.4);
\draw[very thick] (0,0) rectangle (1.2,1.2);
\draw[line, dashed] (0,.6)--(1.2,.6);
\draw[line] (1.35, -.2)--(1.45, -.2) -- (1.45, 1.4)--(1.35, 1.4);
\end{tikzpicture}
\end{gathered} \quad =& \quad \frac{1}{2}\lambda_{\pi,0} + \frac{|\vartheta_3 \vartheta_4|}{2|\eta|^2}\,,\\
    Z_{\text{orb}}
        \,\begin{gathered}
\begin{tikzpicture}[scale=.75]
\node at (0, 1.45) {};
\draw[line] (-.15,-.2)--(-.25, -.2) -- (-.25, 1.4)-- (-.15,1.4);
\draw[very thick] (0,0) rectangle (1.2,1.2);
\draw[line, dashed] (.6,0)--(.6,1.2);
\draw[line] (1.35, -.2)--(1.45, -.2) -- (1.45, 1.4)--(1.35, 1.4);
\end{tikzpicture}
\end{gathered} \quad =& \quad \frac{1}{2} \lambda_{0,\pi}+\frac{|\vartheta_2 \vartheta_3|}{2|\eta|^2}\,,\\
    Z_{\text{orb}}
        \,\begin{gathered}
\begin{tikzpicture}[scale=.75]
\node at (0, 1.45) {};
\draw[line] (-.15,-.2)--(-.25, -.2) -- (-.25, 1.4)-- (-.15,1.4);
\draw[very thick] (0,0) rectangle (1.2,1.2);
\draw[line,dashed]  (.6,0) arc (0:90:.6); 
\draw[line,dashed] (.6,1.2) arc (-180:-90:.6);
\draw[line] (1.35, -.2)--(1.45, -.2) -- (1.45, 1.4)--(1.35, 1.4);
\end{tikzpicture}
\end{gathered} \quad =& \quad \frac{1}{2} \lambda_{\pi,\pi}+\frac{|\vartheta_2 \vartheta_4|}{2|\eta|^2}\,.
\end{split}
\end{equation}
Lastly, we can consider the mixed sector of $\eta_m$ and $L_M$ denoted by
    
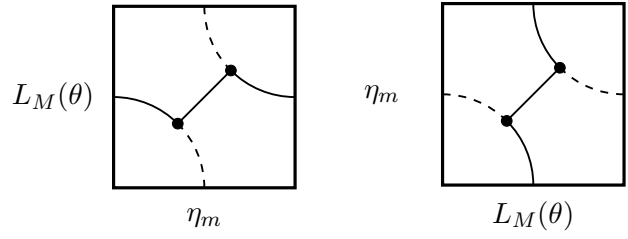
\begin{equation}
        \begin{gathered}
\begin{tikzpicture}[scale=2]
\draw[very thick] (0,0) rectangle (1.2,1.2);
\draw[line,dashed]  (.6,0) arc (0:45:.6); 
\draw[line] (0,0.6) arc (90:45:.6); 
\draw[line,dashed] (.6,1.2) arc (-180:-135:.6);
\draw[line] (1.2,0.6) arc (-90:-135:.6);
\draw[line] (0.6/1.414,0.6/1.414)--(1.2-0.6/1.414,1.2-0.6/1.414);
\node at (-0.4,0.6) {$L_M(\theta)$};
\node at (0.6,-0.2) {$\eta_m$};
\filldraw[black] (0.6/1.414,0.6/1.414) circle (1pt);
\filldraw[black] (1.2-0.6/1.414,1.2-0.6/1.414) circle (1pt);
\end{tikzpicture}
\end{gathered}\qquad
        \begin{gathered}
\begin{tikzpicture}[scale=2]
\draw[very thick] (0,0) rectangle (1.2,1.2);
\draw[line]  (.6,0) arc (0:45:.6); 
\draw[line,dashed]  (0,0.6) arc (90:45:.6); 
\draw[line] (.6,1.2) arc (-180:-135:.6);
\draw[line,dashed] (1.2,0.6) arc (-90:-135:.6);
\draw[line] (0.6/1.414,0.6/1.414)--(1.2-0.6/1.414,1.2-0.6/1.414);
\node at (-0.4,0.6) {$\eta_m$};
\node at (0.6,-0.2) {$L_M(\theta)$};
\filldraw[black] (0.6/1.414,0.6/1.414) circle (1pt);
\filldraw[black] (1.2-0.6/1.414,1.2-0.6/1.414) circle (1pt);
\end{tikzpicture}
\end{gathered}\nonumber
    \end{equation}
and the corresponding partition functions are
    \begin{equation}
    \begin{split}
        Z_{\text{orb}}\,
        \begin{gathered}
\begin{tikzpicture}[scale=.75]
\node at (0, 1.45) {};
\draw[line] (-.15-.4,-.2-.4)--(-.25-.4, -.2-.4) -- (-.25-.4, 1.4)-- (-.15-.4,1.4);
\draw[very thick] (0,0) rectangle (1.2,1.2);
\draw[line,dashed]  (.6,0) arc (0:45:.6); 
\draw[line] (0,0.6) arc (90:45:.6); 
\draw[line,dashed] (.6,1.2) arc (-180:-135:.6);
\draw[line] (1.2,0.6) arc (-90:-135:.6);
\draw[line] (0.6/1.414,0.6/1.414)--(1.2-0.6/1.414,1.2-0.6/1.414);
\draw[line] (1.35, -.2-.4)--(1.45, -.2-.4) -- (1.45, 1.4)--(1.35, 1.4);
\node at (-.15-.2,-.2+0.8) {$\theta$};
\end{tikzpicture}
\end{gathered}\quad =& \quad \frac{1}{2} \left( \lambda_{\theta,\pi} + \lambda_{-\theta,\pi}\right) \,,\\
        Z_{\text{orb}}\,
        \begin{gathered}
\begin{tikzpicture}[scale=.75]
\node at (0, 1.45) {};
\draw[line] (-.15-.4,-.2-.4)--(-.25-.4, -.2-.4) -- (-.25-.4, 1.4)-- (-.15-.4,1.4);
\draw[very thick] (0,0) rectangle (1.2,1.2);
\draw[line]  (.6,0) arc (0:45:.6); 
\draw[line,dashed]  (0,0.6) arc (90:45:.6); 
\draw[line] (.6,1.2) arc (-180:-135:.6);
\draw[line,dashed] (1.2,0.6) arc (-90:-135:.6);
\draw[line] (0.6/1.414,0.6/1.414)--(1.2-0.6/1.414,1.2-0.6/1.414);
\draw[line] (1.35, -.2-.4)--(1.45, -.2-.4) -- (1.45, 1.4)--(1.35, 1.4);
\node at (.6,-.35) {$\theta$};
\end{tikzpicture}
\end{gathered}\quad =& \quad \frac{1}{2} \left( \lambda_{\pi,\theta} +  \lambda_{\pi,-\theta} \right)\,.
\end{split}
\end{equation}

\subsection{Gauging non-invertible symmetry at the level of partition function}

We first summarize the partition with general non-invertible symmetry backgrounds. The partition function with nothing inserted is
\begin{equation}
    Z_{\text{orb}}\,\begin{gathered}
\begin{tikzpicture}[scale=.75]
\node at (0, 1.45) {};
\draw[line] (-.15,-.2)--(-.25, -.2) -- (-.25, 1.4)-- (-.15,1.4);
\draw[very thick] (0,0) rectangle (1.2,1.2);
\draw[line] (1.35, -.2)--(1.45, -.2) -- (1.45, 1.4)--(1.35, 1.4);
\end{tikzpicture}
\end{gathered} \quad = \quad \frac{1}{2} \left(Z(\tau) +\frac{|\vartheta_3 \vartheta_4|}{|\eta|^2} \right) + \frac{1}{2}\left(\frac{|\vartheta_2 \vartheta_3|}{|\eta|^2} + \frac{|\vartheta_2 \vartheta_4|}{|\eta|^2}\right)\,.
\end{equation}
Insert $\eta_m$ along the spatial or/and temporal 1-cycles, we have
\begin{equation}
\begin{split}
    Z_{\text{orb}}
        \,\begin{gathered}
\begin{tikzpicture}[scale=.75]
\node at (0, 1.45) {};
\draw[line] (-.15,-.2)--(-.25, -.2) -- (-.25, 1.4)-- (-.15,1.4);
\draw[very thick] (0,0) rectangle (1.2,1.2);
\draw[line, dashed] (0,.6)--(1.2,.6);
\draw[line] (1.35, -.2)--(1.45, -.2) -- (1.45, 1.4)--(1.35, 1.4);
\end{tikzpicture}
\end{gathered} \quad =& \quad \frac{1}{2}\lambda_{\pi,0} + \frac{|\vartheta_3 \vartheta_4|}{2|\eta|^2}\,,\\
    Z_{\text{orb}}
        \,\begin{gathered}
\begin{tikzpicture}[scale=.75]
\node at (0, 1.45) {};
\draw[line] (-.15,-.2)--(-.25, -.2) -- (-.25, 1.4)-- (-.15,1.4);
\draw[very thick] (0,0) rectangle (1.2,1.2);
\draw[line, dashed] (.6,0)--(.6,1.2);
\draw[line] (1.35, -.2)--(1.45, -.2) -- (1.45, 1.4)--(1.35, 1.4);
\end{tikzpicture}
\end{gathered} \quad =& \quad \frac{1}{2} \lambda_{0,\pi}+\frac{|\vartheta_2 \vartheta_3|}{2|\eta|^2}\,,\\
    Z_{\text{orb}}
        \,\begin{gathered}
\begin{tikzpicture}[scale=.75]
\node at (0, 1.45) {};
\draw[line] (-.15,-.2)--(-.25, -.2) -- (-.25, 1.4)-- (-.15,1.4);
\draw[very thick] (0,0) rectangle (1.2,1.2);
\draw[line,dashed]  (.6,0) arc (0:90:.6); 
\draw[line,dashed] (.6,1.2) arc (-180:-90:.6);
\draw[line] (1.35, -.2)--(1.45, -.2) -- (1.45, 1.4)--(1.35, 1.4);
\end{tikzpicture}
\end{gathered} \quad =& \quad \frac{1}{2} \lambda_{\pi,\pi}+\frac{|\vartheta_2 \vartheta_4|}{2|\eta|^2}\,.
\end{split}
\end{equation}
Insert $L_M(\theta)$ along the spatial or temporal cycle give
    \begin{equation}
    \begin{split}
        Z_{\text{orb}}\,
        \begin{gathered}
\begin{tikzpicture}[scale=.75]
\node at (0, 1.45) {};
\draw[line] (-.15-.4,-.2-.4)--(-.25-.4, -.2-.4) -- (-.25-.4, 1.4)-- (-.15-.4,1.4);
\draw[very thick] (0,0) rectangle (1.2,1.2);
\draw[line, thick] (0,.6)--(1.2,.6);
\draw[line] (1.35, -.2-.4)--(1.45, -.2-.4) -- (1.45, 1.4)--(1.35, 1.4);
\node at (-.15-.2,-.2+0.8) {$\theta$};
\end{tikzpicture}
\end{gathered}
\quad =& \quad \frac{1}{2} \left( \lambda_{\theta,0} + \lambda_{-\theta,0} \right) +\frac{|\vartheta_3 \vartheta_4|}{|\eta|^2}\,,
\\
        Z_{\text{orb}}\,
        \begin{gathered}
\begin{tikzpicture}[scale=.75]
\node at (0, 1.45) {};
\draw[line] (-.15-.4,-.2-.4)--(-.25-.4, -.2-.4) -- (-.25-.4, 1.4)-- (-.15-.4,1.4);
\draw[very thick] (0,0) rectangle (1.2,1.2);
\draw[line, thick] (.6,0)--(.6,1.2);
\draw[line] (1.35, -.2-.4)--(1.45, -.2-.4) -- (1.45, 1.4)--(1.35, 1.4);
\node at (.6,-.35) {$\theta$};
\end{tikzpicture}
\end{gathered}
\quad =& \quad \frac{1}{2}(\lambda_{0,\theta}+\lambda_{0,-\theta})+\frac{|\vartheta_2 \vartheta_3|}{|\eta|^2}\,.
\end{split}
    \end{equation}
Insert $L_M$ along both cycles, depending on the fusion channel we have two different defect networks
        \begin{equation}
        \begin{split}
        Z_{\text{orb}}\,
        \begin{gathered}
\begin{tikzpicture}[scale=.75]
\node at (0, 1.45) {};
\draw[line] (-.15-.4,-.2-.4)--(-.25-.4, -.2-.4) -- (-.25-.4, 1.4)-- (-.15-.4,1.4);
\draw[very thick] (0,0) rectangle (1.2,1.2);
\draw[line]  (.6,0) arc (0:90:.6); 
\draw[line] (.6,1.2) arc (-180:-90:.6);
\draw[line] (0.6/2^0.5,0.6/2^0.5)--(1.2-0.6/2^0.5,1.2-0.6/2^0.5);
\draw[line] (1.35, -.2-.4)--(1.45, -.2-.4) -- (1.45, 1.4)--(1.35, 1.4);
\node at (-.15-.2,-.2+0.8) {$\theta_2$};
\node at (.6,-.35) {$\theta_1$};
\node at (0.6+0.15,0.6-0.15) {$+$};
\end{tikzpicture}
\end{gathered} \quad =& \quad \frac{1}{2} \left(\lambda_{\theta_1,\theta_2}+\lambda_{-\theta_1,-\theta_2} \right)\,,\\
        Z_{\text{orb}}\,
        \begin{gathered}
\begin{tikzpicture}[scale=.75]
\node at (0, 1.45) {};
\draw[line] (-.15-.4,-.2-.4)--(-.25-.4, -.2-.4) -- (-.25-.4, 1.4)-- (-.15-.4,1.4);
\draw[very thick] (0,0) rectangle (1.2,1.2);
\draw[line]  (.6,0) arc (0:90:.6); 
\draw[line] (.6,1.2) arc (-180:-90:.6);
\draw[line] (0.6/2^0.5,0.6/2^0.5)--(1.2-0.6/2^0.5,1.2-0.6/2^0.5);
\draw[line] (1.35, -.2-.4)--(1.45, -.2-.4) -- (1.45, 1.4)--(1.35, 1.4);
\node at (-.15-.2,-.2+0.8) {$\theta_2$};
\node at (.6,-.35) {$\theta_1$};
\node at (0.6+0.15,0.6-0.15) {$-$};
\end{tikzpicture}
\end{gathered} \quad =& \quad \frac{1}{2} \left(\lambda_{-\theta_1,\theta_2}+\lambda_{-\theta_1,\theta_2} \right) + \delta_{\theta_1,\theta_2} \frac{|\vartheta_2 \vartheta_4|}{|\eta|^2}\,,
\end{split}
\end{equation}
where $L_M(\theta_1),L_M(\theta_2)$ fuse to $L_M(\theta_1+\theta_2)$ in the first line, and $L_M(\theta_1-\theta_2)$ in the second line. Finally, we have the mixed sector
    \begin{equation}
    \begin{split}
        Z_{\text{orb}}\,
        \begin{gathered}
\begin{tikzpicture}[scale=.75]
\node at (0, 1.45) {};
\draw[line] (-.15-.4,-.2-.4)--(-.25-.4, -.2-.4) -- (-.25-.4, 1.4)-- (-.15-.4,1.4);
\draw[very thick] (0,0) rectangle (1.2,1.2);
\draw[line,dashed]  (.6,0) arc (0:45:.6); 
\draw[line] (0,0.6) arc (90:45:.6); 
\draw[line,dashed] (.6,1.2) arc (-180:-135:.6);
\draw[line] (1.2,0.6) arc (-90:-135:.6);
\draw[line] (0.6/1.414,0.6/1.414)--(1.2-0.6/1.414,1.2-0.6/1.414);
\draw[line] (1.35, -.2-.4)--(1.45, -.2-.4) -- (1.45, 1.4)--(1.35, 1.4);
\node at (-.15-.2,-.2+0.8) {$\theta$};
\end{tikzpicture}
\end{gathered}\quad =& \quad \frac{1}{2} \left( \lambda_{\theta,\pi} + \lambda_{-\theta,\pi} \right) \,,\\
        Z_{\text{orb}}\,
        \begin{gathered}
\begin{tikzpicture}[scale=.75]
\node at (0, 1.45) {};
\draw[line] (-.15-.4,-.2-.4)--(-.25-.4, -.2-.4) -- (-.25-.4, 1.4)-- (-.15-.4,1.4);
\draw[very thick] (0,0) rectangle (1.2,1.2);
\draw[line]  (.6,0) arc (0:45:.6); 
\draw[line,dashed]  (0,0.6) arc (90:45:.6); 
\draw[line] (.6,1.2) arc (-180:-135:.6);
\draw[line,dashed] (1.2,0.6) arc (-90:-135:.6);
\draw[line] (0.6/1.414,0.6/1.414)--(1.2-0.6/1.414,1.2-0.6/1.414);
\draw[line] (1.35, -.2-.4)--(1.45, -.2-.4) -- (1.45, 1.4)--(1.35, 1.4);
\node at (.6,-.35) {$\theta$};
\end{tikzpicture}
\end{gathered}\quad =& \quad \frac{1}{2} \left(\lambda_{\pi,\theta} +  \lambda_{\pi,-\theta}\right) \,.
\end{split}
\end{equation}

\paragraph{Gauging finite non-invertible symmetry} We first consider gauging the finite non-invertible symmetry, which changes the radius of the theory on the orbifold branch. Let us briefly review the categorical prescription for gauging in 2D. For an ordinary finite group symmetry, gauging can be described by summing over flat gauge configurations labeled by holonomies. In the language of TDLs, this is equivalently described by inserting a mesh of the regular algebra
$\oplus_{g\in G}g$ on the 2D spacetime. For a general finite non-invertible symmetry, the role of the regular algebra is played by a separable Frobenius algebra object $\mathcal{A}$ in the symmetry category. The associativity condition $\mathcal{A}\otimes (\mathcal{A}\otimes \mathcal{A}) = (\mathcal{A}\otimes \mathcal{A})\otimes \mathcal{A}$, together with the separability condition which allows one to shrink the $\mathcal{A}$-loop freely, imply that the $\mathcal{A}$-mesh can be freely deformed, so that the resulting partition function does not depend on the specific choice of the $\mathcal{A}$-network, and can always be simplified to the following on torus
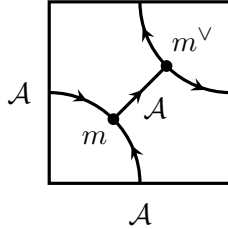
\begin{equation}\label{eq:frobenius-algebra-torus}
        \begin{gathered}
\begin{tikzpicture}[scale=2]
\draw[very thick] (0,0) rectangle (1.2,1.2);
\draw[very thick,->-=.55]  (.6,0) arc (0:45:.6); 
\draw[very thick,->-=.55]  (0,0.6) arc (90:45:.6); 
\draw[very thick,-<-=.55] (.6,1.2) arc (-180:-135:.6);
\draw[very thick,-<-=.55] (1.2,0.6) arc (-90:-135:.6);
\draw[very thick,->-=.55] (0.6/1.414,0.6/1.414)--(1.2-0.6/1.414,1.2-0.6/1.414);
\node at (-0.2,0.6) {$\mathcal{A}$};
\node at (0.6,-0.2) {$\mathcal{A}$};
\node at (0.6+0.1,0.6-0.1) {$\mathcal{A}$};
\filldraw[black] (0.6/1.414,0.6/1.414) circle (1pt);
\filldraw[black] (1.2-0.6/1.414,1.2-0.6/1.414) circle (1pt);
 \node at (0.3,0.3) {$m$};
 \node at (0.95,0.95) {$m^{\vee}$};
\end{tikzpicture}
\end{gathered}\nonumber
    \end{equation}
where the multiplication $m:\mathcal{A}\otimes \mathcal{A}\rightarrow \mathcal{A}$ and comultiplication $m^\vee:\mathcal{A}\rightarrow \mathcal{A}\otimes \mathcal{A}$ are the defining data for $\mathcal{A}$ in order to satisfy the (co)-associativity and separability. This gives the natural generalization of summing over flat gauge fields to the case where the symmetry lines are non-invertible \cite{Diatlyk:2023fwf}.

For the finite non-invertible symmetry generated by the momentum-type lines on the orbifold branch, the relevant algebra objects are
\begin{equation}
    \mathcal{A}_{2N} = 1\oplus \left(\bigoplus_{a=1}^{N-1} L_M(\frac{\pi a}{N}) \right)\oplus \eta_m\,,\quad 
    \mathcal{A}_{2N+1} = 1\oplus \left(\bigoplus_{a=1}^{N} L_M(\frac{\pi a}{N}) \right)\,,
\end{equation}
for $N\in \mathbb{Z}$. Consider the defect networks of $\mathcal{A}$ given by \eqref{eq:frobenius-algebra-torus}, and draw the components of $\mathcal{A}$ explicitly, we have the gauged partition functions
\begin{equation}
    \begin{split}
    Z_{\text{orb},\mathcal{A}_{2N}} =& 
    \frac{1}{2N}\left(Z_{\text{orb}}\,\begin{gathered}
\begin{tikzpicture}[scale=.75]
\node at (0, 1.45) {};
\draw[line] (-.15,-.2)--(-.25, -.2) -- (-.25, 1.4)-- (-.15,1.4);
\draw[very thick] (0,0) rectangle (1.2,1.2);
\draw[line] (1.35, -.2)--(1.45, -.2) -- (1.45, 1.4)--(1.35, 1.4);
\end{tikzpicture}
\end{gathered}
\ +\ 
Z_{\text{orb}}
        \,\begin{gathered}
\begin{tikzpicture}[scale=.75]
\node at (0, 1.45) {};
\draw[line] (-.15,-.2)--(-.25, -.2) -- (-.25, 1.4)-- (-.15,1.4);
\draw[very thick] (0,0) rectangle (1.2,1.2);
\draw[line, dashed] (0,.6)--(1.2,.6);
\draw[line] (1.35, -.2)--(1.45, -.2) -- (1.45, 1.4)--(1.35, 1.4);
\end{tikzpicture}
\end{gathered}
\ +\ 
Z_{\text{orb}}
        \,\begin{gathered}
\begin{tikzpicture}[scale=.75]
\node at (0, 1.45) {};
\draw[line] (-.15,-.2)--(-.25, -.2) -- (-.25, 1.4)-- (-.15,1.4);
\draw[very thick] (0,0) rectangle (1.2,1.2);
\draw[line, dashed] (.6,0)--(.6,1.2);
\draw[line] (1.35, -.2)--(1.45, -.2) -- (1.45, 1.4)--(1.35, 1.4);
\end{tikzpicture}
\end{gathered}
\ +\ 
Z_{\text{orb}}
        \,\begin{gathered}
\begin{tikzpicture}[scale=.75]
\node at (0, 1.45) {};
\draw[line] (-.15,-.2)--(-.25, -.2) -- (-.25, 1.4)-- (-.15,1.4);
\draw[very thick] (0,0) rectangle (1.2,1.2);
\draw[line,dashed]  (.6,0) arc (0:90:.6); 
\draw[line,dashed] (.6,1.2) arc (-180:-90:.6);
\draw[line] (1.35, -.2)--(1.45, -.2) -- (1.45, 1.4)--(1.35, 1.4);
\end{tikzpicture}
\end{gathered}\ 
\right)\\
&+ \frac{1}{2N}\sum_{a=1}^{N-1} \left(Z_{\text{orb}}\,
        \begin{gathered}
\begin{tikzpicture}[scale=.75]
\node at (0, 1.45) {};
\draw[line] (-.15-.4,-.2-.4)--(-.25-.4, -.2-.4) -- (-.25-.4, 1.4)-- (-.15-.4,1.4);
\draw[very thick] (0,0) rectangle (1.2,1.2);
\draw[line, thick] (0,.6)--(1.2,.6);
\draw[line] (1.35, -.2-.4)--(1.45, -.2-.4) -- (1.45, 1.4)--(1.35, 1.4);
\node at (-.15-.2,-.2+0.8) {$\frac{\pi a}{N}$};
\end{tikzpicture}
\end{gathered} \ + \ 
Z_{\text{orb}}\,
        \begin{gathered}
\begin{tikzpicture}[scale=.75]
\node at (0, 1.45) {};
\draw[line] (-.15-.4,-.2-.4)--(-.25-.4, -.2-.4) -- (-.25-.4, 1.4)-- (-.15-.4,1.4);
\draw[very thick] (0,0) rectangle (1.2,1.2);
\draw[line, thick] (.6,0)--(.6,1.2);
\draw[line] (1.35, -.2-.4)--(1.45, -.2-.4) -- (1.45, 1.4)--(1.35, 1.4);
\node at (.6,-.35) {$\frac{\pi a}{N}$};
\end{tikzpicture}
\end{gathered}\ \right)\\
&+ \frac{1}{2N} \sum_{a,b=1}^{N-1} \left( 
Z_{\text{orb}}\,
        \begin{gathered}
\begin{tikzpicture}[scale=.75]
\node at (0, 1.45) {};
\draw[line] (-.15-.4,-.2-.4)--(-.25-.4, -.2-.4) -- (-.25-.4, 1.4)-- (-.15-.4,1.4);
\draw[very thick] (0,0) rectangle (1.2,1.2);
\draw[line]  (.6,0) arc (0:90:.6); 
\draw[line] (.6,1.2) arc (-180:-90:.6);
\draw[line] (0.6/2^0.5,0.6/2^0.5)--(1.2-0.6/2^0.5,1.2-0.6/2^0.5);
\draw[line] (1.35, -.2-.4)--(1.45, -.2-.4) -- (1.45, 1.4)--(1.35, 1.4);
\node at (-.15-.2,-.2+0.8) {$\frac{\pi b}{N}$};
\node at (.6,-.35) {$\frac{\pi a}{N}$};
\node at (0.6+0.15,0.6-0.15) {$+$};
\end{tikzpicture}
\end{gathered} \ + \ 
Z_{\text{orb}}\,
        \begin{gathered}
\begin{tikzpicture}[scale=.75]
\node at (0, 1.45) {};
\draw[line] (-.15-.4,-.2-.4)--(-.25-.4, -.2-.4) -- (-.25-.4, 1.4)-- (-.15-.4,1.4);
\draw[very thick] (0,0) rectangle (1.2,1.2);
\draw[line]  (.6,0) arc (0:90:.6); 
\draw[line] (.6,1.2) arc (-180:-90:.6);
\draw[line] (0.6/2^0.5,0.6/2^0.5)--(1.2-0.6/2^0.5,1.2-0.6/2^0.5);
\draw[line] (1.35, -.2-.4)--(1.45, -.2-.4) -- (1.45, 1.4)--(1.35, 1.4);
\node at (-.15-.2,-.2+0.8) {$\frac{\pi b}{N}$};
\node at (.6,-.35) {$\frac{\pi a }{N}$};
\node at (0.6+0.15,0.6-0.15) {$-$};
\end{tikzpicture}
\end{gathered}\ 
\right)\\
&+\frac{1}{2N} \sum_{a=1}^{N-1} \left(
        Z_{\text{orb}}\,
        \begin{gathered}
\begin{tikzpicture}[scale=.75]
\node at (0, 1.45) {};
\draw[line] (-.15-.4,-.2-.4)--(-.25-.4, -.2-.4) -- (-.25-.4, 1.4)-- (-.15-.4,1.4);
\draw[very thick] (0,0) rectangle (1.2,1.2);
\draw[line,dashed]  (.6,0) arc (0:45:.6); 
\draw[line] (0,0.6) arc (90:45:.6); 
\draw[line,dashed] (.6,1.2) arc (-180:-135:.6);
\draw[line] (1.2,0.6) arc (-90:-135:.6);
\draw[line] (0.6/1.414,0.6/1.414)--(1.2-0.6/1.414,1.2-0.6/1.414);
\draw[line] (1.35, -.2-.4)--(1.45, -.2-.4) -- (1.45, 1.4)--(1.35, 1.4);
\node at (-.15-.2,-.2+0.8) {$\frac{\pi a}{N}$};
\end{tikzpicture}
\end{gathered} \ + \ 
Z_{\text{orb}}\,
        \begin{gathered}
\begin{tikzpicture}[scale=.75]
\node at (0, 1.45) {};
\draw[line] (-.15-.4,-.2-.4)--(-.25-.4, -.2-.4) -- (-.25-.4, 1.4)-- (-.15-.4,1.4);
\draw[very thick] (0,0) rectangle (1.2,1.2);
\draw[line]  (.6,0) arc (0:45:.6); 
\draw[line,dashed]  (0,0.6) arc (90:45:.6); 
\draw[line] (.6,1.2) arc (-180:-135:.6);
\draw[line,dashed] (1.2,0.6) arc (-90:-135:.6);
\draw[line] (0.6/1.414,0.6/1.414)--(1.2-0.6/1.414,1.2-0.6/1.414);
\draw[line] (1.35, -.2-.4)--(1.45, -.2-.4) -- (1.45, 1.4)--(1.35, 1.4);
\node at (.6,-.35) {$\frac{\pi a}N$};
\end{tikzpicture}
\end{gathered}\ 
\right)
\end{split}
\end{equation}
for $A_{2N}$ and
\begin{equation}
    \begin{split}
    Z_{\text{orb},\mathcal{A}_{2N+1}} =& 
    \frac{1}{2N+1}\left(Z_{\text{orb}}\,\begin{gathered}
\begin{tikzpicture}[scale=.75]
\node at (0, 1.45) {};
\draw[line] (-.15,-.2)--(-.25, -.2) -- (-.25, 1.4)-- (-.15,1.4);
\draw[very thick] (0,0) rectangle (1.2,1.2);
\draw[line] (1.35, -.2)--(1.45, -.2) -- (1.45, 1.4)--(1.35, 1.4);
\end{tikzpicture}
\end{gathered}\  
\right)\ 
+\  \frac{1}{2N+1}\sum_{a=1}^{N} \left(Z_{\text{orb}}\,
        \begin{gathered}
\begin{tikzpicture}[scale=.75]
\node at (0, 1.45) {};
\draw[line] (-.15-.4,-.2-.4)--(-.25-.4, -.2-.4) -- (-.25-.4, 1.4)-- (-.15-.4,1.4);
\draw[very thick] (0,0) rectangle (1.2,1.2);
\draw[line, thick] (0,.6)--(1.2,.6);
\draw[line] (1.35, -.2-.4)--(1.45, -.2-.4) -- (1.45, 1.4)--(1.35, 1.4);
\node at (-.15-.2,-.2+0.8) {$\frac{\pi a}{N}$};
\end{tikzpicture}
\end{gathered} \ + \ 
Z_{\text{orb}}\,
        \begin{gathered}
\begin{tikzpicture}[scale=.75]
\node at (0, 1.45) {};
\draw[line] (-.15-.4,-.2-.4)--(-.25-.4, -.2-.4) -- (-.25-.4, 1.4)-- (-.15-.4,1.4);
\draw[very thick] (0,0) rectangle (1.2,1.2);
\draw[line, thick] (.6,0)--(.6,1.2);
\draw[line] (1.35, -.2-.4)--(1.45, -.2-.4) -- (1.45, 1.4)--(1.35, 1.4);
\node at (.6,-.35) {$\frac{\pi a}{N}$};
\end{tikzpicture}
\end{gathered}\ \right)\\
&+ \frac{1}{2N+1} \sum_{a,b=1}^{N} \left( 
Z_{\text{orb}}\,
        \begin{gathered}
\begin{tikzpicture}[scale=.75]
\node at (0, 1.45) {};
\draw[line] (-.15-.4,-.2-.4)--(-.25-.4, -.2-.4) -- (-.25-.4, 1.4)-- (-.15-.4,1.4);
\draw[very thick] (0,0) rectangle (1.2,1.2);
\draw[line]  (.6,0) arc (0:90:.6); 
\draw[line] (.6,1.2) arc (-180:-90:.6);
\draw[line] (0.6/2^0.5,0.6/2^0.5)--(1.2-0.6/2^0.5,1.2-0.6/2^0.5);
\draw[line] (1.35, -.2-.4)--(1.45, -.2-.4) -- (1.45, 1.4)--(1.35, 1.4);
\node at (-.15-.2,-.2+0.8) {$\frac{\pi b}{N}$};
\node at (.6,-.35) {$\frac{\pi a}{N}$};
\node at (0.6+0.15,0.6-0.15) {$+$};
\end{tikzpicture}
\end{gathered} \ + \ 
Z_{\text{orb}}\,
        \begin{gathered}
\begin{tikzpicture}[scale=.75]
\node at (0, 1.45) {};
\draw[line] (-.15-.4,-.2-.4)--(-.25-.4, -.2-.4) -- (-.25-.4, 1.4)-- (-.15-.4,1.4);
\draw[very thick] (0,0) rectangle (1.2,1.2);
\draw[line]  (.6,0) arc (0:90:.6); 
\draw[line] (.6,1.2) arc (-180:-90:.6);
\draw[line] (0.6/2^0.5,0.6/2^0.5)--(1.2-0.6/2^0.5,1.2-0.6/2^0.5);
\draw[line] (1.35, -.2-.4)--(1.45, -.2-.4) -- (1.45, 1.4)--(1.35, 1.4);
\node at (-.15-.2,-.2+0.8) {$\frac{\pi b}{N}$};
\node at (.6,-.35) {$\frac{\pi a }{N}$};
\node at (0.6+0.15,0.6-0.15) {$-$};
\end{tikzpicture}
\end{gathered}\ 
\right)\\
&+\frac{1}{2N+1} \sum_{a=1}^{N} \left(
        Z_{\text{orb}}\,
        \begin{gathered}
\begin{tikzpicture}[scale=.75]
\node at (0, 1.45) {};
\draw[line] (-.15-.4,-.2-.4)--(-.25-.4, -.2-.4) -- (-.25-.4, 1.4)-- (-.15-.4,1.4);
\draw[very thick] (0,0) rectangle (1.2,1.2);
\draw[line,dashed]  (.6,0) arc (0:45:.6); 
\draw[line] (0,0.6) arc (90:45:.6); 
\draw[line,dashed] (.6,1.2) arc (-180:-135:.6);
\draw[line] (1.2,0.6) arc (-90:-135:.6);
\draw[line] (0.6/1.414,0.6/1.414)--(1.2-0.6/1.414,1.2-0.6/1.414);
\draw[line] (1.35, -.2-.4)--(1.45, -.2-.4) -- (1.45, 1.4)--(1.35, 1.4);
\node at (-.15-.2,-.2+0.8) {$\frac{\pi a}{N}$};
\end{tikzpicture}
\end{gathered} \ + \ 
Z_{\text{orb}}\,
        \begin{gathered}
\begin{tikzpicture}[scale=.75]
\node at (0, 1.45) {};
\draw[line] (-.15-.4,-.2-.4)--(-.25-.4, -.2-.4) -- (-.25-.4, 1.4)-- (-.15-.4,1.4);
\draw[very thick] (0,0) rectangle (1.2,1.2);
\draw[line]  (.6,0) arc (0:45:.6); 
\draw[line,dashed]  (0,0.6) arc (90:45:.6); 
\draw[line] (.6,1.2) arc (-180:-135:.6);
\draw[line,dashed] (1.2,0.6) arc (-90:-135:.6);
\draw[line] (0.6/1.414,0.6/1.414)--(1.2-0.6/1.414,1.2-0.6/1.414);
\draw[line] (1.35, -.2-.4)--(1.45, -.2-.4) -- (1.45, 1.4)--(1.35, 1.4);
\node at (.6,-.35) {$\frac{\pi a}N$};
\end{tikzpicture}
\end{gathered}\ 
\right)
\end{split}
\end{equation}
for $A_{2N+1}$, where we have $m=m^{\vee} = \frac{1}{\sqrt{2N}}$ or $\frac{1}{\sqrt{2N+1}}$ for all kind of junctions.

\paragraph{Flat gauging continuous non-invertible symmetry}

Let us finally discuss what happens when the finite algebra is replaced by the continuous family of non-invertible lines. A first natural prescription is to take the finite sums in $\mathcal{A}_{2N}$ or $\mathcal{A}_{2N+1}$ and replace them by an integral over the simple lines $L_M(\theta)$. In the defect partition functions above, the pieces involving the three theta functions are supported only at special loci, for example at the end points $\theta=0,\pi$, or on the diagonal locus in the $L_M$-$L_M$ background. These loci have zero measure in the continuous integral. Therefore the naive continuous average only keeps the regular contribution and gives $\frac{1}{2} Z_{\mathbb{R}_W}$, where $Z_{\mathbb{R}_W}$ is the partition function for the dual non-compact boson. 

However, this cannot be the full answer if the operation is supposed to describe the decompactification of the orbifold branch. Indeed, we may first flat gauge $U(1)_M$ on the circle branch and obtain the non-compact boson. If we then gauge the reflection $C$, the result should be the orbifold $\mathbb{R}_W/\mathbb{Z}_2$, which has a single fixed point at the origin $x=0$. Thus the expected answer should contain not only the regular term $\frac{1}{2}Z_{\mathbb{R}_W}$, but also the oscillator contribution from this fixed point. The naive integral over the simple labels $L_M(\theta)$ does not keep track of this contribution.

This mismatch shows that the measure-zero part should survive the flat gauging of continuous non-invertible symmetry. Moreover, since $\mathbb{R}_W/\mathbb{Z}_2$ has only a single fixed point, it also cannot be obtained by sending $N\rightarrow \infty$ from gauging the finite non-invertible symmetry described by $\mathcal{A}_{2N}$ or $\mathcal{A}_{2N+1}$, because the latter always have two fixed points. To see what happens here, it is useful to retreat to the circle branch and compare the problem with the flat gauging of the semidirect product $U(1)_M\rtimes C$. We will first review the finite group $\mathbb{Z}_{K}\rtimes C$, and then consider the continuous $U(1)_M$.

Let $K=2N+1$ or $2N$, we denote the translation and reflection elements in $\mathbb{Z}_{K}\rtimes C$ by
    \begin{equation}
        t= U_M\left(\frac{2\pi}{K} \right)\,, \quad T_a = t^a\,, \quad R_a = t^a C\,,\quad (a=0,\cdots,K-1)
    \end{equation}
and they obey
    \begin{equation}
        T_aT_bT_a^{-1} = T_b\,,\quad R_aT_b R_a^{-1} = T_{-b}\,,
    \end{equation}
and
    \begin{equation}
        T_a R_b T_a^{-1}=R_{b+2a}\,, \quad R_a R_b R_a^{-1}=R_{-b+2a}\,.
    \end{equation}
For a finite group, the gauged partition function can be written in the standard form
    \begin{equation}
        Z/G = \frac{1}{|G|}\sum_{\substack{g,h\in G\\gh=hg}} Z[g,h]\,,
    \end{equation}
where $g,h$ are holonomies along the two 1-cycles. The contribution from $g,h\in \mathbb{Z}_K$ is universal and gives
    \begin{equation}
        \frac{1}{2K} \sum_{a,b=0}^{K-1} Z[T_a,T_b] = \frac{1}{2} Z_{r/K}\,, 
    \end{equation}
and we will mainly focus on case where one of the holonomies involves $R_a$. 

Since the partition function is invariant under conjugation,
    \begin{equation}
        Z[kgk^{-1},khk^{-1}] = Z[g,h]\,,
    \end{equation}
one can rewrite the gauged partition function, using the orbit-stabilizer theorem, in terms of summation over the conjugacy class $[g,h]$ of commuting pair $(g,h)$
    \begin{equation}
        Z/G = \sum_{\substack{[g,h]\\gh=hg}} \frac{1}{|\text{Stab}(g,h)|} Z_{g,h}\,,
    \end{equation}
where the stabilizer group is
    \begin{equation}
        \text{Stab}(g,h) = \{k\in G|kgk^{-1}=g,khk^{-1}=h \} \,.
    \end{equation}

Because $R_b$ is in the same conjugacy class with $R_{b+2a}$ for any $a$, when $K=2N$ is even, there are two conjugacy classes labeled by $R_0,R_1$; and when $K=2N+1$ is odd, there is only one conjugacy class $R_0$. Let us first consider $K=2N$ is even. For each $i=0,1$, the commuting pair involving $R_i$ are
    \begin{equation}
        (R_i,1)\,, \quad (R_i,T_N)\,, \quad (1,R_i)\,,\quad (T_N,R_i)\,,\quad 
        (R_i,R_i)\,,\quad (T_N R_{i},R_i)\,,
    \end{equation}
where we can always fix one of the element to be $R_i$ by conjugation. Among them, the non-vanishing fixed sector partition functions are
    \begin{equation}
        Z[R_i,1] = \frac{|\vartheta_3 \vartheta_4|}{|\eta|^2}\,,\quad Z[1,R_i] = \frac{|\vartheta_2 \vartheta_3|}{|\eta|^2}\,,\quad Z[R_i,R_i] = \frac{|\vartheta_2 \vartheta_4|}{|\eta|^2}\,.
    \end{equation}
The stabilizer group of such non-vanishing pair is
    \begin{equation}
        \text{Stab}(R_i,1) = \text{Stab}(1,R_i)=\text{Stab}(R_i,R_i) = \{1\,,R_i\,,T_N\,,T_N R_i \}\,,
    \end{equation}
which has order 4. Thus their contribution to the gauged partition functions is
    \begin{equation}
        \sum_{i=0,1}\frac{1}{4} \left(Z[R_i,1]+Z[1,R_i]+Z[R_i,R_i] \right) =  \frac{|\vartheta_3 \vartheta_4|}{2|\eta|^2}+\frac{|\vartheta_2 \vartheta_3|}{2|\eta|^2}+\frac{|\vartheta_2 \vartheta_4|}{2|\eta|^2}\,.
    \end{equation}

The odd case looks different, but eventually gives the same answer. When $K=2N+1$, there is only one conjugacy class in the reflection sector labeled by $R_0=C$, and the commuting pair involving $R_0$ are only
    \begin{equation}
        (R_0,1)\,,\quad (1,R_0)\,,\quad (R_0,R_0)\,.
    \end{equation}
Now the stabilizer groups have order 2
    \begin{equation}
        \text{Stab}(R_0,1) = \text{Stab}(1,R_0)=\text{Stab}(R_0,R_0) = \{1\,,R_0\}\,.
    \end{equation}
Therefore their contribution to the gauged partition functions is again
    \begin{equation}
        \frac{1}{2} \left(Z[R_0,1]+Z[1,R_0]+Z[R_0,R_0] \right) =  \frac{|\vartheta_3 \vartheta_4|}{2|\eta|^2}+\frac{|\vartheta_2 \vartheta_3|}{2|\eta|^2}+\frac{|\vartheta_2 \vartheta_4|}{2|\eta|^2}\,.
    \end{equation}
Thus the two finite cases give the same orbifold partition function. The reason is slightly different in the two cases. For even $K$ there are two reflection conjugacy classes, but each relevant stabilizer group has order $4$. For odd $K$ there is only one reflection conjugacy class, but the stabilizer group has order $2$. These two effects compensate each other, as expected from a finite gauging which only changes the radius of the compact orbifold, leaving the two fixed-point contributions invariant.

Now we can consider the $U(1)$ case. Let
    \begin{equation}
        T_{\alpha}=U_M(\alpha)\,,\qquad R_{\alpha}=U_M(\alpha)C\,,
        \qquad \alpha\in [0,2\pi)\,.
    \end{equation}
Then
    \begin{equation}
        T_{\beta}R_{\alpha}T_{\beta}^{-1}=R_{\alpha+2\beta}\,,
    \end{equation}
so all reflection elements are conjugate to $R_0$. However, the half-period translation $T_{\pi}=U_M(\pi)$ still exists and belongs to the stabilizer group of a reflection sector. Therefore
    \begin{equation}
        \text{Stab}(R_{0},1)=\text{Stab}(1,R_{0})=\text{Stab}(R_{0},R_{0})
        =
        \{1\,,R_{0}\,,T_{\pi}\,,T_{\pi}R_{0}\}\,,
    \end{equation}
which has order $4$. The reflection contribution is then
    \begin{equation}
        \frac{1}{4} \left(Z[R_{0},1]+Z[1,R_{0}]+Z[R_{0},R_{0}] \right)
        =
        \frac{|\vartheta_3 \vartheta_4|}{4|\eta|^2}
        +\frac{|\vartheta_2 \vartheta_3|}{4|\eta|^2}
        +\frac{|\vartheta_2 \vartheta_4|}{4|\eta|^2}\,.
    \end{equation}
Together with the translation-translation contribution, this gives
    \begin{equation}
        Z/(U(1)_M\rtimes C)
        =
        \frac{1}{2}Z_{\mathbb{R}_W}
        +
        \frac{|\vartheta_3 \vartheta_4|}{4|\eta|^2}
        +\frac{|\vartheta_2 \vartheta_3|}{4|\eta|^2}
        +\frac{|\vartheta_2 \vartheta_4|}{4|\eta|^2}\,.
    \end{equation}
This is precisely the partition function of the non-compact orbifold $\mathbb{R}_W/\mathbb{Z}_2$. The coefficient $\frac{1}{4}$ in front of the three theta-function terms means that only a single fixed point remains, as expected for the origin of $\mathbb{R}_W/\mathbb{Z}_2$.

Let us summarize. If we first flat gauge $U(1)_M$ on the circle branch and then gauge $C$, the ordinary group-theoretic computation gives the non-compact orbifold partition function, including the single fixed-point contribution. If we first gauge $C$ and then try to flat gauge the continuous non-invertible symmetry by a direct integral over the simple lines $L_M(\theta)$, all fixed-sector loci have zero measure and only the regular term $\frac{1}{2}Z_{\mathbb{R}}$ is recovered. Therefore the two operations commute only after the second step is equipped with a proper measure, including the stabilizer factors inherited from $U(1)_M\rtimes C$. 

In the rest of this section, we will point out a possible way to gauge the continuous non-invertible symmetry at the level of partition function. In order to keep the contributions of the $\vartheta$-terms, we replace the Kronecker delta functions to the Dirac delta function
    \begin{equation}\label{eq:kronecker_delta_to_dirac_delta}
        \delta_{\widetilde{\alpha}_1,\widetilde{\alpha}_2} \rightarrow \kappa\, \delta( \widetilde{\alpha}_1 - \widetilde{\alpha}_2)\,,
    \end{equation}
where we switch back to the tilde variables, and $\kappa$ is an undetermined factor. To evaluate the integral, let us regularize the integral domain $(\widetilde{\alpha},\widetilde{\beta})\in [0,\frac{r}{2}]\times [0,\frac{r}{2}]$ into nine regions according to the following diagram
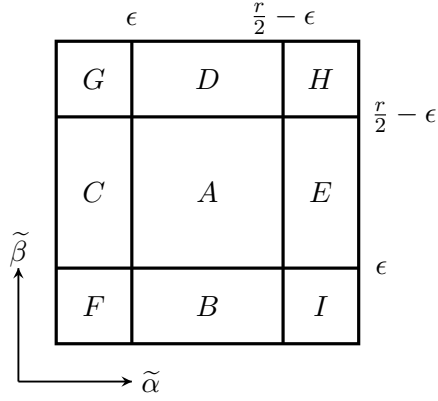
\begin{equation}
    \begin{gathered}
        \begin{tikzpicture}
            \draw[very thick] (0,0) rectangle (4,4);
            \draw[very thick] (0,1)--(4,1);
            \draw[very thick] (1,0)--(1,4);
            \draw[very thick] (0,3)--(4,3);
            \draw[very thick] (3,0)--(3,4);
            \node at (2,2) {$A$};
            \node at (2,0.5) {$B$};
            \node at (.5,2) {$C$};
            \node at (2,3.5) {$D$};
            \node at (3.5,2) {$E$};
            \node at (.5,.5) {$F$};
            \node at (.5,3.5) {$G$};
            \node at (3.5,3.5) {$H$};
            \node at (3.5,.5) {$I$};
            \node at (1,4.3) {$\epsilon$};
            \node at (3,4.3) {$\frac{r}{2}-\epsilon$};
            \node at (4.6,3) {$\frac{r}{2}-\epsilon$};
            \node at (4.3,1) {$\epsilon$};
            \draw[line,->-=1] (-0.5,-0.5)--(-0.5,1);
            \draw[line,->-=1] (-0.5,-0.5)--(1,-0.5);
            \node at (-0.5,1.25) {$\widetilde{\beta}$};
            \node at (1.25,-0.5) {$\widetilde{\alpha}$};
        \end{tikzpicture}
    \end{gathered}\nonumber
\end{equation}
where $\epsilon\rightarrow 0_+$ is a cutoff which should be sent to zero. Then for region $A$, we need to consider the following two diagrams
        \begin{equation}
        \begin{split}
        Z_{\text{orb}}\,
        \begin{gathered}
\begin{tikzpicture}[scale=.75]
\node at (0, 1.45) {};
\draw[line] (-.15-.4,-.2-.4)--(-.25-.4, -.2-.4) -- (-.25-.4, 1.4)-- (-.15-.4,1.4);
\draw[very thick] (0,0) rectangle (1.2,1.2);
\draw[line]  (.6,0) arc (0:90:.6); 
\draw[line] (.6,1.2) arc (-180:-90:.6);
\draw[line] (0.6/2^0.5,0.6/2^0.5)--(1.2-0.6/2^0.5,1.2-0.6/2^0.5);
\draw[line] (1.35, -.2-.4)--(1.45, -.2-.4) -- (1.45, 1.4)--(1.35, 1.4);
\node at (-.15-.2,-.2+0.8) {$\widetilde{\alpha}$};
\node at (.6,-.35) {$\widetilde{\beta}$};
\node at (0.6+0.15,0.6-0.15) {$+$};
\end{tikzpicture}
\end{gathered} \quad =& \quad \frac{1}{2} \left(\lambda_{\widetilde{\beta},\widetilde{\alpha}}+\lambda_{-\widetilde{\beta},-\widetilde{\alpha}} \right)\,,\\
        Z_{\text{orb}}\,
        \begin{gathered}
\begin{tikzpicture}[scale=.75]
\node at (0, 1.45) {};
\draw[line] (-.15-.4,-.2-.4)--(-.25-.4, -.2-.4) -- (-.25-.4, 1.4)-- (-.15-.4,1.4);
\draw[very thick] (0,0) rectangle (1.2,1.2);
\draw[line]  (.6,0) arc (0:90:.6); 
\draw[line] (.6,1.2) arc (-180:-90:.6);
\draw[line] (0.6/2^0.5,0.6/2^0.5)--(1.2-0.6/2^0.5,1.2-0.6/2^0.5);
\draw[line] (1.35, -.2-.4)--(1.45, -.2-.4) -- (1.45, 1.4)--(1.35, 1.4);
\node at (-.15-.2,-.2+0.8) {$\widetilde{\alpha}$};
\node at (.6,-.35) {$\widetilde{\beta}$};
\node at (0.6+0.15,0.6-0.15) {$-$};
\end{tikzpicture}
\end{gathered} \quad =& \quad \frac{1}{2} \left(\lambda_{-\widetilde{\beta},\widetilde{\alpha}}+\lambda_{\widetilde{\beta},-\widetilde{\alpha}} \right) + \kappa \delta(\widetilde{\alpha}-\widetilde{\beta}) \frac{|\vartheta_2 \vartheta_4|}{|\eta|^2}\,,
\end{split}
\end{equation}
where the integral is
\begin{equation}
    \lim_{\epsilon\rightarrow 0_+}\frac{1}{r}\int_{\epsilon}^{\frac{r}{2}-\epsilon} d\widetilde{\alpha}\int_{\epsilon}^{\frac{r}{2}-\epsilon}d\widetilde{\beta} \left(\frac{1}{2} \left(\lambda_{\widetilde{\alpha},\widetilde{\beta}}+\lambda_{-\widetilde{\alpha},\widetilde{\beta}}+\lambda_{\widetilde{\alpha},-\widetilde{\beta}}+\lambda_{-\widetilde{\alpha},-\widetilde{\beta}} \right)+ \kappa \delta(\widetilde{\alpha}-\widetilde{\beta})\frac{|\vartheta_2 \vartheta_4|}{|\eta|^2}\right)
\end{equation}
which gives
    \begin{equation}
        \frac{1}{2}Z_{\mathbb{R}_W} + \frac{\kappa}{2}\frac{|\vartheta_2 \vartheta_4|}{|\eta|^2}\,.
    \end{equation}
The other two $\vartheta$-terms come from the following two sectors
    \begin{equation}
    \begin{split}
        Z_{\text{orb}}\,
        \begin{gathered}
\begin{tikzpicture}[scale=.75]
\node at (0, 1.45) {};
\draw[line] (-.15-.4,-.2-.4)--(-.25-.4, -.2-.4) -- (-.25-.4, 1.4)-- (-.15-.4,1.4);
\draw[very thick] (0,0) rectangle (1.2,1.2);
\draw[line, thick] (0,.6)--(1.2,.6);
\draw[line] (1.35, -.2-.4)--(1.45, -.2-.4) -- (1.45, 1.4)--(1.35, 1.4);
\node at (-.15-.2,-.2+0.8) {$\widetilde{\alpha}$};
\end{tikzpicture}
\end{gathered}
\quad =& \quad \frac{1}{2} \left( \lambda_{\widetilde{\alpha},0} + \lambda_{-\widetilde{\alpha},0} \right) + \frac{\kappa}{2} \delta(\widetilde{\beta}) \frac{|\vartheta_3 \vartheta_4|}{|\eta|^2}\,,
\\
        Z_{\text{orb}}\,
        \begin{gathered}
\begin{tikzpicture}[scale=.75]
\node at (0, 1.45) {};
\draw[line] (-.15-.4,-.2-.4)--(-.25-.4, -.2-.4) -- (-.25-.4, 1.4)-- (-.15-.4,1.4);
\draw[very thick] (0,0) rectangle (1.2,1.2);
\draw[line, thick] (.6,0)--(.6,1.2);
\draw[line] (1.35, -.2-.4)--(1.45, -.2-.4) -- (1.45, 1.4)--(1.35, 1.4);
\node at (.6,-.35) {$\widetilde{\beta}$};
\end{tikzpicture}
\end{gathered}
\quad =& \quad \frac{1}{2}(\lambda_{0,\widetilde{\beta}}+\lambda_{0,-\widetilde{\beta}})+ \frac{\kappa}{2} \delta(\widetilde{\alpha}) \frac{|\vartheta_2 \vartheta_3|}{|\eta|^2}\,,
\end{split}
    \end{equation}
integrated on region $B$ and $C$, where we also include the Dirac delta functions. They respectively give
    \begin{equation}
        \lim_{\epsilon\rightarrow 0} \frac{1}{r} \int_{\epsilon}^{\frac{r}{2}-\epsilon} d \widetilde{\alpha} \int_0^{\epsilon} d\widetilde{\beta} \left(\frac{1}{2} \left( \lambda_{\widetilde{\alpha},0} + \lambda_{-\widetilde{\alpha},0} \right) + \frac{\kappa}{2} \delta(\widetilde{\beta}) \frac{|\vartheta_3 \vartheta_4|}{|\eta|^2}\right) = \frac{\kappa}{2}\frac{|\vartheta_3 \vartheta_4|}{|\eta|^2}\,,
    \end{equation}
and 
    \begin{equation}
        \lim_{\epsilon\rightarrow 0} \frac{1}{r} \int_{0}^{\epsilon} d \widetilde{\alpha} \int_{\epsilon}^{\frac{r}{2}-\epsilon} d\widetilde{\beta} \left(\frac{1}{2}(\lambda_{0,\widetilde{\beta}}+\lambda_{0,-\widetilde{\beta}})+ \frac{\kappa}{2} \delta(\widetilde{\alpha}) \frac{|\vartheta_2 \vartheta_3|}{|\eta|^2}\right) = \frac{\kappa}{2}\frac{|\vartheta_2 \vartheta_3|}{|\eta|^2}\,,
    \end{equation}
where we choose the convention that these delta functions are one-side delta functions on the interval, normalized so that
    \begin{equation}
        \int_{0}^{\epsilon} d \widetilde{\alpha} \delta(\widetilde{\alpha}) = 1\,.
    \end{equation}
For region $D$ and $E$, they receive contributions from 
    \begin{equation}
    \begin{split}
        Z_{\text{orb}}\,
        \begin{gathered}
\begin{tikzpicture}[scale=.75]
\node at (0, 1.45) {};
\draw[line] (-.15-.4,-.2-.4)--(-.25-.4, -.2-.4) -- (-.25-.4, 1.4)-- (-.15-.4,1.4);
\draw[very thick] (0,0) rectangle (1.2,1.2);
\draw[line,dashed]  (.6,0) arc (0:45:.6); 
\draw[line] (0,0.6) arc (90:45:.6); 
\draw[line,dashed] (.6,1.2) arc (-180:-135:.6);
\draw[line] (1.2,0.6) arc (-90:-135:.6);
\draw[line] (0.6/1.414,0.6/1.414)--(1.2-0.6/1.414,1.2-0.6/1.414);
\draw[line] (1.35, -.2-.4)--(1.45, -.2-.4) -- (1.45, 1.4)--(1.35, 1.4);
\node at (-.15-.2,-.2+0.8) {$\widetilde{\alpha}$};
\end{tikzpicture}
\end{gathered}\quad =& \quad \frac{1}{2} \left( \lambda_{\widetilde{\alpha},\pi} + \lambda_{-\widetilde{\alpha},\pi} \right) \,,\\
        Z_{\text{orb}}\,
        \begin{gathered}
\begin{tikzpicture}[scale=.75]
\node at (0, 1.45) {};
\draw[line] (-.15-.4,-.2-.4)--(-.25-.4, -.2-.4) -- (-.25-.4, 1.4)-- (-.15-.4,1.4);
\draw[very thick] (0,0) rectangle (1.2,1.2);
\draw[line]  (.6,0) arc (0:45:.6); 
\draw[line,dashed]  (0,0.6) arc (90:45:.6); 
\draw[line] (.6,1.2) arc (-180:-135:.6);
\draw[line,dashed] (1.2,0.6) arc (-90:-135:.6);
\draw[line] (0.6/1.414,0.6/1.414)--(1.2-0.6/1.414,1.2-0.6/1.414);
\draw[line] (1.35, -.2-.4)--(1.45, -.2-.4) -- (1.45, 1.4)--(1.35, 1.4);
\node at (.6,-.35) {$\widetilde{\beta}$};
\end{tikzpicture}
\end{gathered}\quad =& \quad \frac{1}{2} \left(\lambda_{\pi,\widetilde{\beta}} +  \lambda_{\pi,-\widetilde{\beta}}\right) \,,
\end{split}
\end{equation}
which involve no $\vartheta$-term, and they vanish after taking $\epsilon\rightarrow 0_+$. Lastly, the regions $F,G,H,I$ has order $\epsilon^2$, and they also vanish after taking $\epsilon\rightarrow0_+$ limit. Therefore the gauged partition function reads
    \begin{equation}
        \frac{1}{2}Z_{\mathbb{R}_W} +\frac{\kappa}{2}\frac{|\vartheta_3 \vartheta_4|}{|\eta|^2}+ \frac{\kappa}{2}\frac{|\vartheta_2 \vartheta_3|}{|\eta|^2}+\frac{\kappa}{2}\frac{|\vartheta_2 \vartheta_4|}{|\eta|^2}\,.
    \end{equation}
Here if we simply set $\kappa=1$, then the gauged partition function is the same as the orbifolded compact boson partition function in the $K\rightarrow +\infty$ limit. On the other hand, if we choose $\kappa=1/2$, then the gauged partition function is the same as that of the orbifolded non-compact boson partition function. To determine the correct value of $\kappa$ requires a careful treatment of the measure of the continuous gauging. We leave the intrinsic formulation of this prescription on the orbifold branch as an open question. It would be interesting to understand whether flat gauging of continuous non-invertible symmetries can be defined directly in a way that reproduces the correct partition function.

\section{SymTFT and topological boundaries for $D$ compact bosons}

In \cite{Argurio:2024ewp}, the authors described the circle branch of the compact boson from the point of view of a non-compact SymTFT. The bulk theory is the $\mathbb{R}$-valued BF theory
    \begin{equation}
        S = \frac{1}{2\pi} \int a \wedge db\,,
    \end{equation}
where both $a$ and $b$ are $\mathbb{R}$-valued gauge fields. In this description, the radius of the compact boson is encoded by the choice of topological boundary condition, or equivalently by a Lagrangian algebra of the bulk line operators. See also \cite{Dymarsky:2025agh,Barbar:2025krh} for non-compact BF theory as an infinite level limit of finite BF theory. Changing the radius is therefore interpreted as changing the topological boundary condition. In particular, the non-compact boson and its dual are characterized by the Dirichlet boundary conditions for $a$ and $b$, respectively. From the SymTFT point of view, a 2d gauging operation can also be described by changing the topological boundary condition while keeping the dynamical boundary fixed. Therefore the $\mathbb{R}$-valued BF theory provides a unified framework in which the whole circle branch, together with its decompactification limits, is related by flat gauging finite or continuous symmetries. Related work~\cite{Yu:2026gdf} studies topological boundary averaging in this non-compact BF SymTFT and its connection to Narain moduli.

In this section, we will first review the $\mathbb{R}$-valued BF theory as the description of the circle branch of compact boson. Then we will generalize this picture to a SymTFT description of $D$ components of compact bosons, namely a sigma model whose target space is $T^D$. The non-compact BF theory is replaced by its $D$-component version, with line operators labeled by two real vectors. We will show that the choice of a Lagrangian algebra $\mathcal{L}_{G,B}$ encodes the target-space metric $G$ and antisymmetric $B$-field, and that the overlap between the physical boundary state and this topological boundary state reproduces the Narain partition function. 

\subsection{SymTFT description of circle branch of compact boson}

Consider the three-dimensional BF theory
    \begin{equation}
        S=\frac{1}{2\pi} \int a \wedge db\,,
    \end{equation}
where both $a$ and $b$ are $\mathbb{R}$-valued gauge fields. The general line operators are written as
    \begin{equation}
        W_{x,y}[\Gamma] = \exp \left(i \oint_{\Gamma} x a+ yb \right)\,,
    \end{equation}
with $x,y \in \mathbb{R}$ and $\Gamma$ is a 1-cycle. Given a pair of line operators labeled by $(x_1,y_1)$ and $(x_2,y_2)$, the modular $S$- and $T$-matrix are separately given by
    \begin{equation}
        S_{(x_1,y_1),(x_2,y_2)} = e^{-2\pi i (x_1 y_2+x_2y_1)}\,, \quad T_{(x_1,y_1),(x_2,y_2)} = e^{2\pi i x_1 y_1} \delta(x_1-x_2)\delta(y_1-y_2)\,.
    \end{equation}
In particular, if we quantize the theory on a torus and take $\Gamma_1,\Gamma_2$ to be the two one-cycles, this gives the operator algebra
    \begin{equation}\label{eq:rank-1-algebra}
        W_{x_1,y_1}[\Gamma_1] W_{x_2,y_2}[\Gamma_2] = e^{-2\pi i (x_1 \cdot y_2 + x_2 \cdot y_1)}  W_{x_2,y_2}[\Gamma_2]W_{x_1,y_1}[\Gamma_1]\,.
    \end{equation}
The Lagrangian algebra is a collection of maximally commuting line operators, and is classified by
    \begin{equation}
        \mathcal{L}_{r} := \bigoplus_{n,m\in \mathbb{Z}} W_{nr,mr^{-1}}\,,
    \end{equation}
with $r>0$ identified as the radius of the compact boson. 

The topological boundary $\mathfrak{B}_r$ associated to $\mathcal{L}_{r}$ is defined by condensing $\mathcal{L}_r$ along the codimension-one boundary $\mathfrak{B}_r$. In other words, all components of line operators in $\mathcal{L}_r$ can terminate on $\mathfrak{B}_r$ simultaneously and remain mutually local, and we can identify $W_{nr,mr^{-1}}$ as trivial line operator on the boundary $\mathfrak{B}_r$. In the SymTFT construction, the line operator $W_{nr,mr^{-1}}$ stretching between the topological boundary and physical boundary gives the local operator $\psi_{n,m}$ carrying charge $n,m$ under the $U(1)_M\times U(1)_W$ symmetries, where the latter are generated by
\begin{equation}
    U_M(\theta) = \exp \left(i \oint_{\Gamma} \theta b \right)\,, \quad U_W(\theta) = \exp \left(i \oint_{\Gamma} \varphi a \right)\,,
\end{equation}
supported on $\mathfrak{B}_r$, where $\theta,\varphi$ have period $r$ and $\frac{1}{r}$, respectively, due to the condensing Lagrangian algebra $\mathcal{L}_r$. Moreover, according to \eqref{eq:rank-1-algebra}, $U_M$ and $U_W$ do not commute with each other, which reflects the mixed anomaly between the two $U(1)$ symmetries.

In addition to $\mathcal{L}_r$, we also have another two Lagrangian algebras defined by
    \begin{equation}
        \mathcal{L}_{\text{Dir}} = \oint^{\oplus} W_{x,0}\, dx\,, \quad \mathcal{L}_{\text{Neu}} = \oint^{\oplus} W_{0,y}\, dy\,,
    \end{equation}
where we use the direct integral $\oint^{\oplus}$ as the generalization of direct sum to continuum. The corresponding topological boundary $\mathfrak{B}_{\text{Dir}}$ and $\mathfrak{B}_{\text{Neu}}$ are constructed by condensing all $W_{x,0}$ or $W_{0,y}$ operators, respectively. Then the complement operators will serve as the symmetry generators for the non-compact $\mathbb{R}$-symmetries supported on the two boundaries.

Before generalizing to the multiple compact boson, we should emphasize that the Lagrangian algebra actually describes the pre-moduli space $\mathbb{R}_+$ before $\mathbb{Z}_2$ quotient, where $\mathcal{L}_r$ and $\mathcal{L}_{\frac{1}{2r}}$ are different Lagrangian algebras. It fails to capture the $T$-duality $r\leftrightarrow\frac{1}{2r}$ of the compact boson. Therefore, in order to build the genuine SymTFT describing the circle branch of the compact boson, we should consider the 0-form $\mathbb{Z}_2$ symmetry $\mathcal{T}$ in the SymTFT
    \begin{equation}
        \mathcal{T}(a) = 2b\,, \quad \mathcal{T}(b) = \frac{1}{2}a\,,
    \end{equation}
which switches $\mathcal{L}_r$ with $\mathcal{L}_{\frac{1}{2r}}$. We then lift this transformation to a gauge symmetry.

\subsection{Generalization to rank-$D$ $\mathbb{R}$-valued BF theory }

We then consider the rank $D$ $\mathbb{R}$-valued BF theory formulated as 
    \begin{equation}
        S=\frac{1}{2\pi} \int a^I \wedge db_I\,,\quad (I=1,\cdots,D)
    \end{equation}
where both $a^I$ and $b_I$ are $\mathbb{R}$-valued gauge fields. The line operators are
    \begin{equation}
        W_{x,y}[\Gamma] = \exp \left( i \oint_{\Gamma}  x_I a^I + y^I b_I\right)\,,
    \end{equation}
where $x_I,y^J \in \mathbb{R}^D$. The modular $S$-matrix and $T$-matrix are separately given by
    \begin{equation}\label{eq:Rank_D_modular_data}
        S_{(x_1,y_1),(x_2,y_2)} = e^{-2\pi i (x_1 \cdot y_2 + x_2 \cdot y_1)}\,,\quad T_{(x_1,y_1),(x_2,y_2)} = e^{2\pi i  x\cdot y} \delta^{(D)}(x_1-x_2) \delta^{(D)}(y_1-y_2)\,,
    \end{equation}
where the inner product is defined by $x\cdot y = \sum_I x_I y^I$. If we quantize the theory on a torus and take $\Gamma_1,\Gamma_2$ to be the two one-cycles, this gives the operator algebra
    \begin{equation}\label{eq:rank-D-algebra}
        W_{x_1,y_1}[\Gamma_1] W_{x_2,y_2}[\Gamma_2] = e^{-2\pi i (x_1 \cdot y_2 + x_2 \cdot y_1)}  W_{x_2,y_2}[\Gamma_2]W_{x_1,y_1}[\Gamma_1]\,.
    \end{equation}

Similar to the rank-one case, the geometric data of the $D$-dimensional target space should be encoded in the Lagrangian algebras of the BF theory. Denote $G$ as the constant non-degenerate metric on the $D$-dimensional target torus,
    \begin{equation}
        ds^2 = G_{IJ} d \theta^I d \theta^J\,, \quad (\theta^I \in (0,2\pi])\,,
    \end{equation}
the simplest Lagrangian algebra one can construct using the metric $G_{IJ}$ is the following
    \begin{equation}
        \mathcal{L}_G := \bigoplus_{n,m\in \mathbb{Z}} Q^G_{n,m}\,,
    \end{equation}
where we introduce
    \begin{equation}
        Q^G_{n,m}[\Gamma]: = W_{n\cdot G, m\cdot G^{-1}}[\Gamma]=\exp \left(i \oint_{\Gamma} n^IG_{IJ}a^J + m_I G^{IJ} b_J \right)\,.
    \end{equation}
From \eqref{eq:rank-D-algebra} one can easily check $Q^G_{n,m}[\Gamma_1] Q^G_{n',m'}[\Gamma_2]=Q^G_{n',m'}[\Gamma_2]Q^G_{n,m}[\Gamma_1]$ for integer-valued $n,m,n',m'$.

Compared to the rank-one compact boson where the only modulus is the radius $r$, for higher rank one can also turn on the antisymmetric $B$-field on the target space, which couples to the worldsheet action through the topological term
    \begin{equation}
        \frac{1}{2} \int d^2x\, \epsilon^{\mu \nu}\partial_{\mu}X^I \partial_{\nu} X^J B_{IJ}\,. \quad (\mu=1,2)
    \end{equation}
The full moduli space is parametrized by the $D^2$ components of $(G_{IJ},B_{IJ})$. Therefore, one expects the $B$-moduli should also be integrated into the parametrization of Lagrangian algebras. Indeed, we can turn on an antisymmetric $B$-field by mixing the $a$ and $b$ fields and define the line operators
    \begin{equation}
        Q^{G,B}_{n,m}=\exp \left(i \oint_{\Gamma} n^IG_{IJ}a^J + (m_I G^{IJ}+n^I B_{IK} G^{KJ}) b_J \right)\,,
    \end{equation}
and introduce the general Lagrangian algebra
    \begin{equation}
        \mathcal{L}_{G,B} = \bigoplus_{n,m} Q^{G,B}_{n,m}\,.
    \end{equation}
Similarly, one can check that $Q^{G,B}_{n,m}$ for different $n,m\in \mathbb{Z}^D$ always commute with each other, and thus are mutually local.

It turns out to be more convenient to use the vielbein formalism
    \begin{equation}
        ds^2 = \delta_{AB} e^A e^B\,,\quad G_{IJ}=\delta_{AB}e^A_{I} e^B_J\,,
    \end{equation}
and rewrite the BF action as
    \begin{equation}
        S=\frac{1}{2\pi} \int a^A \wedge db_A\,,\quad (A=1,\cdots,D)
    \end{equation}
where the line operators in the Lagrangian algebra
    \begin{equation}
    \begin{split}
        &Q^{G,B}_{n,m}[\Gamma] = W_{n\cdot G, m\cdot G^{-1}}[\Gamma]=\exp \left(i \oint_{\Gamma} n^Ie_{IA}a^A + m_I e^{IA} b_A +n^IB_{IK}e^{KA}b_A \right)\\
        =&\exp \left[i 
        \left(\begin{array}{cc}        n^I&m_J\end{array}\right) \left(\begin{array}{cc}
            e_{IA} & B_{IK}e^{KB} \\
            0 & e^{JB}
        \end{array} \right) \left(\begin{array}{c}
            a^A\\b_B
        \end{array} \right)\right]\equiv \exp \left[i 
        \left(\begin{array}{cc}        n^I&m_J\end{array}\right) \mathcal{E} \left(\begin{array}{c}
            a^A\\b_B
        \end{array} \right)\right]\,,
    \end{split}
    \end{equation}
where we introduce the $2D\times 2D$ matrix
    \begin{equation}
        \mathcal{E}=\left(\begin{array}{cc}
            e_{IA} & B_{IK}e^{KB} \\
            0 & e^{JB}
        \end{array} \right)\,,
    \end{equation}
which encodes the geometry of the target space.

The BF theory has a large symmetry group acting on $A\equiv (a^A,b_A)^T$. Introduce the matrix
    \begin{equation}
        \Omega = \left(\begin{array}{cc}
            0 & \mathbf{1}_D\\
            \mathbf{1}_D & 0 
        \end{array} \right)\,,
    \end{equation}
where $\mathbf{1}_D$ is a $D\times D$ dimensional identity matrix, and we rewrite the action as
    \begin{equation}
        S = \frac{1}{4\pi} \int A^T\Omega dA\,.
    \end{equation}
This action is invariant under the transformation $\Lambda$
    \begin{equation}
        A\rightarrow \Lambda A\,,\quad \textrm{with}\quad \Lambda^T \Omega \Lambda= \Omega\,,
    \end{equation}
which defines an $O(D,D;\mathbb{R})$ matrix. Moreover, the matrix $\mathcal{E}$ is also an $O(D,D;\mathbb{R})$ matrix satisfying $\mathcal{E}^T \Omega \mathcal{E}=\Omega$. In general, starting from the standard form
    \begin{equation}
        Q_{n,m}[\Gamma] = \exp \left(i \oint_{\Gamma} n^I \delta_{IA} a^A+m_J \delta^{JB} b_B \right)\,,
    \end{equation}
one can apply an arbitrary $O(D,D;\mathbb{R})$ transformation to obtain the other Lagrangian algebras.

We see that the Lagrangian algebra $\mathcal{L}$ is actually parametrized by $O(D,D;\mathbb{R})$, which is not equal to the Narain moduli space
    \begin{equation}
        \mathcal{M}=\left.\left[\frac{O(D,D;\mathbb{R})}{O(D;\mathbb{R})\times O(D;\mathbb{R})}\right]\right/ O(D,D;\mathbb{Z})\,.
    \end{equation}
To see that $O(D;\mathbb{R})\times O(D;\mathbb{R})$ acts on the left- and right-moving momenta, let us introduce $p_L$ and $p_R$ as
\begin{equation}
    (p_L,p_R) = (n,m)\mathcal{E}\left(\begin{array}{cc}
        \sqrt{2} &0  \\
         0& \frac{1}{\sqrt{2}}
    \end{array} \right)  \sqrt{2}U\,, \quad U = U^{-1}=\frac{1}{\sqrt{2}}\left(\begin{array}{cc}
            \mathbf{1}_D & \mathbf{1}_D \\
            \mathbf{1}_D & -\mathbf{1}_D
        \end{array} \right)\,,
\end{equation}
with\footnote{The annoying $\sqrt{2}$ factors and the matrix $\left(\begin{array}{cc}
        \sqrt{2} &0  \\
         0& \frac{1}{\sqrt{2}}
    \end{array} \right)$ in $(p_L,p_R)$ are again related to the choice that self-dual radius is $r=\frac{1}{\sqrt{2}}$. For example, if we set $D=1$ such that $e=r,e^{-1}=r^{-1}$ and $B=0$, then the left and right momenta are
        \begin{equation}
            p_L = \sqrt{2} n r + \frac{m}{\sqrt{2}r}\,,\quad p_R = \sqrt{2} n r - \frac{m}{\sqrt{2}r}\,,
        \end{equation}
and they match our conventions in \eqref{eq:compact-boson-partition-function} and \eqref{eq:left_right_momenta_1}. }
    \begin{equation}
        p_L^A = \sqrt{2} n^Ie_I^A + \frac{(m_I+n^JB_{JI})e^{IA}}{\sqrt{2}}\,, \quad p^A_R = \sqrt{2} n^Ie_I^A -\frac{(m_I+n^JB_{JI})e^{IA}}{\sqrt{2}}\,.
    \end{equation}
Denote $\widetilde{\mathcal{E}}$ as
    \begin{equation}
        \widetilde{\mathcal{E}} = \mathcal{E}\left(\begin{array}{cc}
        \sqrt{2} &0  \\
         0& \frac{1}{\sqrt{2}}
    \end{array} \right)\,,\quad \widetilde{\mathcal{E}} \in O(D,D;\mathbb{R})\,,
    \end{equation}
and consider the transformation $\widetilde{\mathcal{E}}\rightarrow \widetilde{\mathcal{E}} S$ with $S\in O(D,D;\mathbb{R})$, the left and right momenta transform according to
    \begin{equation}
        (p_L,p_R) \rightarrow \sqrt{2}(n,m) \widetilde{\mathcal{E}} SU = (p_L,p_R) S'\,,
    \end{equation}
where $S'=U^{-1}SU$ satisfies
    \begin{equation}\label{eq:restriction_of_S'}
        (S')^T S'=1\,, \quad (S')^T \left(\begin{array}{cc}
            I & 0 \\
            0 & -I
        \end{array} \right)S'=\left(\begin{array}{cc}
            I & 0 \\
            0 & -I
        \end{array} \right)\,.
    \end{equation}
Decompose $S'$ into the block matrix
    \begin{equation}
        S'=\left(\begin{array}{cc}
            A & B \\
            C & D
        \end{array} \right)\,,
    \end{equation}
where $A,B,C,D$ are all $D\times D$ matrices. Then the restriction \eqref{eq:restriction_of_S'} implies
    \begin{equation}
        A^TA=D^TD=I\,,\quad C^TC=B^TB=A^TB=C^TD=0\,,
    \end{equation}
which is solved by $A,D\in O(D;\mathbb{R})$ and $B=C=0$. Therefore the action of $p_L,p_R$ is implemented by
    \begin{equation}
        p_L\rightarrow p_L A\,, \quad p_R \rightarrow p_R D\,.
    \end{equation}

Based on the previous discussion, we should identify the two algebras whose $\widetilde{\mathcal{E}}$ matrices are related by $O(D;\mathbb{R})\times O(D;\mathbb{R})$ transformation. However, that identification is not a built-in property for the SymTFT and depends on the physical system we study. In the following, we shall explicitly check that, from the level of partition function, for rank-D compact bosons different topological boundaries associated to Lagrangian algebras differing by $O(D;\mathbb{R})\times O(D;\mathbb{R})$ transformation indeed give the same physical system. 

Moreover, similar to the rank-one case, we should further mod out by the $T$-duality group by lifting $O(D,D;\mathbb{Z}) \in O(D,D;\mathbb{R})$ acting on $A\equiv (a^A,b_A)^T$ to a gauge symmetry in the SymTFT. After that, we reproduce the Narain moduli space.

In addition to $\mathcal{L}_{e,B}$, we also have the Dirichlet and Neumann Lagrangian algebras defined by
    \begin{equation}
        \mathcal{L}_{\text{Dir}} = \int^{\oplus} W_{x,0} d^Dx\,, \quad \mathcal{L}_{\text{Neu}} = \int^{\oplus} W_{0,y} d^Dy\,,
    \end{equation}
whose target space is described by $\mathbb{R}^D$. More generally, there exist other mixed Lagrangian algebras built from $\mathcal{L}_{G,B}$ and $\mathcal{L}_{\text{Dir}}/\mathcal{L}_{\text{Neu}}$, whose target space is of the form $T^d\times \mathbb{R}^{D-d}$. We will only focus on $\mathcal{L}_{\text{Dir}},\mathcal{L}_{\text{Neu}}$ and the corresponding topological boundaries $\mathfrak{B}_{\text{Dir}},\mathfrak{B}_{\text{Neu}}$ in the following discussion.

\subsection{Analysis of topological boundaries}

In this subsection, we will begin to analyze the Hilbert space $\mathcal{H}$ of the rank-$D$ BF theory quantized on the torus, and the vacuum polarization $|\Omega\rangle_{\mathcal{L}}$ associated to different Lagrangian algebras $\mathcal{L}$ and the corresponding topological boundaries. We will still use the vielbein formalism throughout this subsection.

Given the modular data \eqref{eq:Rank_D_modular_data}, the Hilbert space $\mathcal{H}$ on the torus can be constructed as follows. Begin with the rank-$D$ BF theory living on a solid torus $S^1 \times D^2$, and insert an arbitrary line operator $W_{x,y}$ at the origin of $D^2$ and along $S^1$. The path integral in the bulk produces a state vector $|W_{x,y}\rangle$ on the boundary torus, which satisfies
    \begin{equation}
        W_{x,y}[\Gamma_1] |W_{x',y'}\rangle = \frac{S_{(x',y'),(x,y)}}{S_{(x',y'),(0,0)}}|W_{x',y'}\rangle\,, \quad W_{x,y}[\Gamma_2]|W_{x',y'}\rangle = \sum_{x'',y''} N_{(x,y),(x',y')}^{(x'',y'')}|W_{x'',y''}\rangle\,,
    \end{equation}
where $\Gamma_1$ is the boundary $\partial D_2$ of the disk, and $\Gamma_2$ is the first $S^1$. The fusion coefficients can be read from the fusion rule
    \begin{equation}
        W_{x,y} \times W_{x',y'}=W_{x+x',y+y'}\,,\quad N_{(x,y),(x',y')}^{(x'',y'')}= \delta^{(D)}_{x'',x+x'}\delta_{y'',y+y'}\,.
    \end{equation}
Therefore we have
    \begin{equation}
        W_{x,y}[\Gamma_1] |W_{x',y'}\rangle = e^{-2\pi i (x\cdot y'+x'\cdot y)}|W_{x',y'}\rangle\,, \quad W_{x,y}[\Gamma_2]|W_{x',y'}\rangle = |W_{x+x',y+y'}\rangle\,.
    \end{equation}
In the following, we also impose the normalization
    \begin{equation}\label{eq:rankD_inner_product_W_states}
        \langle W_{x_1,y_1}|W_{x_2,y_2}\rangle = \delta^{(D)}(x_1-x_2) \delta^{(D)}(y_1-y_2)\,,
    \end{equation}
where states $|W_{x,y}\rangle$ with different $(x,y)$ labels are orthogonal to each other, since they have different eigenvalues of $W[\Gamma_1]$. The Dirac $\delta$-function can be expressed as the integral
    \begin{equation}
        \int e^{2\pi i (x-y)\cdot z} dz = (2\pi)^D \delta^{(D)}(2\pi (x-y)) =\delta^{(D)}(x-y)\,.
    \end{equation}

In the SymTFT setup, the BF theory is living on $T^2 \times [0,1]$, where the boundaries $T\times \{0\}$ and $T\times \{1\}$ are symmetry boundary $\mathfrak{B}_{\text{sym}}$ and physical boundary $\mathfrak{B}_{\text{phys}}$, respectively. The former is topological and is determined by a Lagrangian algebra of the theory, and the latter encodes the dynamics of the physical theory. From the Hamiltonian point of view, the two boundaries are associated with the state vectors $|\mathfrak{B}_{\text{sym}}\rangle,|\mathfrak{B}_{\text{phys}}\rangle \in \mathcal{H}$. 

We will give the recipe to construct the vacuum polarization $|\Omega\rangle_{\mathcal{L}}\in \mathcal{H}$ of the symmetry boundary $\mathfrak{B}_{\mathcal{L}}$ associated to the Lagrangian algebra $\mathcal{L}$. Given a Lagrangian algebra $\mathcal{L} = \bigoplus n_{x,y} W_{x,y}$, the corresponding vacuum polarization of the topological boundary is simply~\cite{Chen:2024ulc}
    \begin{equation}
        |\Omega\rangle_{\mathcal{L}} := \sum_{x,y} n_{x,y} |W_{x,y}\rangle\,,
    \end{equation}
which is prepared by inserting the whole algebraic object $\mathcal{L}$ inside the solid torus. In the following, we will mainly focus on three kinds of Lagrangian algebras discussed in the previous section: the Dirichlet and Neumann Lagrangian algebras given by
    \begin{equation}
        \mathcal{L}_{\textrm{Dir}} = \int^{\oplus} dx W_{x,0}\,, \quad \mathcal{L}_{\textrm{Neu}} = \int^{\oplus} dy W_{0,y}\,,
    \end{equation}
and also the one determined by the geometric data of the target space
    \begin{equation}
        \mathcal{L}_{e,B} = \bigoplus_{n,m\in \mathbb{Z}} W_{(n,m) \mathcal{E}}=\bigoplus_{n,m\in \mathbb{Z}}\exp \left(i \oint_{\Gamma} n^Ie_{IA}a^A + m_I e^{IA} b_A +n^IB_{IK}e^{KA}b_A \right)\,,
    \end{equation}
where
\begin{equation}
    (n,m) \mathcal{E} = (n\cdot e, m\cdot e^{-1}+n\cdot Be^{-1})\,.
\end{equation}

\paragraph{Dirichlet Lagrangian algebra $\mathcal{L}_{\text{Dir}}$}

For the Dirichlet Lagrangian algebra, we introduce the vacuum polarization
    \begin{equation}
        |0,0\rangle_{\text{Dir}} \equiv \int_{-\infty}^{+\infty} dz |W_{z,0}\rangle\,,
    \end{equation}
which is the common eigenvector for all $W_{x,0}\in \mathcal{L}_{\text{Dir}}$:
    \begin{equation}
        W_{x,0}[\Gamma_1]|0,0\rangle_{\text{Dir}} = |0,0\rangle_{\textrm{Dir}}\,, \quad W_{x,0}[\Gamma_2]|0,0\rangle_{\text{Dir}} = |0,0\rangle_{\textrm{Dir}}\,.
    \end{equation}
Other states are obtained using the raising operators $W_{0,y}[\Gamma_1]$ and $W_{0,y}[\Gamma_2]$. We define
    \begin{equation}
    \begin{split}
        |x_1,x_2\rangle_{\text{Dir}}=&W_{0,x_2}[\Gamma_1]W_{0,-x_1}[\Gamma_2]|0,0\rangle_{\text{Dir}}\\
        =&W_{0,x_2}[\Gamma_1] \int_{-\infty}^{+\infty} dz |W_{z,-x_1}\rangle= \int_{-\infty}^{+\infty} dz\, e^{-2\pi i x_2\cdot z} |W_{z,-x_1}\rangle\,,
    \end{split}
    \end{equation}
where $x_1=(x_1^1,\cdots,x_1^D)\in\mathbb{R}^D,x_2= (x_2^1,\cdots,x_2^D) \in \mathbb{R}^D$ are $\mathbb{R}$-valued holonomies of the $\mathbb{R}^D$ symmetry along the two one-cycles on the torus. By definition, $|x_1,x_2\rangle_{\text{Dir}}$ satisfies
    \begin{equation}
        W_{0,z}[\Gamma_1]|x_1,x_2\rangle_{\text{Dir}} = |x_1,x_2+z\rangle_{\text{Dir}}\,, \quad W_{0,z}[\Gamma_2]|x_1,x_2\rangle_{\text{Dir}} = |x_1-z,x_2\rangle_{\text{Dir}}\,,
    \end{equation}
and also
    \begin{equation}
        W_{z,0}[\Gamma_1] |x_1,x_2\rangle_{\text{Dir}} = e^{2\pi i z \cdot x_1} |x_1,x_2\rangle_{\text{Dir}}\,, \quad W_{z,0}[\Gamma_2] |x_1,x_2\rangle_{\text{Dir}} = e^{2\pi i z \cdot x_2} |x_1,x_2\rangle_{\text{Dir}}\,. 
    \end{equation}
They form a representation of the operator algebra \eqref{eq:rank-D-algebra}. The inner product between different state vectors are
    \begin{equation}
    \begin{split}
     _{\text{Dir}}\langle x_1,x_2| y_1,y_2\rangle_{\text{Dir}}
     =& \int_{-\infty}^{+\infty}dz \int_{-\infty}^{+\infty}dz' e^{2\pi i x_2 \cdot z} e^{-2\pi i y_2 \cdot z'} \langle W_{{z,-x_1}}|W_{z',-y_1}\rangle\\
     =& \int_{-\infty}^{+\infty} dz\, e^{2\pi i z\cdot (x_2-y_2)} \delta^{(D)}(x_1-y_1)=\delta^{(D)}(x_1-y_1)\delta^{(D)}(x_2-y_2)\,,
    \end{split}
    \end{equation}
where we have used \eqref{eq:rankD_inner_product_W_states}.

\paragraph{Neumann Lagrangian algebra $\mathcal{L}_{\text{Neu}}$}
For the Neumann Lagrangian algebra, we similarly introduce the vacuum polarization
    \begin{equation}
        |0,0\rangle_{\text{Neu}} = \int_{-\infty}^{+\infty} dz |W_{0,z}\rangle\,,
    \end{equation}
which is the common eigenvector for $W_{0,y}\in \mathcal{L}_{\text{Neu}}$:
    \begin{equation}
        W_{0,y}[\Gamma_1]|0,0\rangle_{\text{Neu}} = |0,0\rangle_{\textrm{Neu}}\,, \quad W_{0,y}[\Gamma_2]|0,0\rangle_{\text{Neu}} = |0,0\rangle_{\textrm{Neu}}\,.
    \end{equation}
Other states are obtained using the raising operators $W_{x,0}[\Gamma_1]$ and $W_{x,0}[\Gamma_2]$. We define
    \begin{equation}\label{eq:Neu_boundary_state}
    \begin{split}
        |x_1,x_2\rangle_{\text{Neu}}=&W_{x_2,0}[\Gamma_1]W_{-x_1,0}[\Gamma_2]|0,0\rangle_{\text{Neu}}\\
        =&W_{x_2,0}[\Gamma_1] \int_{-\infty}^{+\infty} dz |W_{-x_1,z}\rangle= \int_{-\infty}^{+\infty} dz\, e^{-2\pi i x_2\cdot z} |W_{-x_1,z}\rangle\,.
    \end{split}
    \end{equation}
By definition, $|x_1,x_2\rangle_{\text{Neu}}$ satisfies
    \begin{equation}
        W_{z,0}[\Gamma_1]|x_1,x_2\rangle_{\text{Neu}} = |x_1,x_2+z\rangle_{\text{Neu}}\,, \quad W_{z,0}[\Gamma_2]|x_1,x_2\rangle_{\text{Neu}} = |x_1-z,x_2\rangle_{\text{Neu}}\,,
    \end{equation}
and also
    \begin{equation}
        W_{0,z}[\Gamma_1] |x_1,x_2\rangle_{\text{Neu}} = e^{2\pi i z \cdot x_1} |x_1,x_2\rangle_{\text{Neu}}\,, \quad W_{0,z}[\Gamma_2] |x_1,x_2\rangle_{\text{Neu}} = e^{2\pi i z \cdot x_2} |x_1,x_2\rangle_{\text{Neu}}\,, 
    \end{equation}
and the inner product between two state vectors are
    \begin{equation}
    \begin{split}
     _{\text{Neu}}\langle x_1,x_2| y_1,y_2\rangle_{\text{Neu}}
     =& \int_{-\infty}^{+\infty} dz \int_{-\infty}^{+\infty} dz' e^{2\pi i x_2 \cdot z} e^{-2\pi i y_2 \cdot z'} \langle W_{{-x_1,z}}|W_{-y_1,z'}\rangle\\
     =& \int_{-\infty}^{+\infty}dz\, e^{2\pi i z\cdot (x_2-y_2)} \delta^{(D)}(x_1-y_1)=\delta^{(D)}(x_2-y_2)\delta^{(D)}(x_1-y_1)\,.
    \end{split}
    \end{equation}
Moverover, inner product between the Dirichlet and Neumann bases is
    \begin{equation}
    \begin{split}
     &_{\text{Dir}}\langle x_1,x_2| y_1,y_2\rangle_{\text{Neu}}\\
     =&\int_{-\infty}^{+\infty}dz \int_{-\infty}^{+\infty}dz' e^{2\pi i x_2 \cdot z} e^{-2\pi i y_2 \cdot z'} \langle W_{{z,-x_1}}|W_{-y_1,z'}\rangle\\
     =& \int_{-\infty}^{+\infty} dz \int_{-\infty}^{+\infty} dz' e^{2\pi i x_2 \cdot z} e^{-2\pi i y_2 \cdot z'} \delta^{(D)}(z+y_1)  \delta^{(D)}(z'+x_1)
     =e^{2\pi i(x_1 \cdot y_2 - x_2 \cdot y_1)}\,,
    \end{split}
    \end{equation}
which gives the Fourier relation
    \begin{equation}\label{eq:relation_Dir_Neu}
        |x_1,x_2\rangle_{\text{Neu}} = \int_{-\infty}^{+\infty} dx'_1 dx'_2\, e^{-2\pi i(x_1\cdot x'_2-x_2\cdot x'_1)} |x'_1,x'_2\rangle_{\text{Dir}}\,.
    \end{equation}

\paragraph{Lagrangian algebra $\mathcal{L}_{e,B}$}

We then turn to the Lagrangian algebra $\mathcal{L}_{(e,B)}$ determined by the vielbein $e$ and the $B$-field
    \begin{equation}
        \mathcal{L}_{e,B} := \bigoplus_{n,m\in \mathbb{Z}} W_{n\cdot e, m\cdot e^{-1}+n\cdot Be^{-1}}=\bigoplus_{n,m\in \mathbb{Z}}\exp \left(i \oint_{\Gamma} n^Ie_{IA}a^A + m_I e^{IA} b_A +n^IB_{IK}e^{KA}b_A \right)\,,
    \end{equation}
where the construction is much more non-trivial than the previous two cases.
The corresponding vacuum polarization of the topological boundary state is
    \begin{equation}\label{eq:topological_boundary_state_e_B}
        |\Omega \rangle_{e,B} = \bigoplus_{n,m} |W_{n\cdot e, m\cdot e^{-1}+n\cdot Be^{-1}}\rangle\,,
    \end{equation}
which consists of the line operators $W_{x,y}$ with
    \begin{equation}
        x_A= n^Je_{JA}\,,\quad y^A= m_J e^{JA}+n^J B_{JK} e^{KA}\,.\quad (n,m\in \mathbb{Z}^D)
    \end{equation}
By definition, the state is invariant under the action of $W_{n'\cdot e, m'\cdot e^{-1}+n'\cdot Be^{-1}}[\Gamma_2]\in \mathcal{L}_{e,B}$ by shifting the $n,m$ labels. It is also invariant under the action along the other cycle:
    \begin{equation}
    \begin{split}
        &W_{n\cdot e,m\cdot e^{-1}+n\cdot Be^{-1}}[\Gamma_1] |\Omega\rangle_{G,B}\\
        =&\bigoplus_{n',m'} W_{n\cdot e,m\cdot e^{-1}+n\cdot Be^{-1}}[\Gamma_1] |W_{n'\cdot e,m'\cdot e^{-1}+n'\cdot Be^{-1}}\rangle\\
        =& \bigoplus_{n',m'} \exp\left\{-2\pi i  \left[ n^Je_{JA} (m'_K e^{KA}+{n'}^K B_{KL}e^{LA}) + (n\leftrightarrow n'\,, m\leftrightarrow m')\right]\right\}|W_{n'\cdot e,m'\cdot e^{-1}+n'\cdot Be^{-1}}\rangle\\
        =& \bigoplus_{n',m'}\exp\left\{-2\pi i  \left[ (n^J m'_J + B_{KJ} {n'}^K n^J ) + (n\leftrightarrow n'\,, m\leftrightarrow m')\right]\right\}|W_{n'\cdot e,m'\cdot e^{-1}+n'\cdot Be^{-1}}\rangle\\
        =& \bigoplus_{n',m'}\exp\left[-2\pi i  \left( n \cdot m' + n' \cdot m\right)\right]|W_{n'\cdot e,m'\cdot e^{-1}+n'\cdot Be^{-1}}\rangle\\
        =&\bigoplus_{n',m'}|W_{n'\cdot e,m'\cdot e^{-1}+n'\cdot Be^{-1}}\rangle=|\Omega\rangle_{e,B}\,.
    \end{split}
    \end{equation}
For convenience, let us introduce
    \begin{equation}
        U_{\theta,\varphi} = W_{\theta \cdot e,\varphi\cdot e^{-1}+\theta\cdot Be^{-1}}\,,
    \end{equation}
with $\theta^I = (\theta^1,\cdots,\theta^D)$ and $\varphi_I=(\varphi_1,\cdots,\varphi_D)$, and they have unit period. These are the generators of the $U(1)^D_{\theta}\times U(1)^D_{\varphi}$ symmetry. Let us denote
    \begin{equation}
        |(0,0);(0,0)\rangle_{e,B} \equiv |\Omega\rangle_{e,B}\,,
    \end{equation}
where the two pairs stand for the holonomies for $U(1)_{\theta}$ and $ U(1)_{\varphi}$, respectively. Use the symmetry generator $U_{\theta,\varphi}$, let us define the state with holonomies along the $\Gamma_1$-cycle
    \begin{equation}
    \begin{split}
        &|(\theta_1,0);(\varphi_1,0)\rangle_{e,B} := U_{-\theta_1,-\varphi_1}[\Gamma_2] |(0,0);(0,0)\rangle_{e,B}\,,\\
        =& \bigoplus_{n,m} |W_{(n-\theta_1)\cdot e,(m-\varphi_1)\cdot e^{-1}+(n-\theta_1)\cdot Be^{-1}}\rangle\,,
    \end{split}
    \end{equation}
and also with holonomies along $\Gamma_2$-cycle
    \begin{equation}
    \begin{split}
        &|(0,\theta_2);(0,\varphi_2)\rangle_{e,B} := U_{\theta_2,\varphi_2}[\Gamma_1] |(0,0);(0,0)\rangle_{e,B}\,,\\
        =&\bigoplus_{n',m'}\exp\left\{-2\pi i  \left[ (\theta_2^J m'_J + B_{KJ} {n'}^K \theta_2^J ) + ({n'}^J \varphi_{2;J} + B_{KJ} \theta_2^K {n'}^J) \right]\right\}|W_{n'\cdot e,m'\cdot e^{-1}+n'\cdot Be^{-1}}\rangle\\
        =&\bigoplus_{n,m}\exp\left[-2\pi i  \left( \theta_2^J m_J   + \varphi_{2;J}n^J  \right)\right]|W_{n\cdot e,m\cdot e^{-1}+n\cdot Be^{-1}}\rangle\,.
    \end{split}
    \end{equation}

However, it is ambiguous to write down a state $|(\theta_1,\theta_2);(\varphi_1,\varphi_2)\rangle_{e,B}$ with general holonomies due to the mixed anomaly between $U(1)^D_{\theta}$ and $U(1)^D_{\varphi}$. Indeed, if we act $U_{-\theta_1,-\varphi_1}[\Gamma_2]$ and $U_{\theta_2,\varphi_2}[\Gamma_1]$ simultaneously, there is a phase ambiguity because the two operators do not commute
    \begin{equation}
        U_{\theta_2,\varphi_2}[\Gamma_1]U_{-\theta_1,-\varphi_1}[\Gamma_2]  = e^{2\pi i(\theta_1 \varphi_2+\theta_2\varphi_1)} U_{-\theta_1,-\varphi_1}[\Gamma_2] U_{\theta_2,\varphi_2}[\Gamma_1]\,.
    \end{equation}
That implies the state $|(\theta_1,\theta_2);(\varphi_1,\varphi_2)\rangle_{e,B}$ is a section of the line bundle on the moduli space of flat connection twisted by the mixed anomaly. To build a state with good properties, we will use the composite operator
    \begin{equation}
    \begin{split}
        U_{(\theta_1,\varphi_1);(\theta_2,\varphi_2)}:=& \exp \left(i \oint_{\Gamma_1} \theta_2 e a+ (\varphi_2 e^{-1}+\theta_2 Be^{-1}) b - i \oint_{\Gamma_2} \theta_1 e a+ (\varphi_1 e^{-1}+\theta_1 Be^{-1}) b \right)\,,\\
        =& e^{-\pi i(\theta_1 \varphi_2+\theta_2\varphi_1)} U_{\theta_2,\varphi_2}[\Gamma_1]U_{-\theta_1,-\varphi_1}[\Gamma_2]\,,
    \end{split}
    \end{equation}
where the phase factor follows from the BCH formula. We then define the state with general holonomies according to
    \begin{equation}
    \begin{split}
        &|(\theta_1,\theta_2);(\varphi_1,\varphi_2)\rangle_{e,B} := U_{(\theta_1,\varphi_1);(\theta_2,\varphi_2)} |(0,0);(0,0)\rangle_{e,B}\\
        =&e^{-\pi i(\theta_1 \varphi_2+\theta_2\varphi_1)} U_{\theta_2,\varphi_2}[\Gamma_1]U_{-\theta_1,-\varphi_1}[\Gamma_2] |(0,0);(0,0)\rangle_{e,B}\\
        =&\bigoplus_{n,m} e^{-\pi i(\theta_1 \varphi_2+\theta_2\varphi_1)} U_{\theta_2,\varphi_2}[\Gamma_1] |W_{(n-\theta_1)\cdot e,(m-\varphi_1)\cdot e^{-1}+(n-\theta_1)\cdot Be^{-1}}\rangle\\
        =&\bigoplus_{n,m} e^{-\pi i(\theta_1 \varphi_2+\theta_2\varphi_1)} \exp\left\{-2\pi i  \left[ (\theta_2^J (m_J-\varphi_{1;J}) + B_{KJ} (n^K-\theta_1^K) \theta_2^J ) \right.\right.\\
        &\left.\left.+ ((n^J-\theta_1^J) \varphi_{2;J} + B_{KJ} \theta_2^K (n^J-\theta_1^J)) \right]\right\}|W_{(n-\theta_1)\cdot e,(m-\varphi_1)\cdot e^{-1}+(n-\theta_1)\cdot Be^{-1}}\rangle\\
        =&\bigoplus_{n,m}e^{-2\pi i (\theta_2 \cdot (m-\frac{1}{2}\varphi_1)+\varphi_2 \cdot (n-\frac{1}{2}\theta_1))}|W_{(n-\theta_1)\cdot e,(m-\varphi_1)\cdot e^{-1}+(n-\theta_1)\cdot Be^{-1}}\rangle\,,
    \end{split}
    \end{equation}
and one can check they satisfy the twisted boundary conditions
    \begin{equation}
    \begin{split}
        &|(\theta_1+1,\theta_2);(\varphi_1,\varphi_2)\rangle_{e,B}=e^{-\pi i \varphi_2}|(\theta_1,\theta_2);(\varphi_1,\varphi_2)\rangle_{e,B}\,,\\
        &|(\theta_1,\theta_2+1);(\varphi_1,\varphi_2)\rangle_{e,B}=e^{\pi i \varphi_1}|(\theta_1,\theta_2);(\varphi_1,\varphi_2)\rangle_{e,B}\,,
    \end{split}
    \end{equation}
and also
    \begin{equation}
    \begin{split}
        &|(\theta_1,\theta_2);(\varphi_1+1,\varphi_2)\rangle_{e,B}=e^{-\pi i \theta_2}|(\theta_1,\theta_2);(\varphi_1,\varphi_2)\rangle_{e,B}\,,\\
        &|(\theta_1,\theta_2);(\varphi_1,\varphi_2+1)\rangle_{e,B}=e^{\pi i \theta_1}|(\theta_1,\theta_2);(\varphi_1,\varphi_2)\rangle_{e,B}\,,
    \end{split}
    \end{equation}
which are parallel to the anomalous phase satisfied by the rank-one partition functions as discussed in \eqref{eq:moduli-boundary-condition-1} and \eqref{eq:moduli-boundary-condition-2}.

Finally, let us compute the inner product between two general states. Begin with
    \begin{equation}
    \begin{split}
        &_{e,B}\langle(\theta'_1,\theta'_2);(\varphi'_1,\varphi'_2)| (\theta_1,\theta_2),(\varphi_1,\varphi_2)\rangle_{e,B}\\
        =&\bigoplus_{n,m,n',m'}e^{2\pi i (\theta'_2 \cdot (m'-\frac{1}{2}\varphi'_1)+\varphi'_2 \cdot (n'-\frac{1}{2}\theta'_1))}e^{-2\pi i (\theta_2 \cdot (m-\frac{1}{2}\varphi_1)+\varphi_2 \cdot (n-\frac{1}{2}\theta_1))}\\
        &\langle W_{(n'-\theta'_1)\cdot e,(m'-\varphi'_1)\cdot e^{-1}+(n'-\theta'_1)\cdot Be^{-1}} |W_{(n-\theta_1)\cdot e,(m-\varphi_1)\cdot e^{-1}+(n-\theta_1)\cdot Be^{-1}}\rangle\\
        =&\sum_{n,m,n',m'}e^{2\pi i (\theta'_2 \cdot (m'-\frac{1}{2}\varphi'_1)+\varphi'_2 \cdot (n'-\frac{1}{2}\theta'_1))}e^{-2\pi i (\theta_2 \cdot (m-\frac{1}{2}\varphi_1)+\varphi_2 \cdot (n-\frac{1}{2}\theta_1))}\\
        & \times \delta^{(D)}\left((n'-n)\cdot e-(\theta'_1-\theta_1)\cdot e\right)\\
        &\times \delta^{(D)}\left((m'-\varphi'_1)\cdot e^{-1}-(m-\varphi_1)\cdot e^{-1}+(n'-\theta'_1)\cdot Be^{-1}-(n-\theta_1)\cdot Be^{-1}\right)\\
        =&\sum_{n,m,n',m'}e^{2\pi i (\theta'_2 \cdot (m'-\frac{1}{2}\varphi'_1)+\varphi'_2 \cdot (n'-\frac{1}{2}\theta'_1))}e^{-2\pi i (\theta_2 \cdot (m-\frac{1}{2}\varphi_1)+\varphi_2 \cdot (n-\frac{1}{2}\theta_1))}\\
        &\times \delta^{(D)}\left((n'-n)-(\theta'_1-\theta_1)\right)\delta^{(D)}\left((m'-\varphi'_1)-(m-\varphi_1)+(n'-\theta'_1)\cdot B-(n-\theta_1)\cdot B\right)\,,
    \end{split}
    \end{equation}
where the Jacobians from the two Dirac delta functions cancel each others. Shifting $n'\rightarrow n'+n$ and $m'\rightarrow m'+m$, we have
\begin{equation}
    \begin{split}
        &\sum_{n,m,n',n'}e^{2\pi i (\theta'_2 \cdot (m'+m-\frac{1}{2}\varphi'_1)+\varphi'_2 \cdot (n'+n-\frac{1}{2}\theta'_1))}e^{-2\pi i (\theta_2 \cdot (m-\frac{1}{2}\varphi_1)+\varphi_2 \cdot (n-\frac{1}{2}\theta_1))}\\
        &\times \delta^{(D)}\left(n'-(\theta'_1-\theta_1)\right)\delta^{(D)}\left(m'-(\varphi'_1-\varphi_1)+(n'-(\theta'_1-\theta_1))\cdot B\right)\,,\\
        =&\sum_{n,m,n',n'}e^{2\pi i (\theta'_2 \cdot (m'+m-\frac{1}{2}\varphi'_1)+\varphi'_2 \cdot (n'+n-\frac{1}{2}\theta'_1))}e^{-2\pi i (\theta_2 \cdot (m-\frac{1}{2}\varphi_1)+\varphi_2 \cdot (n-\frac{1}{2}\theta_1))}\\
        &\times \delta^{(D)}\left(n'-(\theta'_1-\theta_1)\right)\delta^{(D)}\left(m'-(\varphi'_1-\varphi_1)\right)\,,
    \end{split}
\end{equation}
where the Dirac $\delta$-functions are independent of $n$ and $m$. The exponential term is written as
    \begin{equation}
    \begin{split}
        &e^{2\pi i (\theta'_2 \cdot (m'+m-\frac{1}{2}\varphi'_1)+\varphi'_2 \cdot (n'+n-\frac{1}{2}\theta'_1))}e^{-2\pi i (\theta_2 \cdot (m-\frac{1}{2}\varphi_1)+\varphi_2 \cdot (n-\frac{1}{2}\theta_1))}\\
        =&e^{2\pi i (\theta'_2 \cdot (m'-\frac{1}{2}\varphi'_1)+\varphi'_2 \cdot (n'-\frac{1}{2}\theta'_1))} e^{\pi i(\theta_1 \varphi_2 + \theta_2\varphi_1)} e^{2\pi i m(\theta'_2-\theta_2)} e^{2\pi i n(\varphi'_2-\varphi_2)}\,,
    \end{split}
    \end{equation}
and we can first sum over $m,n$ using
    \begin{equation}
        \sum_{m,n}e^{2\pi i m(\theta'_2-\theta_2)} e^{2\pi i n(\varphi'_2-\varphi_2)} = \delta^{(D)}_{P}(\theta'_2-\theta_2) \delta^{(D)}_{P}(\varphi'_2-\varphi_2)\,,
    \end{equation}
where we introduce the periodic Dirac $\delta$-function $\delta_P$, which is related to the ordinary Dirac $\delta$-function as
    \begin{equation}
        \delta^{(D)}_{P} (\theta) = \sum_{k\in \mathbb{Z}} \delta^{(D)}(k+\theta)\,.
    \end{equation}
Thus we write
    \begin{equation}
    \begin{split}
        &\sum_{n',m'}e^{2\pi i (\theta'_2 \cdot (m'-\frac{1}{2}\varphi'_1)+\varphi'_2 \cdot (n'-\frac{1}{2}\theta'_1))}e^{\pi i(\theta_1 \varphi_2 + \theta_2\varphi_1)}\delta^{(D)}_{P}(\theta'_2-\theta_2) \delta^{(D)}_{P}(\varphi'_2-\varphi_2)\\
        &\times \delta^{(D)}\left(n'-(\theta'_1-\theta_1)\right)\delta^{(D)}\left(m'-(\varphi'_1-\varphi_1))\right)\,,\\
        =&\sum_{n',m'}e^{2\pi i (m' \cdot \theta'_2+n'\cdot \varphi'_2)} e^{\pi i(\theta_1 \varphi_2 + \theta_2\varphi_1)-\pi i(\theta'_1 \varphi'_2 + \theta'_2\varphi'_1)}\delta^{(D)}_{P}(\theta'_2-\theta_2) \delta^{(D)}_{P}(\varphi'_2-\varphi_2)\\
        &\times \delta^{(D)}\left(n'-(\theta'_1-\theta_1)\right)\delta^{(D)}\left(m'-(\varphi'_1-\varphi_1))\right)\,.
    \end{split}
    \end{equation}
To evaluate the remaining sum, let us rewrite the $\delta$-functions into the integral form
    \begin{equation}
    \begin{split}
        &\sum_{n',m'} e^{2\pi i (m' \cdot \theta_2+n'\cdot \varphi_2)}\delta^{(D)}\left(n'-(\theta'_1-\theta_1)\right)\delta^{(D)}\left(m'-(\varphi'_1-\varphi_1))\right)\\
        =&\sum_{n',m'} e^{2\pi i (m' \cdot \theta_2+n'\cdot \varphi_2)} \int_{-\infty}^{+\infty} dz \int_{-\infty}^{+\infty} dz' e^{2\pi i z\cdot(n'-(\theta'_1-\theta_1))} e^{2\pi i z' \cdot (m'-(\varphi'_1-\varphi_1))}\\
        =&\int_{-\infty}^{+\infty} dz \int_{-\infty}^{+\infty} dz' \sum_{n',m'} e^{2\pi i n'\cdot(z+\varphi_2)} e^{2\pi im' \cdot(z'+\theta_2)}e^{2\pi i z\cdot(\theta_1-\theta'_1)+2\pi i z' \cdot (\varphi_1-\varphi'_1)}\\
        =&\int_{-\infty}^{+\infty} dz \int_{-\infty}^{+\infty} dz' \delta^{(D)}_{P} (z+\varphi_2) \delta^{(D)}_{P} (z'+\theta_2) e^{2\pi i z\cdot(\theta_1-\theta'_1)+2\pi i z' \cdot (\varphi_1-\varphi'_1)}\,,
    \end{split}
    \end{equation}
where the periodic Dirac $\delta$-function will restrict $z=k-\varphi_2$ and $z'=k'-\theta_2$ for any $k,k'\in \mathbb{Z}$, and we have
    \begin{equation}
    \begin{split}
        &\sum_{k,k'} e^{2\pi i (k-\varphi_2)\cdot(\theta_1-\theta'_1)} e^{2\pi i (k'-\theta_2)\cdot(\varphi_1-\varphi'_1)}\\
        =&\delta^{(D)}_P(\theta_1-\theta'_1) \delta^{(D)}_P(\varphi_1-\varphi'_1) e^{-2\pi i \varphi_2(\theta_1-\theta'_1)-2\pi i \theta_2(\varphi_1-\varphi'_1)}\,.
    \end{split}
    \end{equation}
Therefore we have
    \begin{equation}
        \begin{split}
        &_{e,B}\langle(\theta'_1,\theta'_2);(\varphi'_1,\varphi'_2)| (\theta_1,\theta_2),(\varphi_1,\varphi_2)\rangle_{e,B}\\
        =&e^{\pi i(\theta_1 \varphi_2 + \theta_2\varphi_1)-\pi i(\theta'_1 \varphi'_2 + \theta'_2\varphi'_1)} e^{-2\pi i \varphi_2(\theta_1-\theta'_1)-2\pi i \theta_2(\varphi_1-\varphi'_1)}\\
        &\times\delta^{(D)}_P(\theta_1-\theta'_1) \delta^{(D)}_P(\varphi_1-\varphi'_1)\delta^{(D)}_{P}(\theta_2-\theta'_2) \delta^{(D)}_{P}(\varphi_2-\varphi'_2)\,.
        \end{split}
    \end{equation}    
Since all holonomies are chosen within $[0,1)$, we have
    \begin{equation}
        -1<\theta_i -\theta'_i<1\,, \quad -1 <\varphi_i - \varphi'_i<1\,, \quad (i=1,2)
    \end{equation}
and $\delta$-functions impose
    \begin{equation}
        \theta_i=\theta'_i\,, \quad \varphi_i=\varphi'_i\,,
    \end{equation}
so that the additional phase factor vanishes. Therefore we obtain the desired orthogonality relation
    \begin{equation}
    \begin{split}
    &_{e,B}\langle(\theta'_1,\theta'_2);(\varphi'_1,\varphi'_2)| (\theta_1,\theta_2),(\varphi_1,\varphi_2)\rangle_{e,B}\\
    =&\delta^{(D)}(\theta_1-\theta'_1) \delta^{(D)}(\varphi_1-\varphi'_1)\delta^{(D)}(\theta'_2-\theta_2) \delta^{(D)}(\varphi'_2-\varphi_2)\,,
    \end{split}
    \end{equation} 
with $\theta_i,\varphi_i,\theta'_i,\varphi'_i \in [0,1)$.

\subsection{Physical boundary and partition functions}

In the last subsection, we will turn to the discussion of the physical boundary state $|\mathfrak{B}_{\text{phys}}\rangle$ and to recover the partition functions corresponding to different choices of symmetry boundary states.

Let us define the physical boundary state via the partition vector as
\begin{equation}
    \begin{split}
    |\mathfrak{B}_{\text{phys}}\rangle =& \int_{-\infty}^{+\infty}dx_1 dx_2 Z_{\text{Dir}}[x_1,x_2] |x_1,x_2\rangle_{\text{Dir}}\\
    =&\int_{-\infty}^{+\infty} dx_1 dx_2 dz Z_{\text{Dir}}[x_1,x_2] e^{-2\pi i x_2 \cdot z}| W_{z,-x_1}\rangle\,,
    \end{split}
\end{equation}
where we expand using the Dirichlet basis, and the coefficient $Z_{\text{Dir}}[x_1,x_2]$ is the partition function of physical theory with a general $\mathbb{R}^D$ flat backgrounds, namely the $\mathbb{R}^D$-valued holonomies $x_1,x_2$ along the two 1-cycles of the worldsheet torus. Here we choose the partition function for the Dirichlet boundary to be
    \begin{equation}\label{eq:partition_function_Dir}
        Z_{\text{Dir}}[x_1,x_2]= \frac{1}{(\eta\bar{\eta})^D}\left(\frac{1}{2\tau_2}\right)^{\frac{D}{2}} e^{-\frac{\pi}{2\tau_2}|x_1+\tau x_2|^2}\,,
    \end{equation}
which is the higher rank generalization of \eqref{eq:partition_function_gauge_U1}.

First, let us consider the projection onto the Neumann topological boundary
    \begin{equation}
        _{\text{Neu}}\langle y_1,y_2| \mathfrak{B}_{\text{phys}}\rangle\,,
    \end{equation}
where $|y_1,y_2\rangle_{\text{Neu}}$ is defined in \eqref{eq:Neu_boundary_state}. Using the relation \eqref{eq:relation_Dir_Neu}, we have
    \begin{equation}
        \begin{split}
        &_{\text{Neu}}\langle y_1,y_2| \mathfrak{B}_{\text{phys}}\rangle =\int dx_1 dx_2  e^{-2\pi i (x_1 \cdot y_2 - x_2 \cdot y_1)}Z[x_1,x_2]\\
        =&\frac{1}{(\eta\bar{\eta})^D} \left(\frac{1}{2\tau_2} \right)^{\frac{D}{2}}\int dx_1 d x_2  e^{-2\pi i (x_1 \cdot y_2 - x_2 \cdot y_1)} e^{-\frac{\pi}{2\tau_2}|x_1+\tau x_2|^2}\\
        =&\frac{1}{(\eta\bar{\eta})^D} \left(\frac{1}{2\tau_2} \right)^{\frac{D}{2}}\int dx_1 d x_2  e^{-2\pi i (x_1 \cdot y_2 - x_2 \cdot y_1)} e^{-\frac{\pi}{2\tau_2}(x_1^2+|\tau|^2 x_2^2 +2 \tau_1 x_1 \cdot x_2)}\,,
        \end{split}
    \end{equation}
where the exponential terms are
    \begin{equation}
        \exp\left[-\frac{\pi}{2\tau_2} \left(x_1 +2i\tau_2 y_2+\tau_1 x_2 \right)^2 + \frac{\pi}{2\tau_2}\left(2i \tau_2 y_2 + \tau_1 x_2\right)^2 -\frac{\pi}{2\tau_2} |\tau|^2 x_2^2  \right] \exp\left(2\pi i x_2 \cdot y_1 \right)\,.
    \end{equation}
Perform Gauss integral on $x_1$ gives
    \begin{equation}
        \frac{1}{(\eta\bar{\eta})^D} \int  d x_2  \exp\left[ \frac{\pi}{2\tau_2}(2i \tau_2 y_2 + \tau_1 x_2)^2 -\frac{\pi}{2\tau_2} |\tau|^2 x_2^2  \right] \exp\left(2\pi i x_2 \cdot y_1 \right)\,,
    \end{equation}
where the exponential term is further written as
    \begin{equation}
    \begin{split}
        &\exp\left[\frac{\pi}{2\tau_2}\left(-4 \tau_2^2 y_2^2+ 4i \tau_2 \tau_1 x_2\cdot y_2 -\tau_2^2 x_2^2 +4i\tau_2 x_2 \cdot y_1 \right)  \right]\\
        =&\exp\left[-\frac{2\pi}{\tau_2}\left( \tau_2^2 y_2^2- i \tau_2 \tau_1 x_2\cdot y_2+\frac{1}{4}\tau_2^2 x_2^2 -i\tau_2 x_2 \cdot y_1 \right)  \right]\\
        =&\exp\left[-\frac{2\pi}{\tau_2}\left(\frac{1}{2}\tau_2 x_2-i y_1-i \tau_1 y_2\right)^2-\frac{2\pi}{\tau_2}\left(  y_1+\tau_1 y_2\right)^2-\frac{2\pi}{\tau_2} (\tau_2^2 y_2^2)\right]\,.
    \end{split}
    \end{equation}
After integrating $x_2$, we have
    \begin{equation}
    \begin{split}
        _{\text{Neu}}\langle y_1,y_2| \mathfrak{B}_{\text{phys}}\rangle=&\frac{1}{(\eta\bar{\eta})^D}\left(\frac{2}{\tau_2}\right)^{\frac{D}{2}} \exp\left[-\frac{2\pi}{\tau_2}\left(  y_1+\tau_1 y_2\right)^2-\frac{2\pi}{\tau_2} (\tau_2^2 y_2^2)\right]\,,\\
        =&\frac{1}{(\eta\bar{\eta})^D}\left(\frac{2}{\tau_2}\right)^{\frac{D}{2}} \exp \left[-\frac{2\pi}{\tau_2}|y_1+\tau y_2|^2 \right]\,,
    \end{split}
    \end{equation}
which is dual partition function after gauging the $\mathbb{R}^D$ symmetry.

We would also like to compute $_{e,B}\langle \Omega | \mathfrak{B}_{\text{phys}}\rangle$, where $_{e,B}\langle \Omega |$ is the vacuum polarization defined in \eqref{eq:topological_boundary_state_e_B}. Here we consider the more general state $|\Omega\rangle_{\mathcal{E}}$ where the geometric data is encoded in $\mathcal{E}$ such that
    \begin{equation}
        |\Omega\rangle_{\mathcal{E}} := \sum_{n,m} | W_{(n,m) \mathcal{E}} \rangle \equiv \sum_{n,m} | W_{(\alpha,\beta)} \rangle \,,
    \end{equation}
where we denote $(\alpha,\beta) = (n,m) \mathcal{E}$. Then one has
    \begin{equation}
        \begin{split}
            &_{\mathcal{E}}\langle \Omega | \chi\rangle = \sum_{m,n}\int dx_1dx_2dz Z_{\text{Dir}}[x_1,x_2] e^{-2\pi i x_2 \cdot z}\langle W_{\alpha,\beta}| W_{z,-x_1}\rangle\\
            =&\sum_{m,n}\int dx_1dx_2dz Z_{\text{Dir}}[x_1,x_2] e^{-2\pi i x_2 \cdot z} \delta^{(D)}(x_1+\beta) \delta^{(D)}(z-\alpha)\\
            =&\sum_{m,n}\int dx_2 Z_{\text{Dir}}[-\beta,x_2] e^{-2\pi i x_2 \cdot \alpha}\,.
        \end{split}
    \end{equation}
Now substitute \eqref{eq:partition_function_Dir}, for a fixed pair of $(m,n)$, we need to compute
    \begin{equation}
    \begin{split}
        I(\alpha,\beta) =& \left(\frac{1}{2\tau_2}\right)^{\frac{D}{2}}  \int dx_2  \exp \left(-\frac{\pi}{2\tau_2} |\beta-\tau x_2|^2 -2\pi i x_2 \cdot \alpha  \right)\\
        =& \left(\frac{1}{2\tau_2}\right)^{\frac{D}{2}}  \int dx_2  \exp \left(-\frac{\pi}{2\tau_2} (\beta^2-2\tau_1 \beta\cdot x_2+|\tau|^2 x_2^2) -2\pi i x_2 \cdot \alpha  \right)\\
        =&\left(\frac{1}{2\tau_2}\right)^{\frac{D}{2}}  \int dx_2  \exp \left(-\frac{\pi|\tau|^2}{2\tau_2} x_2^2 + \frac{\pi \tau_1}{\tau_2} \beta \cdot x_2  -2\pi i \alpha \cdot x_2-\frac{\pi}{2\tau_2}\beta^2 \right)\\
        =&\left(\frac{1}{2\tau_2}\right)^{\frac{D}{2}} \int dx_2  \exp \left[-\frac{\pi|\tau|^2}{2\tau_2}\left(x_2 - \frac{\tau_1}{|\tau|^2}\beta + \frac{2i \tau_2}{|\tau|^2} \alpha \right)^2\right]\\
        &\times \exp \left[\frac{\pi |\tau|^2}{2\tau_2}\left(-\frac{\tau_1}{|\tau|^2}\beta + \frac{2i \tau_2}{|\tau|^2} \alpha \right)^2-\frac{\pi}{2\tau_2}\beta^2 \right]\,.
    \end{split}
    \end{equation}
Performing the Gauss integral
    \begin{equation}
        \int dx_2  \exp \left[-\frac{\pi|\tau|^2}{2\tau_2}\left(x_2 - \frac{\tau_1}{|\tau|^2}\beta + \frac{2i \tau_2}{|\tau|^2} \alpha \right)^2\right] = \left(\frac{2\tau_2}{ |\tau|^2} \right)^{\frac{D}{2}}\,,
    \end{equation}
we arrive at
    \begin{equation}
        I(\alpha,\beta)=\frac{1}{|\tau|^D} \exp \left(-\frac{2\pi \tau_2}{|\tau|^2}\alpha^2 -\frac{\pi \tau_2}{2|\tau|^2}\beta^2-\frac{2\pi i \tau_1}{|\tau|^2} \alpha \cdot \beta \right)\,.
    \end{equation}
Denote the $\tau'=-1/\tau$ and using
    \begin{equation}
        \tau'_1 = -\frac{\tau_1}{|\tau|^2}\,,\quad \tau'_2 = \frac{\tau_2}{|\tau|^2}\,, \quad |\eta(\tau')|^{2D} = |\eta(\tau)|^{2D} |\tau|^{-D}\,,
    \end{equation}
and we have
    \begin{equation}
        _{\mathcal{E}}\langle \Omega | \chi\rangle  = \frac{1}{|\eta(\tau')|^{2D}} \exp \left(-2\pi \tau'_2 \alpha^2 - \frac{\pi}{2} \tau'_2 \beta^2 +2\pi i \tau'_1 \alpha \cdot \beta \right)\,.
    \end{equation}
Let us introduce the left and right momenta as
    \begin{equation}
        p_L = \sqrt{2} \alpha + \frac{1}{\sqrt{2}} \beta\,, \quad p_R = \sqrt{2} \alpha - \frac{1}{\sqrt{2}} \beta\,,
    \end{equation}
and the inner product is then 
    \begin{equation}
        _{\mathcal{E}}\langle \Omega | \chi\rangle  = \frac{1}{|\eta(\tau')|^{2D}}  \sum_{n,m} q^{\frac{1}{4} p_L^2} \bar{q}^{\frac{1}{4}p_R^2}\,,
    \end{equation}
with $q=e^{2\pi i \tau'}$, and we recover the partition function of the rank-$D$ compact boson. Moreover, the partition function is manifestly invariant under the $O_L(D;\mathbb{R})\times O_R(D;\mathbb{R})$ rotation acting on $p_L$ and $p_R$, respectively. Therefore the different Lagrangian algebras related by the $O_L(D;\mathbb{R})\times O_R(D;\mathbb{R})$ rotations indeed give the equivalent physical theory. 

\section{Conclusion}
In this paper, we studied flat gauging of continuous symmetries in two-dimensional conformal field theory at the level of torus partition functions. For the compact boson, we constructed the partition function with general flat $U(1)_M\times U(1)_W$ backgrounds and exhibited its anomalous transformations and modular properties. We then showed that flat gauging either $U(1)_M$ or $U(1)_W$ decompactifies the theory: the $U(1)$ projected sector together with the continuous twisted sectors gives the non-compact free boson, and the remaining compact $U(1)$ background combines with the dual $\mathbb{Z}$ background into a non-compact $\mathbb{R}$ symmetry background. At the self-dual radius, where the compact boson is equivalent to the $\widehat{\mathfrak{su}(2)}_1$ WZW model, we revisited the flat gauging of the diagonal $SO(3)$ subgroup, which leads to a theory outside the usual $c=1$ moduli space. Although the contribution from the twisted sectors resembles that of the orbifolded non-compact boson, the untwisted sector is different. We also studied the orbifold branch, where the continuous $U(1)$ symmetry defects become continuous families of non-invertible defects. For finite non-invertible symmetries, the corresponding Frobenius algebra gauging changes the radius of the orbifold theory. On the other hand, the continuous flat gauging is more subtle and receives a zero-measure contribution from the singular loci. Finally, we formulated the compact boson and its higher-rank generalization in terms of a non-compact BF SymTFT, where the geometric data of the target space are encoded by Lagrangian algebras/topological boundary states, and the Narain moduli space together with the $O(D,D;\mathbb{Z})$ T-duality action can be recovered from the redundancy of these boundary data.

There are also several future directions. In section~\ref{sec:non-invertible}, we introduced a regularization scheme for the singular loci in the moduli space of continuous non-invertible defect networks. It would be interesting to understand the origin of this prescription more intrinsically, and to formulate flat gauging for more general continuous non-invertible symmetries. Just as discrete torsion in finite orbifolds amounts to a choice of topological phases weighting the twisted sectors and can change the resulting gauged theory, one may ask whether continuous orbifolds, or flat gaugings, admit analogous torsion-like phases or measure refinements, now constrained by the anomaly interplay between the gauged continuous subgroup and the ambient symmetry.
In this paper we mainly focused on the two-dimensional compact boson. It would be useful to study flat gauging in other two-dimensional CFTs, or higher-dimensional quantum field theories, and to see whether this operation can produce new theories in a systematic way. Another natural direction is to extend the SymTFT description so that the circle branch and the orbifold branch of the compact boson are treated in a unified framework.

\acknowledgments

The authors would like to thank Yuan Miao, Yuji Tachikawa, Yi-Nan Wang, Xingyang Yu and Hao Y. Zhang for discussions. QJ is supported by National Research Foundation of Korea (NRF) Grant No. RS-2024-00405629 and Jang Young-Sil Fellow Program at the Korea Advanced Institute of Science and Technology. YZ is supported by WPI Initiative, MEXT, Japan at Kavli IPMU, the University of Tokyo.

\appendix

\section{Modular property of the partition function $Z[(\alpha,\beta),(\mu,\nu)]$}

In this appendix, we will check the modular covariance \eqref{eq:modular-covariance}. In terms of the unit-period variables, the same partition function is
\begin{equation}
    \begin{split}
    Z[(\widetilde{\alpha},\widetilde{\beta}),(\widetilde{\mu},\widetilde{\nu})]=&\frac{1}{\eta \bar{\eta}} \sum_{n,m={-\infty}}^{\infty} e^{-\pi i (\alpha \nu+\mu \beta)} e^{2\pi i \alpha (m+\nu)+2\pi i \mu(n+\beta)}\\
    &\times q^{\frac{1}{2}(\frac{(m+\nu)}{2r}+(n+\beta)r)^2}\bar{q}^{\frac{1}{2}(\frac{(m+\nu)}{2r}-(n+\beta)r)^2}\,,
    \end{split}
\end{equation}
For modular transformation, it is more useful to apply Poisson resummation to the momentum number. We write
\begin{equation}
\begin{split}
    &\frac{1}{\eta \bar{\eta}} \int dx \sum_{n,m} \delta(x-m-\nu) e^{-\pi i (\alpha \nu+\mu \beta)}e^{2\pi i \alpha x+2\pi i \mu(n+\beta)} q^{\frac{1}{2}(\frac{x}{2r}+(n+\beta)r)^2}\bar{q}^{\frac{1}{2}(\frac{x}{2r}-(n+\beta)r)^2}\\
    =&\frac{1}{\eta \bar{\eta}}e^{-\pi i (\alpha \nu+\mu \beta)} \int dx \sum_{n,s}  e^{2\pi i (x-\nu)s} e^{2\pi i \alpha x+2\pi i \mu(n+\beta)} q^{\frac{1}{2}(\frac{x}{2r}+(n+\beta)r)^2}\bar{q}^{\frac{1}{2}(\frac{x}{2r}-(n+\beta)r)^2}
\end{split}
\end{equation}
where
    \begin{equation}
        \sum_{m} \delta(x-m-\nu) = \sum_s e^{2\pi i (x-\nu)s}\,.
    \end{equation}
The exponent, apart from the first phase factor $e^{-\pi i (\alpha \nu+\mu \beta)}$, can be reorganized as a Gaussian term in $x$,
    \begin{equation}
    \begin{split}
        &\exp \left[2\pi i\left((x-\nu)s+\alpha x+\mu(n+\beta) +\frac{1}{2}(\tau -\bar{\tau})\left(\frac{x^2}{4r^2}+(n+\beta)^2r^2 \right)+\frac{1}{2}(\tau+\bar{\tau})(n+\beta)x\right) \right]\\
        =&\exp \left[2\pi i\left((x-\nu)s+\alpha x+\mu(n+\beta) +i \tau_2 \left(\frac{x^2}{4r^2}+(n+\beta)^2r^2 \right)+\tau_1(n+\beta)x\right) \right]\\
        =&\exp \left[-\frac{\pi \tau_2}{2 r^2}x^2 + 2\pi i (s+\alpha+\tau_1(n+\beta))x + 2\pi i(\mu (n+\beta)-s \nu + i \tau_2 (n+\beta)^2 r^2)  \right]\\
        =&\exp \left[-\frac{\pi \tau_2}{2 r^2} \left(x- \frac{\pi i (s+\alpha+\tau_1(n+\beta))}{\frac{\pi \tau_2}{2 r^2}} \right)^2 + \frac{\pi \tau_2}{2 r^2}\left(\frac{\pi i (s+\alpha+\tau_1(n+\beta))}{\frac{\pi \tau_2}{2 r^2}}\right)^2\right.\\
        &\qquad\left.+ 2\pi i(\mu (n+\beta)-s \nu + i \tau_2 (n+\beta)^2 r^2)\right]\\
        =&\exp \left[-\frac{\pi \tau_2}{2 r^2} \left(x- \frac{\pi i (s+\alpha+\tau_1(n+\beta))}{\frac{\pi \tau_2}{2 r^2}} \right)^2\right]\exp\left[-\frac{2\pi r^2}{ \tau_2}|s+\alpha + (n+\beta)\tau|^2 \right]\\
        &\qquad \times \exp (2\pi i \left((n+\beta)\mu - s \nu \right))\,.
    \end{split}
    \end{equation}
After integrating over $x$, one obtains
    \begin{equation}
        Z[(\widetilde{\alpha},\widetilde{\beta}),(\widetilde{\mu},\widetilde{\nu})]=\frac{\sqrt{2}r}{\sqrt{\tau_2}\eta \bar{\eta}} \sum_{s,n\in \mathbb{Z}}\exp\left[-\frac{2\pi r^2}{ \tau_2}|(s+\alpha) + (n+\beta)\tau|^2 \right] e^{2\pi i (n+\frac{1}{2}\beta)\mu - 2\pi i (s+\frac{1}{2}\alpha)\nu} \,.
    \end{equation}
    
This is the form in which the modular covariance is almost manifest. For the $T$-transformation $\tau\rightarrow\tau +1$, one has
    \begin{equation}
    \begin{split}
        &Z[(\widetilde{\alpha},\widetilde{\beta}),(\widetilde{\mu},\widetilde{\nu})](\tau+1)\\
        =&\frac{\sqrt{2}r}{\sqrt{\tau_2}\eta \bar{\eta}} \sum_{s,n\in \mathbb{Z}}\exp\left[-\frac{2\pi r^2}{ \tau_2}|(s+n+\alpha+\beta) + (n+\beta)\tau|^2 \right] e^{2\pi i (n+\frac{1}{2}\beta)\mu - 2\pi i (s+\frac{1}{2}\alpha)\nu}\\
        =&\frac{\sqrt{2}r}{\sqrt{\tau_2}\eta \bar{\eta}} \sum_{s,n\in \mathbb{Z}}\exp\left[-\frac{2\pi r^2}{ \tau_2}|(s+\alpha+\beta) + (n+\beta)\tau|^2 \right] e^{2\pi i (n+\frac{1}{2}\beta)\mu - 2\pi i (s-n+\frac{1}{2}\alpha)\nu}\\
        =&\frac{\sqrt{2}r}{\sqrt{\tau_2}\eta \bar{\eta}} \sum_{s,n\in \mathbb{Z}}\exp\left[-\frac{2\pi r^2}{ \tau_2}|(s+\alpha+\beta) + (n+\beta)\tau|^2 \right] e^{2\pi i (n+\frac{1}{2}\beta)(\mu+\nu) - 2\pi i (s+\frac{1}{2}\alpha + \frac{1}{2}\beta)\nu}\\
        =& Z[(\widetilde{\alpha}+\widetilde{\beta},\widetilde{\beta}),(\widetilde{\mu}+\widetilde{\nu},\widetilde{\nu})](\tau)\,.
    \end{split}
    \end{equation}
For the $S$-transformation $\tau\rightarrow -1/\tau$, we use
    \begin{equation}
        \eta(-1/\tau) = \sqrt{-i\tau} \eta(\tau)\,, \quad \tau_2(-1/\tau) = \frac{\tau_2}{|\tau|^2}\,,
    \end{equation}
then
    \begin{equation}
        -\frac{2\pi r^2}{ \tau_2}|(s+\alpha) + (n+\beta)\tau|^2 = -\frac{2\pi r^2}{\tau_2}|(n+\beta)-(s+\alpha)\tau|^2\,,
    \end{equation}
and after relabeling $n\rightarrow s$ and $s\rightarrow-n$, one gets
    \begin{equation}
    \begin{split}
&Z[(\widetilde{\alpha},\widetilde{\beta}),(\widetilde{\mu},\widetilde{\nu})](-1/\tau)\\
        =&\frac{\sqrt{2}r}{\sqrt{\tau_2}\eta \bar{\eta}} \sum_{s,n\in \mathbb{Z}}\exp\left[-\frac{2\pi r^2}{ \tau_2}|(s+\beta) + (n-\alpha)\tau|^2 \right] e^{2\pi i (s+\frac{1}{2}\beta)\mu - 2\pi i (-n+\frac{1}{2}\alpha)\nu}\\
        =&\frac{\sqrt{2}r}{\sqrt{\tau_2}\eta \bar{\eta}} \sum_{s,n\in \mathbb{Z}}\exp\left[-\frac{2\pi r^2}{ \tau_2}|(s+\beta) + (n-\alpha)\tau|^2 \right] e^{2\pi i (n - (-\frac{1}{2}\alpha))\nu - 2\pi i (s+\frac{1}{2}\beta)(-\mu)}\\
        =&Z[(\widetilde{\beta},-\widetilde{\alpha}),(\widetilde{\nu},-\widetilde{\mu})](\tau)\,.
    \end{split}        
    \end{equation}
In summary, the partition function obeys
    \begin{equation}
    \begin{split}
        Z[(\widetilde{\alpha},\widetilde{\beta}),(\widetilde{\mu},\widetilde{\nu})](\tau+1)=&Z[(\widetilde{\alpha}+\widetilde{\beta},\widetilde{\beta}),(\widetilde{\mu}+\widetilde{\nu},\widetilde{\nu})](\tau)\,,\\
        Z[(\widetilde{\alpha},\widetilde{\beta}),(\widetilde{\mu},\widetilde{\nu})](-1/\tau)=&Z[(\widetilde{\beta},-\widetilde{\alpha}),(\widetilde{\nu},-\widetilde{\mu})](\tau)\,.
    \end{split}
    \end{equation}

\bibliographystyle{JHEP}
\bibliography{biblio.bib}

\end{document}